\documentclass[11pt,a4paper]{article}
\usepackage[margin=20mm]{geometry}
\usepackage{graphicx,xcolor,tikz}
\usepackage{amsmath,amssymb}
\usepackage{microtype}
\usepackage[numbers,sort&compress]{natbib}
\usepackage{xurl}
\usepackage[hidelinks]{hyperref}
\hypersetup{pdftitle={Blast freezing a black hole},pdfauthor={Shoaib Akhtar and Xiao-Liang Qi}}
\allowdisplaybreaks
\title{Blast freezing a black hole}
\author{Shoaib Akhtar\textsuperscript{1}\thanks{Corresponding author: \href{mailto:shoaib@stanford.edu}{shoaib@stanford.edu}.}
\quad Xiao-Liang Qi\textsuperscript{1,2}\thanks{Corresponding author: \href{mailto:xlqi@stanford.edu}{xlqi@stanford.edu}.}\\[0.5em]
{\small\textsuperscript{1}Leinweber Institute for Theoretical Physics, Stanford University}\\
{\small Stanford, CA 94305, USA}\\
{\small\textsuperscript{2}OpenAI}}
\date{}
\begin{document}
\maketitle
\begin{abstract}
What happens to information carried through an evaporating black-hole horizon? We introduce a solvable model of evaporation built from coupled Sachdev-Ye-Kitaev systems, in which an initially two-sided black hole is coupled at a finite time to a larger, colder bath. Evaporation is rapid in this model, so we refer to the process as ``blast freezing'' of a black hole. In an appropriate large-$N$ and large-$p$ limit, the two-point functions and certain four-point probes can be computed analytically. Using the two-point functions as input to a generalized HKLL reconstruction, we obtain the emergent bulk geometry of the evaporation process. We then track the information carried by an infalling particle using operator size and Renyi-2 mutual information, showing how it is preserved in nonlocal many-body degrees of freedom after the blast-freezing transition.
\end{abstract}

\section{Black-hole information and evaporation}

One of the deepest open questions in science is how to unify the two pillars of modern physics: quantum mechanics and general relativity. Black holes sharpen this tension. Quantum mechanics predicts unitary time evolution, while classical general relativity predicts that the black hole horizon is a one-way surface that information enters and never leaves. This is known as the black hole information paradox. There are different types of information paradoxes. The ``old'' paradox proposed by Hawking\cite{hawking1976breakdown} concerns the fate of infalling information after the black hole evaporates: quantum mechanics says it should be preserved in the Hawking radiation, whereas the semiclassical radiation appears thermal. This paradox is resolved by a modified gravitational path integral approach\cite{penington2020entanglement,penington2022replica,almheiri2019entropy,almheiri2020replica}. However, the firewall paradox\cite{almheiri2013black,almheiri2013apologia}, which concerns the experience of an infalling observer, remains open. General relativity predicts that the observer feels nothing special when crossing the horizon, while quantum mechanics predicts that, at a late enough time, information in the interior is already known to the exterior Hawking radiation. If an interior particle is completely entangled with a nearby partner, it cannot also be entangled with the radiation, so the local quantum field theory description appears to fail. To resolve this paradox, a more explicit description of quantum dynamics in the black hole interior is needed.

\begin{figure}[htbp]
	\centering
	\includegraphics[width=0.30\textwidth]{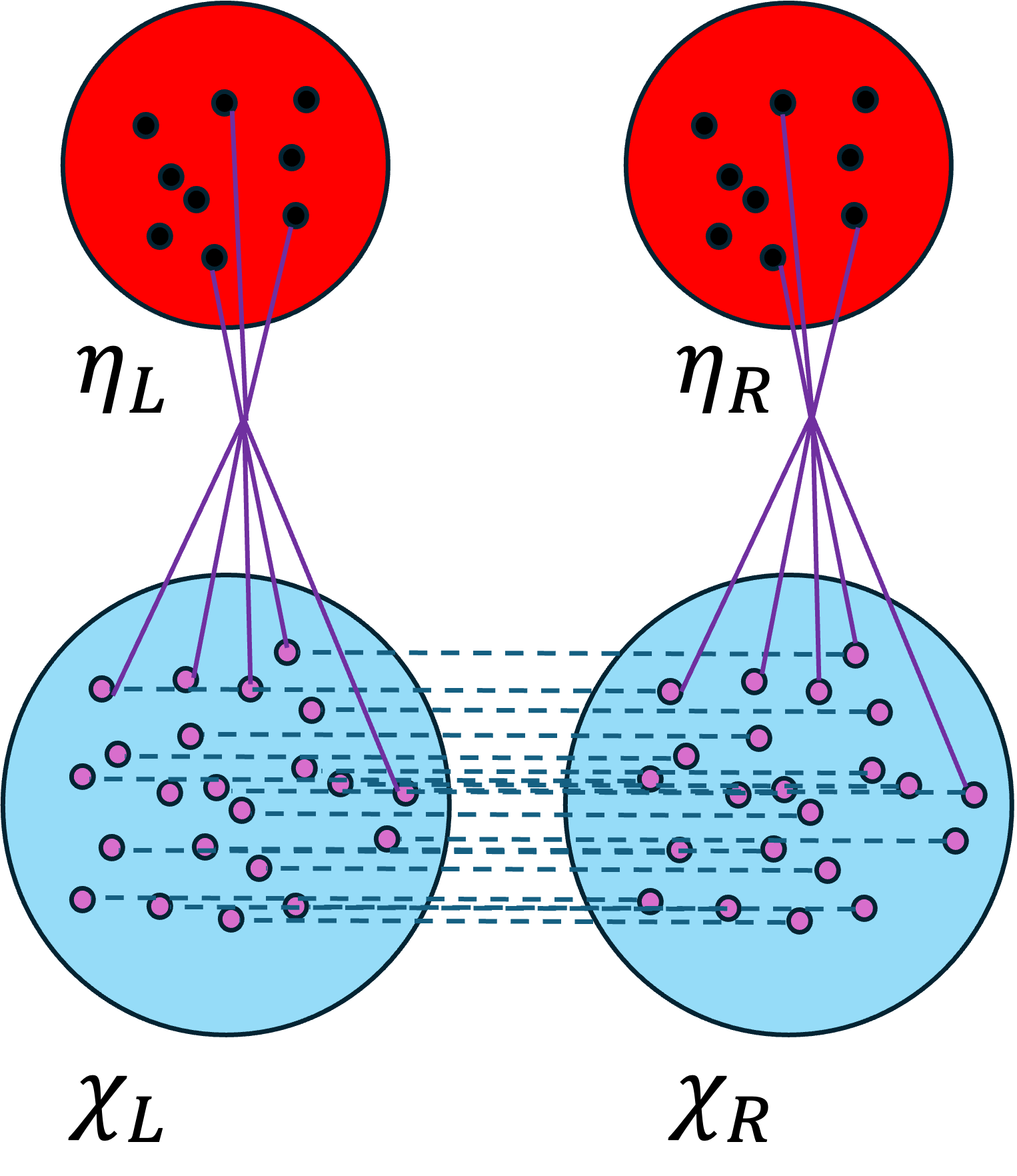}
	\caption{Illustration of the model setup. The dashed lines illustrate the bilinear coupling between $\chi$ fermions, while the solid lines illustrate the SYK coupling between $\eta$ and $\chi$ fermions, which is only turned on after the evaporation time $t_{\rm ev}$.}\label{fig:setup}
\end{figure}

In this paper, we provide new insight into the black hole information paradox by proposing a new model of a $1+1$-dimensional evaporating black hole. Our proposal is based on holographic duality, also known as the Anti-de Sitter/Conformal field theory (AdS/CFT) correspondence\cite{maldacena1999large}. Holographic duality is the mapping between a theory of gravity in $d+1$ dimensional spacetime and a quantum field theory in $d$ dimensional spacetime.
The Sachdev-Ye-Kitaev (SYK) model\cite{sachdev1993gapless,kitaev2015simple,maldacena2016remarks} is proposed as a $d=0$ example of this duality, where the boundary model describes $N$ interacting Majorana fermions, and the dual bulk theory is $1+1$-d Jackiw-Teitelboim (JT) gravity\cite{jackiw1985lower,teitelboim1983gravitation} (coupled with matter fields). In holographic duality, a thermal state of the boundary theory corresponds to a black hole in the bulk theory. Thermal correlation functions of fermions in the low temperature SYK model are reproduced by the fermion two-point function in the background of a $1+1$-dimensional AdS black hole. However, large black holes in AdS space do not evaporate, so to study black hole evaporation, we need to couple the SYK model to a bath system. The model we consider contains four SYK islands (Fig. \ref{fig:setup}). Two of them are entangled in a thermal-field double state of $N_\eta$ Majorana fermions $\eta_{iL},\eta_{iR}$, but they are not coupled. The other two islands each contain $N_\chi$ Majorana fermions $\chi_{iL},\chi_{iR}$. We consider $\eta$ fermions as a model of the black hole and the $\chi$ fermions as a model of the bath, so we are interested in the region $N_\chi >N_\eta$. The $\chi$ fermions are coupled by a bilinear coupling, which defines a Hamiltonian with a unique ground state and gapped excitations. It was shown that\cite{maldacena2018eternal} ground state physics in this coupled system is dual to a global AdS$_2$ geometry, which provides a good candidate for a cool bath system. Then we turn on a coupling between $\eta$ and $\chi$ fermions at some finite time $t_{\rm ev}$. The coupling leads to a transfer of energy and entropy from $\eta$ fermions to the cooler $\chi$ fermion system, which describes the black hole evaporation process. We will show that this model is solvable in the large $N_\eta,N_\chi$ limit. In particular, when all SYK islands have $p$-body interactions and \(s=N_\eta/(N_\eta+N_\chi)\), we consider the limit $p\rightarrow \infty, N_\eta\rightarrow \infty, N_\chi\rightarrow \infty, p^2/N_\eta\rightarrow 0, sp\rightarrow 0$. In this limit two-point functions can be solved analytically, which enables us to study many aspects of this model and its gravity dual. By applying the bulk reconstruction algorithm developed in Ref.\cite{nebabu2024bulk}, we obtain the $1+1$-d bulk fermion dynamics and the infalling bulk operators. We analyze the bulk geometry and show how the $\eta$ fermion experiences a quick transition from the black hole geometry to a vacuum with an end-of-the-world brane. This analytic solution also allows us to obtain further information about the fate of infalling operators at the horizon, which provides some new insight about the black hole information paradox. Similar models of the SYK model coupled with a bath have been studied before\cite{zhang2019evaporation,chen2020replica,gaikwad2023microscopic,bragagnolo2026probing}, but bulk reconstruction in such models has not been constructed from the boundary theory.

In the following, we will first define the model and then introduce its key properties.

\section{Definition of the Model}
The model we study has the following Hamiltonian:
\begin{equation}
        \begin{aligned}
H_{\rm tot}(t)&=H^L+H^R\\
&\quad+i\frac{\mu_\eta(t)}{p}\sum_{k=1}^{N_\eta}\eta_{Lk}\eta_{Rk}+i\frac{\mu_\chi(t)}{p}\sum_{k=1}^{N_\chi}\chi_{Lk}\chi_{Rk}.
\end{aligned}
\end{equation}
where $H^R=H[\eta^R,\chi^R]$ and $H^L=(-1)^{p/2}H[\eta^L,\chi^L]$ are time-dependent Hamiltonians, each of which describes two independent SYK islands that are merged into one bigger SYK island at time $t=t_{\rm ev}$. More precisely,
\begin{align}
	H[\eta,\chi]=\left\{\begin{array}{cc}H_{\rm SYK}[\eta]+H_{\rm SYK}[\chi],&t<t_{\rm ev}\\
										H_{\rm SYK}[\eta\oplus \chi],&t\geq t_{\rm ev}\end{array}\right.
\end{align}
Here $H_{\rm SYK}[\eta\oplus \chi]$ means we treat the $N_\eta+N_\chi$ fermions $\eta$ and $\chi$ together as a bigger SYK island. More details about the model definition are provided in Methods and Supplementary Text, Sec.~S1. The bilinear coupling distinguishes $\eta$ and $\chi$ dynamics. In most of the paper, we will consider $\mu_\eta(t)\equiv 0$ so that the $\eta$ fermions are in a two-sided black hole state before they are coupled with $\chi$, while $\mu_\chi(t)=\mu_0$ is a constant which allows the $\chi$ fermions to be in the traversable wormhole phase\cite{maldacena2018eternal}. Before the coupling is turned on, the $\eta$ fermions are in a time-evolved thermal field double state, which in the energy eigenstate basis has the form $\left|TFD(t)\right\rangle=Z_\beta^{-1/2}\sum_n e^{-\beta E_n/2}e^{-2iE_nt}\left|n\right\rangle_L\left|n\right\rangle_R$. The $\chi$ fermions are in the ground state of the traversable wormhole.

\section{The solvable limit}
We first take the large $N$ limit
\begin{align}
N_\eta\rightarrow \infty,\quad N_\chi\rightarrow \infty,\quad s=\frac{N_\eta}{N_\eta+N_\chi}={\rm const}.
\end{align}
In this limit, the dynamics of the system can be described by the Kadanoff-Baym (KB) equation, which is a self-consistent equation for the two-point functions of $\eta$ and $\chi$ fermions. For a simple SYK site, the self-energy is determined by its two-point function. Schematically $\Sigma\propto \mathcal{J}^2G^{p-1}$. In our model, the self-energy becomes a weighted sum of $G_\eta^kG_{\chi}^{p-1-k}$ kinds of terms, with a time-dependent weight.

Details about the KB equations are summarized in Methods and derived in Supplementary Text, Sec.~S2. Here we specialize to the simplest limit that retains the essential physics of black-hole evaporation:
\begin{align}
p\rightarrow \infty,\quad \frac{p^2}{N_\eta}\rightarrow 0,\quad sp\rightarrow 0
\end{align}
This is the large bath limit when $N_\chi \gg N_\eta p$. In this limit, the effect of $\chi$ fermions dominates the $\eta$ self-energy, while the back-reaction of $\eta$ fermions on $\chi$ fermions is negligible. This allows us to generalize the large-$p$ solution of the SYK model\cite{maldacena2016remarks}. In this limit, the two-point function has the following form:
\begin{displaymath}
	G^{\psi,>}(t_1,t_2)=-\frac{1}{2}
	\left(\begin{array}{cc}
	i e^{g^\psi_R(t_1,t_2)/p} & e^{g^\psi_L(t_1,t_2)/p}\\
	-e^{g^\psi_L(t_1,t_2)/p} & +i e^{g^\psi_R(t_1,t_2)/p}
	\end{array}\right).
\end{displaymath}
Here $\psi$ represents $\eta$ or $\chi$ fermions. The KB equation reduces to differential equations for $g^\psi_{L,R}$. $g^\chi_{R,L}$ satisfies the Liouville equation, which is the same as a stand-alone large-$p$ solution:
\begin{align}
	 \partial_{t_1}\partial_{t_2}g^\chi_{R,L}=\pm 2\mathcal{J}^2e^{g^\chi_{R,L}}\label{eq:large_p_chi}
\end{align}
For $\eta$ fermions, the right-hand side is a bit more complicated:
\begin{equation}
\partial_{t_1}\partial_{t_2}g^\eta_{R,L}=\left\{\begin{array}{cc}\pm2\mathcal{J}^2e^{g^\eta_{R,L}},&t_1,t_2<t_{\rm ev}\\\pm2\mathcal{J}^2e^{g^\chi_{R,L}},&t_1,t_2>t_{\rm ev}\\0,&\text{otherwise}\end{array}\right.\label{eq:large_p_eta}
\end{equation}
This equation illustrates that the self-energy of $\eta$ fermions is entirely determined by $\chi$ after the evaporation time. Furthermore, the self-energy becomes block-diagonal: the self-energy between the $t<t_{\rm ev}$ region and the $t>t_{\rm ev}$ region vanishes. Both are consequences of the fact that the most important terms in the Hamiltonian switched from $\eta^p$ types of terms to $\eta\chi^{p-1}$ types of terms. The solution of the $\chi$ equation has been studied before for generic couplings $\mu_\chi(t)$ with a thermal field double initial state\cite{lensky2021rescuing}. The $\eta$ equation (\ref{eq:large_p_eta}) can also be solved.

The probe-limit solutions for \(g^\chi\) and \(g^\eta\), including their matching across \(t_{\rm ev}\), are summarized in Methods and derived in Supplementary Text, Sec.~S3. The two-point function has the form
\begin{align}
	 g^\eta_{R(L)}(t_1,t_2)&=f_{1R(L)}(t_1)+f_{2R(L)}(t_2),\nonumber\\
&\qquad t_2<t_{\rm ev}<t_1,\label{eq:regionII}\\
g^\eta_{R(L)}(t_1,t_2)&=g^\chi_{R,L}(t_1,t_2)+f_{3R(L)}(t_1)\nonumber\\
&\quad+f^*_{3R(L)}(t_2),\quad t_1,t_2>t_{\rm ev}.\label{eq:regionIII}
\end{align}
With the ordering shown in Eq.~\eqref{eq:regionII}, \(f_2\) is fixed by the pre-evaporation $\eta$ solution, while \(f_1\) and \(f_3\) are fixed by the post-evaporation $\chi$ evolution.

The form of Eq. (\ref{eq:regionII}) has a deep consequence. The two-point function has the factorized form $e^{g^\eta_{R(L)}(t_1,t_2)/p}=e^{f_{1R(L)}(t_1)/p}e^{f_{2R(L)}(t_2)/p}$ in the region $t_2<t_{\rm ev}<t_1$. Consider the linear spaces $\mathbb{V}_{\rm past}$ spanned by $\eta(t),t<t_{\rm ev}$ and $\mathbb{V}_{\rm fut}$ spanned by $\eta(t),t>t_{\rm ev}$. The overlap between these two spaces is defined by the expectation value of the anti-commutator, $A_{ab}(t_1,t_2)=\langle\{\eta_a(t_1),\eta_b(t_2)\}\rangle=-2\operatorname{Im}G^{\eta,>}_{ab}(t_1,t_2)$ (Supplementary Text, Sec.~S3). It has rank $\leq4$ per flavor due to the factorized correlator. This means that the single particle states from the past are almost orthogonal to those detectable in the future, except at most four independent modes per flavor. This is consistent with the fact that information that falls into the black hole in the form of simple states appears to be lost in the large $N$ limit. When we consider finite $N$ effects, we see that such information actually returns in complex operators, which will be discussed below.
We believe this is the first black hole evaporation model where the information loss in the large-$N$ limit is analytically tractable from a boundary calculation.

\section{Bulk Reconstruction}

To turn such intuition into sharp quantitative statements, we would like to study the holographic dual of this model. In the large $N$ limit, $\eta$ is a generalized free field, and we can apply the bulk reconstruction algorithm developed in Refs.~\cite{nebabu2024bulk,nebabu2026two}. The algorithm constructs a $1+1$-d bulk free fermion theory with the boundary correlators identical to the boundary Wightman function \(C_{ab}(t_1,t_2)=\left\langle \eta_a(t_1)\eta_b(t_2)\right\rangle\). Importantly, up to a local basis choice, the bulk theory is uniquely determined by the boundary Wightman function. The bulk fermion is related to the boundary fermion by a linear kernel, which is a generalization of the Hamilton-Kabat-Lifschytz-Lowe (HKLL) construction\cite{hamilton2006holographic}. The boundary anti-commutator defines the inner product on the linear space of simple boundary operators, and the bulk reconstruction kernel is determined by the requirement of canonical anti-commutation relations for the bulk fermions. In general, this procedure has to be implemented numerically, but
analytic results can be obtained for a large class of time-translation invariant correlation functions\cite{nebabu2026two}. Interestingly, the reconstruction does not refer to an AdS background, and the bulk geometry obtained may have either sign of curvature. As a special case, this algorithm correctly reproduces the duality between the low temperature SYK model and AdS$_2$ geometry. Details of the construction are given in Methods.

\begin{figure}[htbp]
	\centering
	\includegraphics[width=\textwidth]{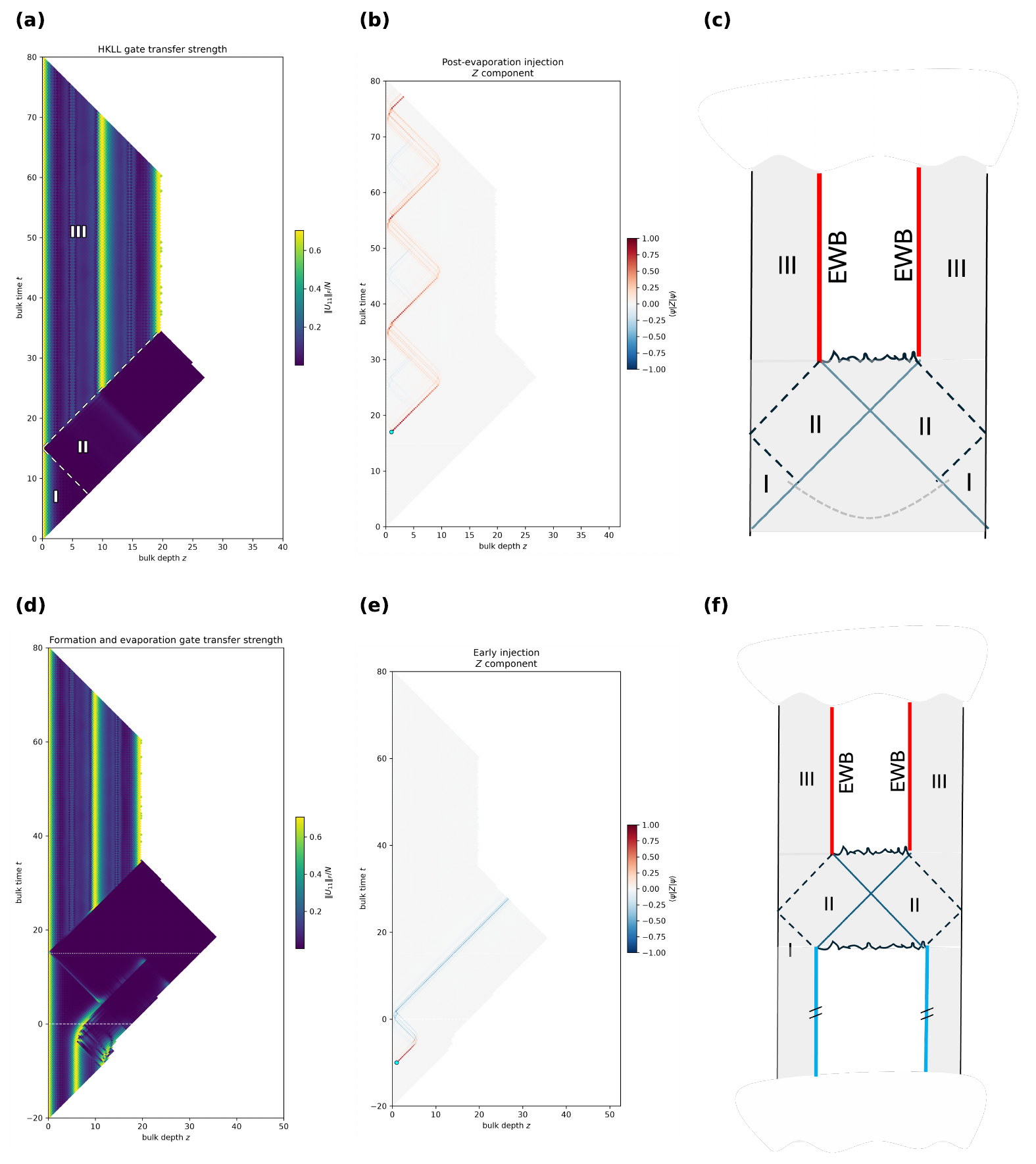}
	\caption{Bulk reconstruction of the blast freezing model. The color map in (a) represents reflection probability at the bulk coordinate $(u,v)$ when the state before evaporation is an eternal black hole. (b) shows the propagation of a wavepacket in this geometry when a particle is sent in after evaporation. The color is the polarization in the L, R components. The figure shows that a particle from the left does not reach the right boundary. (c) is the unfolded Penrose diagram illustration based on the results in (a) and (b). (d), (e), and (f) are the same plots for a different state. The initial state was a traversable wormhole and a black hole is formed at $t=0$.}\label{fig:bulk_geometry}
\end{figure}

One interesting aspect of this reconstruction algorithm that is relevant to the current work is that the boundary anti-commutator determines not only the bulk geometry but also its topology. When the anti-commutator matrix on a boundary time interval becomes singular, further expanding the interval does not generate new orthogonal bulk modes, and the reconstructed geometry ends. Applying this reconstruction algorithm to $\eta_a(t)$ fermions in our model leads to the geometry of Fig.~\ref{fig:bulk_geometry} (a). (Parameters used to plot all figures are provided in the last subsection of Methods.) Region I corresponds to the two-sided black hole geometry, region II to the transition region, and region III to the geometry after evaporation. Because the post-evaporation $\eta$ correlator inherits the oscillatory bath dynamics, the reconstructed geometry ends at finite depth. However, unlike the $\chi$ traversable wormhole, the $\eta$ geometry does not connect the left and right boundaries with appreciable probability. The evolution of a fermion wavepacket in Fig.~\ref{fig:bulk_geometry} (b) shows that the color never changes from red to blue, which implies that the left-boundary fermion never propagates to the right boundary. This is consistent with the analytic left-right correlator bound derived in Supplementary Text, Secs.~S3 and S6. Physically, this is because the coupling with $\chi$ cannot restore the already scrambled left-right correlation of $\eta$. Schematically, the unfolded geometry looks like Fig.~\ref{fig:bulk_geometry} (c), with a pair of end-of-the-world branes.

To obtain a more complete picture of the black hole evaporation procedure, we can also introduce a black hole formation process before the evaporation. This can be achieved by considering a nonzero constant $\mu_\eta$ bilinear coupling for $t<0$. At time $t=0$ we turn off the coupling, which corresponds to an in-falling light-like shell of matter that creates a black hole. Then at a later time $t_{\rm ev}$ we turn on the coupling with the bath. In this case, the bulk geometry has an unusual topology illustrated in Fig.~\ref{fig:bulk_geometry} (d-f). If no bath coupling is introduced, the geometry is a traversable wormhole turning into a black hole (which can be verified by the wavepacket propagation picture in subfigure (e)), with a future horizon but no past horizon. The coupling with the bath requires a modified geometry with both a future horizon and a past horizon. Physically we can consider the modes at the past horizon as auxiliary degrees of freedom which are introduced to represent the effect of the bath. It is interesting to note that a similar spacetime geometry has been discussed but then excluded for a different type of model that couples JT gravity with a bath\cite{almheiri2025}.

\section{Fate of Infalling Information}

\begin{figure}[htbp]
	\centering
	\includegraphics[width=0.49\linewidth]{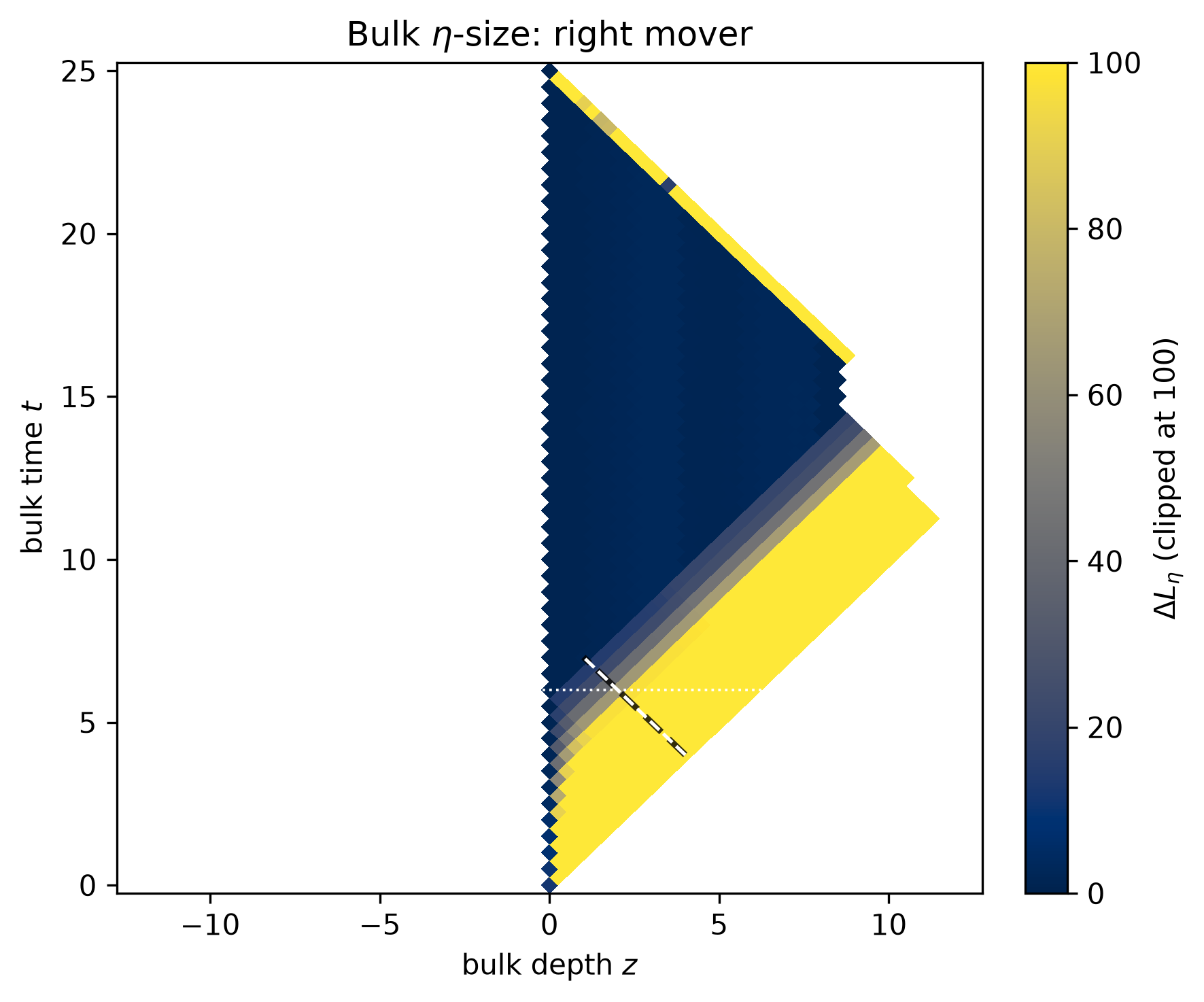}\includegraphics[width=0.49\linewidth]{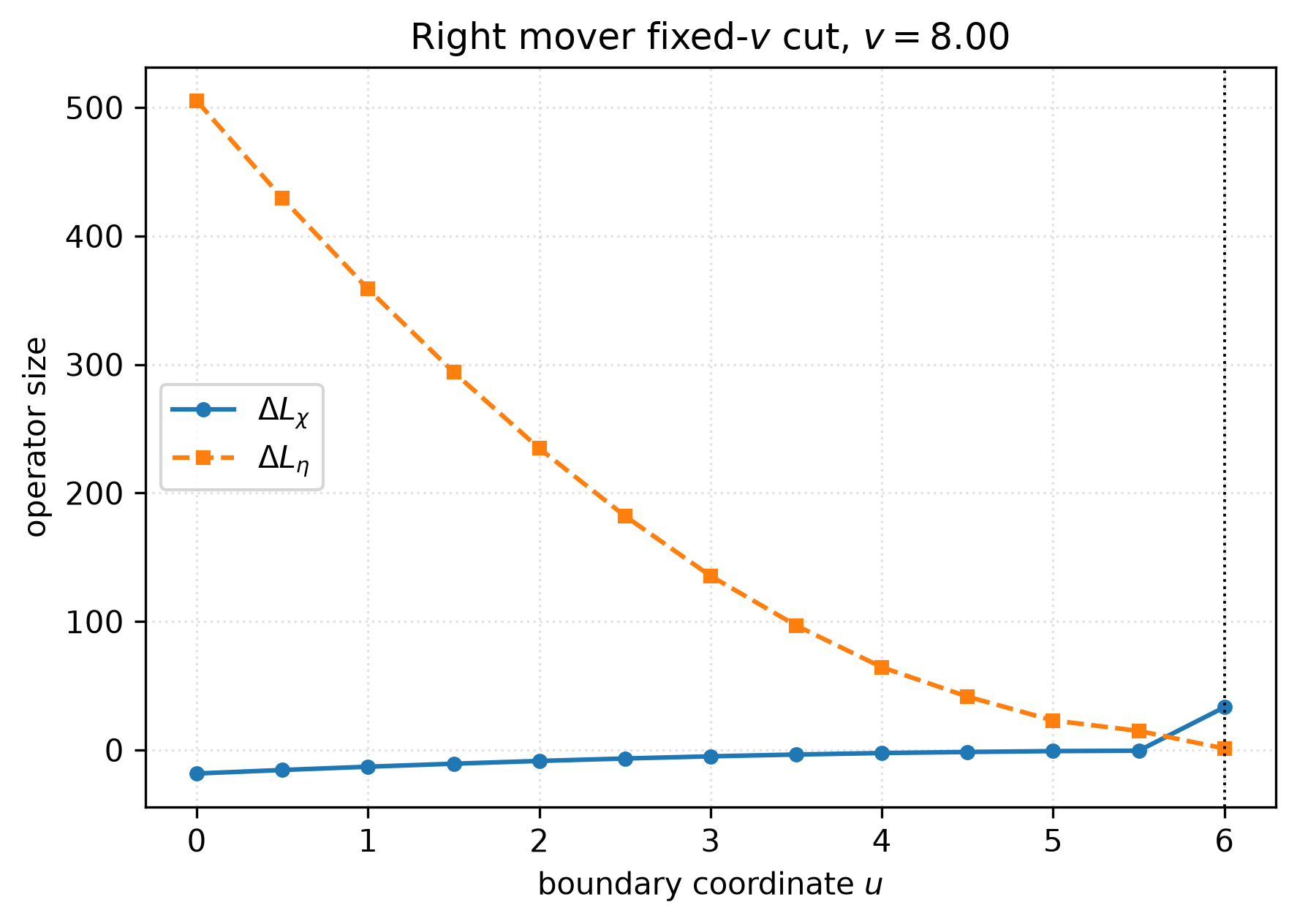}
	\caption{(a) $\Delta L_\eta$ for a right-moving fermion operator $\psi_f(u,v)$. (b) $\Delta L_\eta$ and $\Delta L_\chi$ for a fixed $v$ as a function of $u$.}\label{fig:operator_size}
\end{figure}

The emergence of the past and future horizons in region II is a direct consequence of the factorized two-point function in Eq. (\ref{eq:regionII}), which shows that $\eta_a(t)$ before and after $t=t_{\rm ev}$ are almost linearly independent from each other. In the large-$N$ limit, the bulk modes at the future horizon are independent from the future simple boundary operators. In contrast, at finite $N$, unitarity of the boundary dynamics requires that no information is lost; the information carried by an infalling mode should return in a complex multi-particle excitation rather than a simple fermion mode. Our model allows us to study this return quantitatively.

In the evaporation geometry, there is a time $t_{\rm max}$ when the anti-commutator matrix $A(t,t')$ for the interval $\left[t_{\rm ev},t_{\rm max}\right]$ becomes singular. This translates to the fact that the reconstructed geometry ends at the future horizon $v=t_{\rm max}$. Consider a future-horizon infalling mode $\psi_{a,f}(u,v=t_{\rm max})$. By construction, $\left\{\psi_{a,f}(u,t_{\rm max}),\eta_b(t)\right\}=0$ for any $t\in\left(u,t_{\rm max}\right]$. Since the post-evaporation geometry has an end, all future $\eta_a(t),~t>t_{\rm max}$ only depend on $\eta_a(t),~t\in[t_{\rm ev},t_{\rm max}]$. Therefore, $\left\{\psi_{a,f}(u,t_{\rm max}),\eta_b(t)\right\}=0$ for any $t>u$. This clarifies that in the large-$N$ limit such an infalling mode is invisible in the future. Denote $\left|\Phi_\eta\right\rangle$ and $\left|\Phi_\chi\right\rangle$ as the initial states of $\eta$ and $\chi$ fermions respectively. Define
\begin{align}
	\left|\Phi(t)\right\rangle&=U(t,0)\left|\Phi_\eta\right\rangle\otimes \left|\Phi_\chi\right\rangle\\
	\left|\Psi(t)\right\rangle&=\sqrt{2}\,U(t,0)\psi_{R,f}(u,t_{\rm max})\left|\Phi_\eta\right\rangle\otimes \left|\Phi_\chi\right\rangle
\end{align}
with $U(t,0)=T\exp\left[-i\int_0^tdt'H(t')\right]$ the time evolution operator of the coupled system. $\left|\Psi(t)\right\rangle$ and $\left|\Phi(t)\right\rangle$ are the states with and without the extra infalling fermion. According to the anti-commutation computation and Wick's theorem, these two states have identical correlation functions at all future times if we choose $t>t_{\rm ev}$. (Actually it works for all $t>u$ but we will focus on $t>t_{\rm ev}$.) However, we know these two states are orthogonal (since $\left|\Phi(t)\right\rangle$ has an even number of fermions while $\left|\Psi(t)\right\rangle$ has an odd number of fermions). If we consider the presence of the additional infalling fermion as a single bit of quantum information, this information is preserved, but becomes invisible in simple correlation functions in the large-$N$ limit.

Our solution of the coupled SYK model allows us to study how the difference between these two states is detectable when we consider finite $N$ corrections. The first diagnostic is the change in the size operator
\(\hat L_\psi=i\sum_j\psi^L_j\psi^R_j\). For the infalling mode,
\begin{align}
	\Delta L_\eta(u;t_0)&=\left\langle \Psi(t_0)\right|L_\eta\left|\Psi(t_0)\right\rangle-\left\langle \Phi(t_0)\right|L_\eta\left|\Phi(t_0)\right\rangle\label{eq:DeltaLf}
\end{align}
The computation is achieved by adding a $\delta$-function perturbation to the bilinear coupling, evaluating the corresponding folded-contour four-point response, and convolving it with the bulk reconstruction kernel. The method is summarized in Methods and derived in Supplementary Text, Sec.~S4. For a pure state that entangles $\eta_{aL}$ and $\eta_{aR}$, the expectation value of $L_\eta$ has the interpretation of the operator size for the Schmidt operator, i.e. the single-sided operator whose Choi state is $\left|\Phi(t_0)\right\rangle$ or $\left|\Psi(t_0)\right\rangle$\cite{roberts2018operator,qi2019quantum}. In our model $\eta$ is entangled with $\chi$, so we can view the state of $\eta$ as an ensemble of orthogonal pure states, and $L_\eta$ is the average operator size in this ensemble. Since $L_\eta\propto N_\eta$ but $\Delta L_\eta(u;t_0)$ is of order $1$, the effect is a $\frac1{N_\eta}$ correction. The behavior of $\Delta L_{\eta,\chi}(u;t_0)$ is shown in Fig.~\ref{fig:operator_size}. We see that the operator size grows exponentially as a function of $t_{\rm ev}-u$, due to the scrambling dynamics before evaporation, and oscillates in the detection time $t_0$ due to the coupling with the bath.

\begin{figure}[htbp]
	\centering
	\includegraphics[width=\linewidth]{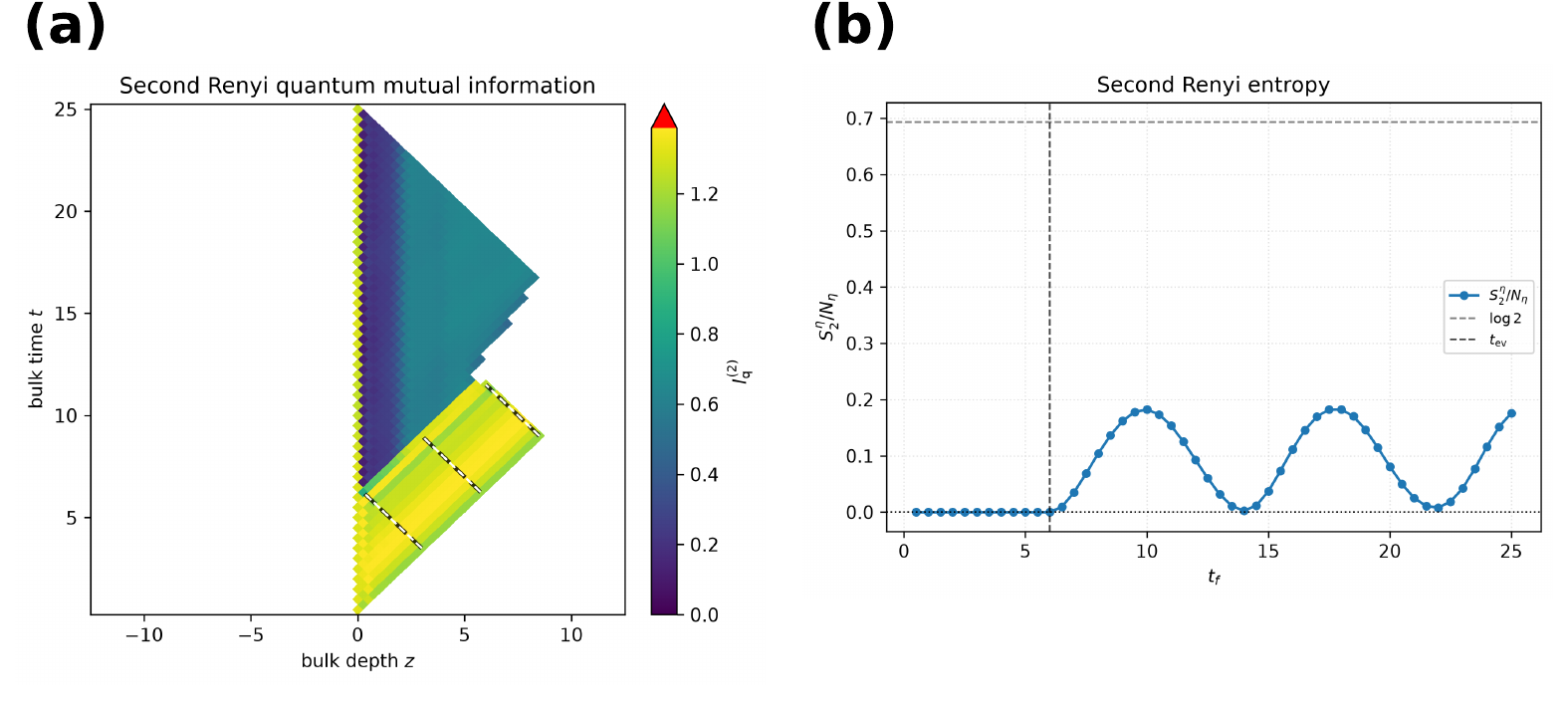}
	\caption{Finite $p$ numerical results: (a) Renyi-$2$ mutual information of an infalling bulk fermion $\psi_f(u,v)$ with the $\eta$ system at $t_f=25$. (b) Second Renyi entropy of the $\eta$ system (without the probe particle) due to entanglement with $\chi$.}\label{fig:finite_p}
\end{figure}

In addition to the operator size, we introduce a more direct measure of the fate of the quantum information carried by the infalling fermion. For that purpose, we introduce an auxiliary fermion $P$ which is entangled with the infalling mode. This leads to the entangled state
\begin{align}
	\left|E\right\rangle=\frac1{\sqrt{2}}\left(|0\rangle_P\otimes \left|\Phi(t)\right\rangle+|1\rangle_P\otimes\left|\Psi(t)\right\rangle\right)
\end{align}
To see whether the input information is preserved in the $\eta$ or $\chi$ systems at time $t$, we can study the mutual information between $P$ and $\eta$ fermions $I(P:\eta)=S_\eta+S_P-S_{\eta P}=S_\eta+S_P-S_\chi$. The last step is true because the entire system is in a pure state. Since it is difficult to compute the von Neumann entropy, we instead study the Renyi-$2$ mutual information:
\begin{align}
		I^{(2)}(P:\eta)&=S_\eta^{(2)}+S_P^{(2)}-S_\chi^{(2)}\\
		e^{-I^{(2)}(P:\eta)}&=\frac{{\rm tr}\left(\left(\sigma_\eta+\rho_\eta\right)^2\right)}{2{\rm tr}\left(\left(\sigma_\chi+\rho_\chi\right)^2\right)}\label{eq:renyi2_MI}
\end{align}
where $\rho_\eta$ and $\sigma_\eta$ are the reduced density matrices of $\eta$ in the states $\left|\Phi(t)\right\rangle$ and $\left|\Psi(t)\right\rangle$ respectively, and similarly for $\rho_\chi,\sigma_\chi$. We have also used $S_P^{(2)}=\log 2$. Eq. (\ref{eq:renyi2_MI}) can be computed by a two-replica generalization of the Kadanoff-Baym equation, with a twisted boundary condition that swaps the two replicas of $\eta$ or $\chi$ at the final time $t$ (see Ref.~\cite{chen2020replica} for a related calculation in a different model). This method is summarized in Methods and detailed in Supplementary Text, Sec.~S5. The result is shown in Fig.~\ref{fig:finite_p}. We find that the mutual information between $P$ and $\eta$ fermions is oscillating in time. For the parameters shown, the mutual information can be close to its maximum, which means that the infalling information is mostly preserved in $\eta$ fermions. For other finite \(p\) parameter regions, \(I^{(2)}(P:\eta)\) can be small. Information moves between the $\eta$ and $\chi$ systems periodically.

The same two-replica calculation can also be used to compute the second Renyi entropy $S^{(2)}_\eta(t)=-\log{\rm tr}\left(\rho_\eta(t)^2\right)$. In the large-$p$ limit we find $S^{(2)}_\eta(t)\propto N_\eta/p$. The behavior of $S^{(2)}$ is shown in Fig.~\ref{fig:finite_p}(b). Periodically at time $t=t_{\rm ev}+nT$ with $T$ the oscillation period, the $\eta$-$\chi$ entanglement vanishes. Therefore what plays the role of Hawking radiation is mainly $\eta$ fermions, the dynamics of which becomes non-scrambling after $t_{\rm ev}$.
Part of the information goes to $\chi$ and returns to $\eta$ periodically.

\section{Interior dynamics and future directions}

In conclusion, we have proposed a new model for black hole evaporation that enables a new quantitative characterization of how information appears to be lost in the large $N$ limit from simple probes, but actually gets preserved in complex operators. The bulk $1+1$-d geometry and bulk fermions crossing the black hole horizon can be determined entirely from the boundary fermion correlation functions. Operator size and two-replica calculations allow us to directly study the fate of quantum information falling across the horizon.

We would like to emphasize that our approach can be generalized to many other models. The bulk reconstruction applies to all generalized free fields. For example, we can go beyond the probe limit and study systems with finite $s=N_\eta/(N_\eta+N_\chi)$. We can also turn on the coupling between $\eta$ and $\chi$ gradually, which provides a smoother evaporation process rather than blast freezing.

There are many open questions for future work. To obtain further information on the smoothness of the horizon and address problems such as the firewall paradox, we need to study the interior dynamics. Our bulk reconstruction algorithm only reconstructs the causal wedge of the boundary, which does not directly determine the interior dynamics. A new principle is needed to determine the interior dynamics from boundary data. Another open question is whether we can compute the entropy (or second Renyi entropy) of part of the Hawking radiation, and determine whether part of the infalling modes at the future horizon are in its entanglement island.

\section{Methods}
\label{sec:methods}

\subsection{Microscopic model and evaporation protocol}
\label{sec:methods-model}

We take \(p/2\) to be even. On one side of the doubled system the microscopic Hamiltonian is
\begin{equation}
\begin{aligned}
H[\eta,\chi;J]
&=
i^{p/2}s^{-p/2}l_\eta(t)
\sum_{I_p}J_{I_p}\eta_{I_p}
\\
&\quad+
i^{p/2}(1-s)^{-p/2}l_\chi(t)
\sum_{K_p}J_{K_p}\chi_{K_p}
\\
&\quad
+i^{p/2}l_c(t)
\sum_{m=1}^{p-1}
\sum_{I_m,K_{p-m}}
J_{I_m;K_{p-m}}\eta_{I_m}\chi_{K_{p-m}} ,
\end{aligned}
\label{eq:methods-one-sided-H}
\end{equation}
where \(s=N_\eta/(N_\eta+N_\chi)\). The multi-index \(I_m=(i_1,\ldots,i_m)\) is an increasing \(m\)-tuple of \(\eta\)-indices and \(\eta_{I_m}\equiv\eta_{i_1}\cdots\eta_{i_m}\); \(K_n\) and \(\chi_{K_n}\) are defined similarly. All couplings with distinct index sets are independent Gaussian variables with zero mean and common variance
\begin{equation}
\overline{J_{I_p}^{\,2}}
=
\overline{J_{K_p}^{\,2}}
=
\overline{J_{I_m;K_{p-m}}^{\,2}}
=
\mathcal J^2\frac{2^{p-1}}{p}
\frac{(p-1)!}{(N_\eta+N_\chi)^{p-1}} .
\label{eq:methods-disorder}
\end{equation}
The left and right copies use the same disorder realization, with the reflected sign convention stated in the main text.

The sharp protocol used in the calculation is
\begin{equation}
\begin{aligned}
l_\eta(t)
&=
a\,s^{1/2}\theta(t_{\rm ev}-t)
+s^{p/2}\theta(t-t_{\rm ev}),
\\
l_\chi(t)
&=
b\,(1-s)^{1/2}\theta(t_{\rm ev}-t)
+(1-s)^{p/2}\theta(t-t_{\rm ev}),
\\
l_c(t)
&=
\theta(t-t_{\rm ev}) .
\end{aligned}
\label{eq:methods-l-protocol}
\end{equation}
For \(t<t_{\rm ev}\), \(l_c=0\), so the \(\eta\) and \(\chi\) systems are independent SYK islands of sizes \(N_\eta\) and \(N_\chi\). For \(t>t_{\rm ev}\), \(s^{-p/2}l_\eta=(1-s)^{-p/2}l_\chi=l_c=1\), so the pure and mixed \(p\)-body monomials assemble into a single SYK Hamiltonian on \(N_\eta+N_\chi\) fermions.

The bilinear couplings can be parameterized as
\begin{equation}
\begin{aligned}
\mu_\eta(t)
&=
\mu\left[
c\,\theta(-t)+\nu\,\theta(t-t_{\rm ev})
\right],
\\
\mu_\chi(t)
&=
\mu\left[
d\,\theta(t_{\rm ev}-t)+\theta(t-t_{\rm ev})
\right].
\end{aligned}
\label{eq:methods-bilinear-protocol}
\end{equation}
The evaporation protocol emphasized in the main calculation is \(a=b=1\), \(c=\nu=0\), \(d=1\), and \(\mu=\mu_0\), for which \(\mu_\eta(t)=0\) and \(\mu_\chi(t)=\mu_0\). The formation-and-evaporation extension uses \(c\neq0\), \(\nu=0\), and \(d=1\), so that the \(\eta\) bilinear is present before \(t=0\) and switched off at \(t=0\), and the \(\eta\) sector is then coupled to the bath at \(t_{\rm ev}\). Beyond this case, Supplementary Text, Sec.~S3 gives an analytical solution in the large $p$ limit for \(c=0\), \(d=1\) and \(\nu\neq0\), for which both
post-evaporation bilinear couplings are nonzero. The same formalism also allows smooth versions of these step functions.

\subsection{Large-\texorpdfstring{\(p\)}{p} two-point functions}
\label{sec:methods-large-p}

Thermofield-double initial data can be prepared by two Euclidean segments of length \(\beta/4\) on a Schwinger-Keldysh contour. The analytic figures choose the thermal \(\eta\) data and stationary coupled bath independently; the finite-\(p\) calculation uses a common preparation temperature (Supplementary Text, Secs.~S2, S3, and S5). The large-\(N\) saddle is written in terms of contour two-point functions \(G^\psi_{ab}(z_1,z_2)=-i\langle{\cal T}_{\cal C}\psi_a(z_1)\psi_b(z_2)\rangle\), where \(\psi=\eta,\chi\) and \(a,b\in\{R,L\}\) are copy indices, distinct from the protocol constants in Eq.~\eqref{eq:methods-l-protocol}. The large-\(p\) expansion separates a universal free-fermion part from order-one functions \(g^\psi_R\) and \(g^\psi_L\):
\begin{equation}
G^{\psi,>}(t_1,t_2)=-\frac{1}{2}
\left(\begin{array}{cc}
i e^{g^\psi_R(t_1,t_2)/p} & e^{g^\psi_L(t_1,t_2)/p}\\
-e^{g^\psi_L(t_1,t_2)/p} & +i e^{g^\psi_R(t_1,t_2)/p}
\end{array}\right).
\label{eq:methods-large-p-ansatz}
\end{equation}
In the general large-\(p\) equations, the interaction self-energy is determined by the kernels
\begin{equation}
\mathcal K^\psi_X
=
\alpha_\psi e^{g^\psi_X}
+l_c(t_1)l_c(t_2)
e^{s g^\eta_X+(1-s)g^\chi_X},
\qquad X=R,L ,
\label{eq:methods-large-p-kernel}
\end{equation}
where the arguments \((t_1,t_2)\) are suppressed and
\(\alpha_\eta=s^{-1}l_\eta(t_1)l_\eta(t_2)-s^{p-1}l_c(t_1)l_c(t_2)\),
\(\alpha_\chi=(1-s)^{-1}l_\chi(t_1)l_\chi(t_2)-(1-s)^{p-1}l_c(t_1)l_c(t_2)\).
Taking a second time derivative of the first-order KB equations gives
\begin{equation}
\partial_{t_1}\partial_{t_2}g^\psi_R=2\mathcal J^2\mathcal K^\psi_R,
\qquad
\partial_{t_1}\partial_{t_2}g^\psi_L=-2\mathcal J^2\mathcal K^\psi_L .
\label{eq:methods-large-p-liouville}
\end{equation}
The first-order KB equations supply the equal-time and quench matching data that are lost in this second-order form.

\begin{figure}[htbp]
	\centering
	\includegraphics[width=0.60\textwidth]{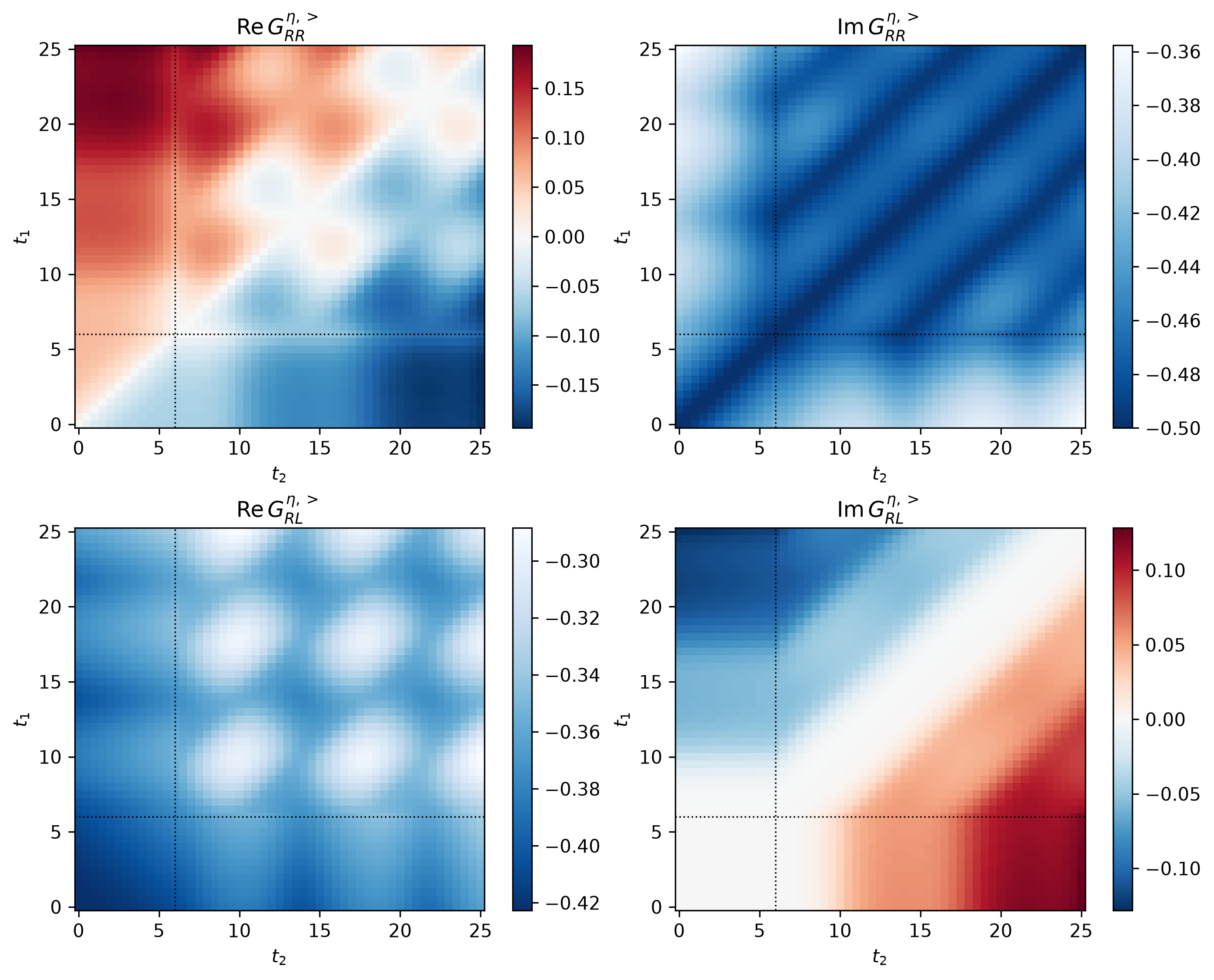}
	\caption{The heatmap of the two-point function in the large-\(p\) probe limit. The cross-shaped structure follows from the factorized form of the mixed-region solution.}\label{fig:2pt_func}
\end{figure}

In the probe limit \(p\rightarrow\infty\), \(s\rightarrow0\), \(sp\rightarrow0\), \(p^2/N_\eta\rightarrow0\), the \(\chi\) sector is fixed by the traversable-wormhole solution. Defining \(r_G,\phi_G,V_G\) by
\begin{equation}
\begin{aligned}
r_G^2=e^{-2\phi_G}
&=
\frac{\mu_0}{2\mathcal J}
\left[
\sqrt{1+\left(\frac{\mu_0}{4\mathcal J}\right)^2}
-\frac{\mu_0}{4\mathcal J}
\right],
\\
V_G&=\frac{r_G}{\sqrt{1-r_G^2}},
\label{eq:methods-bath-parameters}
\end{aligned}
\end{equation}
the bath correlators are
\begin{equation}
\begin{aligned}
e^{g^\chi_R(t_1,t_2)}
&=
e^{-i\pi}
\left[
\frac{V_G}{
\sin\left(
\mathcal J V_G(t_1-t_2)-i\tanh^{-1}r_G
\right)}
\right]^2,
\\
e^{g^\chi_L(t_1,t_2)}
&=
\left[
\frac{V_G}{
\cos\left(
\mathcal J V_G(t_1-t_2)-i\tanh^{-1}r_G
\right)}
\right]^2 .
\end{aligned}
\label{eq:methods-bath-solution}
\end{equation}
Before evaporation, the \(\eta\) solution with \(\mu_\eta=0\) is
\begin{equation}
\begin{aligned}
e^{g^\eta_R(t_1,t_2)}
&=
e^{-i\pi}
\left[
\frac{\sin\vartheta}{
\sinh\left(
a\mathcal J\sin\vartheta\,(t_1-t_2)-i\vartheta
\right)}
\right]^2,
\\
e^{g^\eta_L(t_1,t_2)}
&=
\left[
\frac{\sin\vartheta}{
\cosh\left(
a\mathcal J\sin\vartheta\,(t_1+t_2)
\right)}
\right]^2,
\end{aligned}
\label{eq:methods-eta-pre}
\end{equation}
where \(\beta a\mathcal J\sin\vartheta=\pi-2\vartheta\) and $0 < \vartheta < \pi/2$. The first-order matching across \(t_{\rm ev}\) gives Eqs.~\eqref{eq:regionII} and \eqref{eq:regionIII}. The cross-region factorization in Eq.~\eqref{eq:regionII} is visible in the heatmap of the two-point function as shown in Fig.~\ref{fig:2pt_func}.

\subsection{Bulk reconstruction algorithm}
\label{sec:methods-bulk-reconstruction}

We use the bulk reconstruction algorithm developed in Refs.~\cite{nebabu2024bulk,nebabu2026two}. For a generalized free boundary fermion, the input is the Wightman function \(C_{ab}(t_1,t_2)=\langle\eta_a(t_1)\eta_b(t_2)\rangle\). The reconstruction kernel is fixed by the anti-commutator
\begin{equation}
A_{ab}(t_1,t_2)
=
\langle\{\eta_a(t_1),\eta_b(t_2)\}\rangle
=2\,\operatorname{Re}C_{ab}(t_1,t_2).
\label{eq:methods-anticommutator}
\end{equation}
For a boundary time interval \([u,v]\), the outgoing and incoming bulk fermions are represented as linear combinations of boundary operators,
\begin{align}
\psi_{a,f(p)}(u,v)&=\int_u^v dt K_{f(p)}(u,v;t)\eta_a(t).\label{eq:HKLL_kernel}
\end{align}
After discretizing the boundary interval, write this relation as \(\psi_f=K_f\eta\) and \(\psi_p=K_p\eta\). The canonical anti-commutation relation requires
\begin{equation}
K_fAK_f^\dagger=\mathbb I,\qquad
K_pAK_p^\dagger=\mathbb I .
\label{eq:methods-kernel-orthonormal}
\end{equation}
The two kernels differ by their causal support: \(K_f\) is upper triangular in the boundary-time discretization and \(K_p\) is lower triangular. Equivalently, their inverses are obtained from Cholesky decompositions of \(A\). Once the kernels are known, the Gaussian bulk evolution between Cauchy slices \(\Sigma\) and \(\Sigma'\) is
\begin{equation}
\psi_{\Sigma'}=U_{\Sigma'\Sigma}\psi_\Sigma,
\qquad
U_{\Sigma'\Sigma}=\{\psi_{\Sigma'},\psi_\Sigma^\dagger\}.
\label{eq:methods-bulk-unitary}
\end{equation}
This construction also diagnoses topology. If \(A\) becomes singular when the boundary interval is enlarged, the bulk Hilbert space stops acquiring new orthogonal modes, which is interpreted as the end of the reconstructed geometry.

\subsection{Operator-size response}
\label{sec:methods-operator-size}

For each sector \(\psi=\eta,\chi\), define
\begin{equation}
\hat n^\psi_j
=
\left(c^\psi_j\right)^\dagger c^\psi_j
=
\frac{1}{2}
+i\psi^L_j\psi^R_j,
\qquad
\hat L_\psi=i\sum_{j=1}^{N_\psi}\psi^L_j\psi^R_j .
\label{eq:methods-size-operator}
\end{equation}
The constant \(N_\psi/2\) in the total number operator cancels in the size differences. A normalized simple boundary excitation is
\begin{equation}
\lvert\Phi_{b,j}\rangle=\sqrt{2}\,\eta^R_j(t_b)\lvert\Phi\rangle ,
\label{eq:methods-boundary-excitation}
\end{equation}
and its background-subtracted size response is
\begin{equation}
\begin{aligned}
&\Delta L^{\rm bdry}_\psi(t_b;t_0)
\\
&=
\frac{1}{N_\eta}
\sum_{j=1}^{N_\eta}
\left[
\langle\Phi_{b,j}\rvert
\hat L_\psi(t_0)
\lvert\Phi_{b,j}\rangle
-
\langle\Phi\rvert
\hat L_\psi(t_0)
\lvert\Phi\rangle
\right].
\end{aligned}
\label{eq:methods-boundary-size}
\end{equation}
To compute it, introduce
\begin{equation}
\begin{aligned}
&f_\psi(t_1,t_2;t_0;\lambda)
\\
&=
-\frac{2i}{N_\eta}
\sum_{j=1}^{N_\eta}
\langle\Phi\rvert
e^{i\lambda\hat L_\psi(t_0)/p}
\eta^R_j(t_1)
e^{-i\lambda\hat L_\psi(t_0)/p}
\eta^R_j(t_2)
\lvert\Phi\rangle .
\end{aligned}
\label{eq:methods-generating-correlator}
\end{equation}
Then
\begin{equation}
\Delta L^{\rm bdry}_\psi(t_b;t_0)
=
-p\left.\partial_\lambda
f_\psi(t_b,t_b;t_0;\lambda)
\right|_{\lambda=0}.
\label{eq:methods-response-from-generating}
\end{equation}
The conjugations in Eq.~\eqref{eq:methods-generating-correlator} are represented by a folded real-time contour with a local source \(\lambda\delta(t-t_0)\hat L_\psi/p\). In the large-\(p\) variables, define
\begin{equation}
\mathfrak D_\psi(t_1,t_2;t_0)
\equiv
i\left.
\partial_\lambda
g^\eta_{R,\psi}(\widetilde t_1,t_2;t_0,\lambda)
\right|_{\lambda=0},
\label{eq:methods-D-definition}
\end{equation}
Here \(\widetilde t_1=2t_0-t_1\). The bilocal response is
\begin{equation}
\Delta L_\psi(t_1,t_2;t_0)
=
\exp\left[
\frac{g^\eta_{R,0}(t_1,t_2)}{p}
\right]
\mathfrak D_\psi(t_1,t_2;t_0),
\label{eq:methods-bilocal-size}
\end{equation}
and for a physical boundary insertion \(\Delta L^{\rm bdry}_\psi(t_b;t_0)=\mathfrak D_\psi(t_b,t_b;t_0)\). For a reconstructed infalling mode, the same bilocal response is projected with the reconstruction row \(K_f(u,t_{\rm max})\) from Eq.~\eqref{eq:HKLL_kernel}. This kernel projection gives the \(\Delta L_\eta\) and \(\Delta L_\chi\) curves in Fig.~\ref{fig:operator_size}; the explicit region-by-region response functions are in Supplementary Text, Sec.~S4.

\subsection{Two-replica KB equation and Renyi diagnostics}
\label{sec:methods-two-replica}

The Renyi-2 mutual information is computed by a two-replica version of the KB equation. Let \(\alpha,\beta\in\{1,2\}\) denote replica indices. The final-time trace over one subsystem is implemented by a signed fermionic swap between replicas. The copy labels \(a,b\) below should again be distinguished from the protocol constant \(a\) multiplying the pre-evaporation \(\eta\) interaction. In the \(\chi\)-twisted representation, the endpoint condition can be moved into a branch-dependent replica frame, producing the kernel
\begin{equation}
\Pi(z_1,z_2)=\mathsf U^\dagger(z_1)\mathsf U(z_2),
\qquad
\mathsf S=
\begin{pmatrix}
0&-1\\
1&0
\end{pmatrix},
\label{eq:methods-twist-kernel}
\end{equation}
where \(\mathsf U=\mathbf 1\) on the upper fold and \(\mathsf U=\mathsf S\) on the lower fold. In the finite-\(p\) probe limit, the two-replica \(\eta\) self-energy is

\begin{equation}
\begin{aligned}
(\Sigma^\eta_{ab})_{\alpha\beta}(z_1,z_2)
&=
(-1)^{1+p\delta_{ab}/2}
\frac{\mathcal J^2}{p}
\begin{cases}
a^2
\left[
2(G^\eta_{ab})_{\alpha\beta}
\right]^{p-1},
&z_1,z_2\in\mathcal C_<,
\\
\left[
2G^\chi_{ab}
\right]^{p-1}
\Pi_{\alpha\beta}(z_1,z_2),
&z_1,z_2\in\mathcal C_>,
\\
0,
&z_1,z_2\ {\rm on\ opposite\ sides\ of}\ t_{\rm ev},
\end{cases}
\\
&\quad
-i\epsilon_{ab}\delta_{\alpha\beta}
\frac{\mu_\eta(z_1)}{p}
\delta_{\mathcal C}(z_1,z_2).
\end{aligned}
\label{eq:methods-two-replica-self-energy}
\end{equation}

Here \(\mathcal C_<\) and \(\mathcal C_>\) denote the portions of the contour before and after evaporation. The bath correlator \(G^\chi\) is first obtained from the ordinary one-replica equation and then held fixed while solving the two-replica \(\eta\) Dyson equation. After discretizing the contour, we solve the nonlinear equation by a damped Newton-Krylov method with a GMRES inner solve.

For \(\lambda_1,\lambda_2\in\{+,-\}\), define branch-resolved contractions
\begin{equation}
C_{\alpha\beta}^{\lambda_1\lambda_2}
\equiv
2i\,
G^{(2),\eta;\chi\text{-tw}}_{RR;\alpha\beta}
(t_{b,\lambda_1},t_{b,\lambda_2};t_f),
\label{eq:methods-branch-contractions}
\end{equation}
At leading order in large \(N_\eta\), the contractions entering the mutual information are
\begin{equation}
\begin{aligned}
F_2&=C_{12}^{-+},\qquad
K_2=C_{11}^{-+},\\
F_4&=
C_{11}^{-+}C_{22}^{-+}
-C_{12}^{-+}C_{21}^{-+}
-C_{12}^{--}C_{12}^{++}.
\end{aligned}
\label{eq:methods-F-contractions}
\end{equation}
For a reconstructed infalling mode, each \(C_{\alpha\beta}^{\lambda_1\lambda_2}\) is replaced by its projection with the reconstruction row \(K_f(u,t_{\rm max})\). The Renyi-2 mutual information is then
\begin{equation}
e^{-I^{(2)}(P:\eta)}
=
\frac{1}{2}
\frac{1+2F_2+F_4}{1+2K_2+F_4}.
\label{eq:methods-MI-ratio}
\end{equation}
The same two-replica saddles give the second Renyi entropy of the \(\eta\) sector. If \(i_{\eta,\chi\text{-tw}}\) and \(i_{\eta,\rm disc}\) are the probe contributions to the statistical on-shell action in the twisted and disconnected saddles, then
\begin{equation}
\frac{S_\eta^{(2)}(t_f)}{N_\eta}
=
\operatorname{Re}\left[
i_{\eta,\chi\text{-tw}}(t_f)
-i_{\eta,\rm disc}(t_f)
\right].
\label{eq:methods-entropy-normalization}
\end{equation}
The detailed contour gluing, solver discretization, and on-shell-action formula are given in Supplementary Text, Sec.~S5.

\subsection{Parameters used in the figures}
\label{sec:methods-figure-parameters}

All numerical figures use \(\mathcal J=0.5\), \(a=1\), \(p=16\), and a real-time grid spacing \(\Delta t=0.5\). Figures~\ref{fig:bulk_geometry}, \ref{fig:operator_size}, and \ref{fig:2pt_func} use the large-\(p\) probe-limit correlators, evaluated at \(p=16\), with the \(\eta\) black-hole temperature parameter \(\beta_\eta=20\). In the convention of Eq.~\eqref{eq:methods-eta-pre}, this corresponds to \(\vartheta=0.264356318425\). The \(\eta\) bilinear coupling is zero except during the initial wormhole stage of Fig.~\ref{fig:bulk_geometry}(d--f). Figure~\ref{fig:setup} is a schematic and does not specify numerical parameter values.

\subsubsection{Bulk geometry and wavepackets (Fig.~\ref{fig:bulk_geometry}).}
The bath coupling is \(\mu_\chi=0.1\) and evaporation begins at \(t_{\rm ev}=15\). Panels (a--c) use an initially time-evolved thermofield-double state and the reconstruction interval \(0\leq t\leq80\). Panels (d--f) use the interval \(-20\leq t\leq80\), with an initial wormhole coupling \(\mu_\eta=2\mathcal J\sin^2\vartheta/\cos\vartheta\simeq0.0707285\) switched off at \(t=0\); here \(\beta_\eta\) characterizes the black-hole segment after formation. The right-moving wavepackets have initial component amplitudes \((1,0)\) and are injected at \((t,z)=(17,1)\), or \((u,v)=(16,18)\), in panel (b), and at \((t,z)=(-10,1)\), or \((u,v)=(-11,-9)\), in panel (e).

\subsubsection{Operator size and two-point functions (Figs.~\ref{fig:operator_size} and \ref{fig:2pt_func}).}
Both figures use \(\mu_\chi=0.5\), \(t_{\rm ev}=6\), and the boundary-time interval \(0\leq t\leq25\). The operator sizes in Fig.~\ref{fig:operator_size} are evaluated at detection time \(t_0=25\) for a right-moving bulk fermion. The heatmap color scale in panel (a) is clipped at \(\Delta L_\eta=100\), and the line cut in panel (b) has \(v=8\). Figure~\ref{fig:2pt_func} displays the correlators over \(0\leq t_1,t_2\leq25\).

\subsubsection{Finite-\texorpdfstring{\(p\)}{p} mutual information and entropy (Fig.~\ref{fig:finite_p}).}
The two-replica calculation uses the preparation inverse temperature \(\beta_\eta=\beta_\chi=5.7355465869\), \(\mu_\chi=0.5\), and \(t_{\rm ev}=6\). We use \(N_\beta=3\), real-time spacing \(\Delta t=0.5\), and Euclidean spacing \(\Delta\tau=\beta/(4N_\beta)\simeq0.4779622156\), so that \(\beta_{\rm grid}=4N_\beta\Delta\tau=\beta\) (Supplementary Text, Sec.~S5). Panel (a) has final measurement time \(t_f=25\) and uses the first component of the right-moving reconstructed fermion. Reconstruction kernels with norm greater than 50 are excluded; the dashed lines mark \(v=6.5,12,17.5\). Panel (b) shows \(S_\eta^{(2)}/N_\eta\) for \(t_f=0.5,1.0,\ldots,25\).

\subsection{AI assistance}
We used an AI agent (Codex) to assist in coding, checking derivations, and proofreading the manuscript.

\section*{Data availability}
The numerical data used to generate the figures are available at \url{https://github.com/shoaibphysics/blast-freezing-black-hole}.

\section*{Code availability}
The open-source code, figure-generation scripts and notebooks, reusable AI-agent skills, and manuscript source are available at \url{https://github.com/shoaibphysics/blast-freezing-black-hole}.

\section*{Acknowledgements}
We would like to thank Yiming Chen and Zhenbin Yang for helpful discussions.

\section*{Funding}
This work is supported by the Simons Foundation.

% Shared bibliography embedded for source submission.
% BEGIN REFERENCES

% END REFERENCES

% Supplementary Information uses independent S-numbering and unique PDF anchors.
\clearpage
\setcounter{equation}{0}
\setcounter{figure}{0}
\setcounter{table}{0}
\renewcommand{\theequation}{S\arabic{equation}}
\renewcommand{\thefigure}{S\arabic{figure}}
\renewcommand{\thetable}{S\arabic{table}}
\renewcommand{\theHequation}{supp.\arabic{equation}}
\renewcommand{\theHfigure}{supp.\arabic{figure}}
\renewcommand{\theHtable}{supp.\arabic{table}}
\setcounter{secnumdepth}{0}
\allowdisplaybreaks
\setlength{\emergencystretch}{1em}

\begin{center}
{\Large\bfseries Supplementary Information}\par\medskip
{\large Blast freezing a black hole}\par\medskip
Shoaib Akhtar and Xiao-Liang Qi
\end{center}
\bigskip
\section*{S1. Microscopic model and evaporation protocol}
The main text and Methods describe the quench schematically as initially
decoupled \(\eta\) and \(\chi\) SYK sectors that merge at \(t_{\rm ev}\). Here
we give the microscopic realization of that shorthand model, including the
relative normalization of the pure and mixed interactions and a slightly more
general formation-and-evaporation protocol. Using the parameter \(s\) defined
in Methods, the one-sided Hamiltonian is
\begin{equation}
\begin{aligned}
H[\eta,\chi;J]
&=
i^{p/2}s^{-p/2}l_\eta(t)
\sum_{I_p}J_{I_p}\eta_{I_p}
+
i^{p/2}(1-s)^{-p/2}l_\chi(t)
\sum_{K_p}J_{K_p}\chi_{K_p}
\\
&\quad
+i^{p/2}l_c(t)
\sum_{m=1}^{p-1}
\sum_{I_m,K_{p-m}}
J_{I_m;K_{p-m}}\eta_{I_m}\chi_{K_{p-m}} .
\end{aligned}
\label{eq:S-one-sided-H}
\end{equation}
Here \(p/2\) is even, matching the reflected-copy convention used below. The multi-index
\(I_m=(i_1,\ldots,i_m)\) is an increasing \(m\)-tuple of
\(\eta\)-indices, and
\(\eta_{I_m}\equiv\eta_{i_1}\cdots\eta_{i_m}\).
Similarly, \(K_n=(k_1,\ldots,k_n)\) is an increasing \(n\)-tuple of
\(\chi\)-indices, with
\(\chi_{K_n}\equiv\chi_{k_1}\cdots\chi_{k_n}\).

All couplings associated with distinct index sets are independent Gaussian
random variables with zero mean and common variance
\begin{equation}
\begin{aligned}
\overline{J_{I_p}^{\,2}}
=
\overline{J_{K_p}^{\,2}}
=
\overline{J_{I_m;K_{p-m}}^{\,2}}
&=
J^2\frac{(p-1)!}{(N_\eta+N_\chi)^{p-1}}
\\
&=
\mathcal J^2\frac{2^{p-1}}{p}
\frac{(p-1)!}{(N_\eta+N_\chi)^{p-1}},
\qquad 1\leq m\leq p-1 .
\end{aligned}
\label{eq:S-disorder}
\end{equation}
The left and right copies use the same disorder realization.

A sharp evaporation protocol that realizes the shorthand Hamiltonian in
Methods is
\begin{equation}
\begin{aligned}
l_\eta(t)
&=
a\,s^{1/2}\theta(t_{\rm ev}-t)
+s^{p/2}\theta(t-t_{\rm ev}),
\\
l_\chi(t)
&=
b\,(1-s)^{1/2}\theta(t_{\rm ev}-t)
+(1-s)^{p/2}\theta(t-t_{\rm ev}),
\\
l_c(t)
&=
\theta(t-t_{\rm ev}) .
\end{aligned}
\label{eq:S-l-protocol}
\end{equation}
For \(t<t_{\rm ev}\), \(l_c=0\), while the total coefficients of the pure
\(\eta\) and \(\chi\) interactions are
\(a\,s^{(1-p)/2}\) and \(b\,(1-s)^{(1-p)/2}\), respectively. Together with
Eq.~\eqref{eq:S-disorder}, these give the standard SYK normalization for
systems of sizes \(N_\eta\) and \(N_\chi\), with interaction scales controlled
by \(a\) and \(b\); the main calculation uses \(a=b=1\). For
\(t>t_{\rm ev}\),
\(s^{-p/2}l_\eta=(1-s)^{-p/2}l_\chi=l_c=1\), so the pure and mixed
\(p\)-body monomials have the same variance and assemble into
\(H_{\rm SYK}[\eta\oplus\chi]\).

The left-right bilinear couplings in the doubled Hamiltonian can be
parameterized more generally as
\begin{equation}
\begin{aligned}
\mu_\eta(t)
&=
\mu\left[
c\,\theta(-t)+\nu\,\theta(t-t_{\rm ev})
\right],
\\
\mu_\chi(t)
&=
\mu\left[
d\,\theta(t_{\rm ev}-t)+\theta(t-t_{\rm ev})
\right].
\end{aligned}
\label{eq:S-bilinear-protocol}
\end{equation}
Here \(c,d,\nu\) are dimensionless constants. The evaporation protocol used
in the main calculation is \(c=\nu=0\), \(d=1\), and \(\mu=\mu_0\), for which
\(\mu_\eta(t)=0\) and \(\mu_\chi(t)=\mu_0\). The
formation-and-evaporation extension discussed in the main text and Methods instead takes
\(c\neq0\), \(\nu=0\), and \(d=1\): the \(\eta\) bilinear is present for
\(t<0\), is switched off at \(t=0\), and the \(\eta\)-\(\chi\) interaction is
subsequently switched on at \(t_{\rm ev}\). The more general choices
\(d\neq1\) and \(\nu\neq0\) would allow, respectively, a change of the bath
bilinear at \(t_{\rm ev}\) and an \(\eta\) bilinear after evaporation.

More generally, the changes in \(l_\eta,l_\chi,l_c,\mu_\eta\), and
\(\mu_\chi\) may occur at distinct times, and the step functions may be
replaced by smooth profiles, including the gradual evaporation mentioned in
the main text and Methods. The calculations presented here use the sharp
profiles above.

\section*{S2. Bilocal effective theory}

\noindent
Methods summarizes the large-\(N\) and large-\(p\) equations used in the main
text. Here we derive those equations from the disorder-averaged bilocal action
and spell out the contour conventions.

\paragraph*{Large-\(N\) scaling and real-time contour.}

We first take \(N_\eta,N_\chi\to\infty\) at fixed \(s\) and fixed \(p\).
With the normalization introduced in Sec.~S1, the decoupled \(\eta\) and
\(\chi\) sectors contribute at order \(N_\eta\) and \(N_\chi\), respectively,
whereas the post-evaporation action is of order \(N_\eta+N_\chi\). Throughout
this section, \(a,b\in\{R,L\}\) denote copy indices and should not be confused
with the protocol constants \(a\) and \(b\) in Sec.~S1.

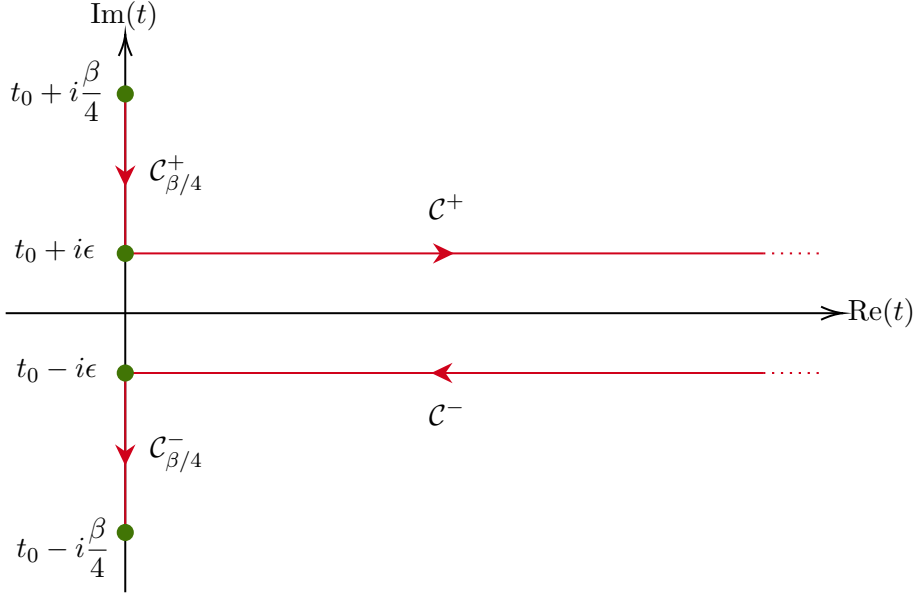
\begin{figure}
    \centering

\tikzset{every picture/.style={line width=0.75pt}}

\begin{tikzpicture}[x=0.75pt,y=0.75pt,yscale=-1,xscale=1]

\draw    (100,220) -- (518,220) ;
\draw [shift={(520,220)}, rotate = 180] [color={rgb, 255:red, 0; green, 0; blue, 0 }  ][line width=0.75]    (10.93,-3.29) .. controls (6.95,-1.4) and (3.31,-0.3) .. (0,0) .. controls (3.31,0.3) and (6.95,1.4) .. (10.93,3.29)   ;

\draw    (160,360) -- (160,82) ;
\draw [shift={(160,80)}, rotate = 90] [color={rgb, 255:red, 0; green, 0; blue, 0 }  ][line width=0.75]    (10.93,-3.29) .. controls (6.95,-1.4) and (3.31,-0.3) .. (0,0) .. controls (3.31,0.3) and (6.95,1.4) .. (10.93,3.29)   ;

\draw [color={rgb, 255:red, 208; green, 2; blue, 27 }  ,draw opacity=1 ]   (160,190) -- (480,190) ;
\draw [shift={(325,190)}, rotate = 180] [fill={rgb, 255:red, 208; green, 2; blue, 27 }  ,fill opacity=1 ][line width=0.08]  [draw opacity=0] (10.72,-5.15) -- (0,0) -- (10.72,5.15) -- (7.12,0) -- cycle    ;

\draw [color={rgb, 255:red, 208; green, 2; blue, 27 }  ,draw opacity=1 ] [dash pattern={on 0.84pt off 2.51pt}]  (480,190) -- (510,190) ;

\draw [color={rgb, 255:red, 208; green, 2; blue, 27 }  ,draw opacity=1 ]   (160,250) -- (480,250) ;
\draw [shift={(313.5,250)}, rotate = 0] [fill={rgb, 255:red, 208; green, 2; blue, 27 }  ,fill opacity=1 ][line width=0.08]  [draw opacity=0] (10.72,-5.15) -- (0,0) -- (10.72,5.15) -- (7.12,0) -- cycle    ;

\draw [color={rgb, 255:red, 208; green, 2; blue, 27 }  ,draw opacity=1 ] [dash pattern={on 0.84pt off 2.51pt}]  (480,250) -- (510,250) ;

\draw [color={rgb, 255:red, 208; green, 2; blue, 27 }  ,draw opacity=1 ]   (160,330) -- (160,250) ;
\draw [shift={(160,296.5)}, rotate = 270] [fill={rgb, 255:red, 208; green, 2; blue, 27 }  ,fill opacity=1 ][line width=0.08]  [draw opacity=0] (10.72,-5.15) -- (0,0) -- (10.72,5.15) -- (7.12,0) -- cycle    ;

\draw  [color={rgb, 255:red, 65; green, 117; blue, 5 }  ,draw opacity=1 ][fill={rgb, 255:red, 65; green, 117; blue, 5 }  ,fill opacity=1 ] (156.21,250) .. controls (156.21,247.91) and (157.91,246.21) .. (160,246.21) .. controls (162.09,246.21) and (163.79,247.91) .. (163.79,250) .. controls (163.79,252.09) and (162.09,253.79) .. (160,253.79) .. controls (157.91,253.79) and (156.21,252.09) .. (156.21,250) -- cycle ;

\draw  [color={rgb, 255:red, 65; green, 117; blue, 5 }  ,draw opacity=1 ][fill={rgb, 255:red, 65; green, 117; blue, 5 }  ,fill opacity=1 ] (156.21,330) .. controls (156.21,327.91) and (157.91,326.21) .. (160,326.21) .. controls (162.09,326.21) and (163.79,327.91) .. (163.79,330) .. controls (163.79,332.09) and (162.09,333.79) .. (160,333.79) .. controls (157.91,333.79) and (156.21,332.09) .. (156.21,330) -- cycle ;

\draw [color={rgb, 255:red, 208; green, 2; blue, 27 }  ,draw opacity=1 ]   (160,190) -- (160,110) ;
\draw [shift={(160,156.5)}, rotate = 270] [fill={rgb, 255:red, 208; green, 2; blue, 27 }  ,fill opacity=1 ][line width=0.08]  [draw opacity=0] (10.72,-5.15) -- (0,0) -- (10.72,5.15) -- (7.12,0) -- cycle    ;

\draw  [color={rgb, 255:red, 65; green, 117; blue, 5 }  ,draw opacity=1 ][fill={rgb, 255:red, 65; green, 117; blue, 5 }  ,fill opacity=1 ] (156.21,110) .. controls (156.21,107.91) and (157.91,106.21) .. (160,106.21) .. controls (162.09,106.21) and (163.79,107.91) .. (163.79,110) .. controls (163.79,112.09) and (162.09,113.79) .. (160,113.79) .. controls (157.91,113.79) and (156.21,112.09) .. (156.21,110) -- cycle ;

\draw  [color={rgb, 255:red, 65; green, 117; blue, 5 }  ,draw opacity=1 ][fill={rgb, 255:red, 65; green, 117; blue, 5 }  ,fill opacity=1 ] (156.21,190) .. controls (156.21,187.91) and (157.91,186.21) .. (160,186.21) .. controls (162.09,186.21) and (163.79,187.91) .. (163.79,190) .. controls (163.79,192.09) and (162.09,193.79) .. (160,193.79) .. controls (157.91,193.79) and (156.21,192.09) .. (156.21,190) -- cycle ;

\draw (103,181) node [anchor=north west][inner sep=0.75pt]   [align=left] {$\displaystyle t_{0} +i\epsilon $};

\draw (104,241) node [anchor=north west][inner sep=0.75pt]   [align=left] {$\displaystyle t_{0} -i\epsilon $};

\draw (521,211) node [anchor=north west][inner sep=0.75pt]   [align=left] {$\displaystyle \text{Re}( t)$};

\draw (141,62) node [anchor=north west][inner sep=0.75pt]   [align=left] {$\displaystyle \text{Im}( t)$};

\draw (104,321) node [anchor=north west][inner sep=0.75pt]   [align=left] {$\displaystyle t_{0} -i\frac{\beta }{4}$};

\draw (101,92) node [anchor=north west][inner sep=0.75pt]   [align=left] {$\displaystyle t_{0} +i\frac{\beta }{4}$};

\draw (311,159) node [anchor=north west][inner sep=0.75pt]   [align=left] {$\displaystyle \mathcal{C}^{+}$};

\draw (311,262) node [anchor=north west][inner sep=0.75pt]   [align=left] {$\displaystyle \mathcal{C}^{-}$};

\draw (171,139) node [anchor=north west][inner sep=0.75pt]   [align=left] {$\displaystyle \mathcal{C}_{\beta /4}^{+}$};

\draw (171,278) node [anchor=north west][inner sep=0.75pt]   [align=left] {$\displaystyle \mathcal{C}_{\beta /4}^{-}$};

\end{tikzpicture}
\caption{\textbf{Contour for real-time evolution and TFD preparation.}
The horizontal branches \(\mathcal C^+\) and \(\mathcal C^-\) describe
forward and backward real-time evolution, while
\(\mathcal C^+_{\beta/4}\) and \(\mathcal C^-_{\beta/4}\) prepare the initial
thermofield-double data. The endpoint conditions are given in
Eq.~\eqref{eq:S-tfd-gluing}.}
\label{fig:S-kb-contour}
\end{figure}

The calculation is performed on the Schwinger--Keldysh contour shown in
Fig.~\ref{fig:S-kb-contour}, defined by
\begin{equation}
\mathcal C
=
\mathcal C^+_{\beta/4}
\cup
\mathcal C^+
\cup
\mathcal C^-
\cup
\mathcal C^-_{\beta/4}.
\label{eq:S-kb-contour}
\end{equation}
The forward and backward real-time branches \(\mathcal C^+\) and
\(\mathcal C^-\) compute in-in observables, while the vertical segments
prepare the common-inverse-temperature \(\beta\) thermofield-double data
used in the finite-\(p\) calculation of Sec.~S5. The analytic solutions in
Sec.~S3 instead allow independent initial data for the two decoupled sectors:
a thermal \(\eta\) state and a stationary coupled \(\chi\) bath. The
real-time equations below apply to both preparations, with the corresponding
Euclidean boundary data.

For each species \(\psi=\eta,\chi\), define
\begin{equation}
c^\psi_j
=
\frac{\psi^L_j+i\psi^R_j}{\sqrt{2}},
\qquad
\left(c^\psi_j\right)^\dagger
=
\frac{\psi^L_j-i\psi^R_j}{\sqrt{2}} .
\label{eq:S-c-def}
\end{equation}
The reference state \(\lvert\mathbf 1\rangle\) is annihilated by all
\(c^\psi_j\). Equivalently,
\begin{equation}
\psi^L_j\lvert\mathbf 1\rangle
=
-i\psi^R_j\lvert\mathbf 1\rangle,
\qquad
\langle\mathbf 1\rvert\psi^L_j
=
i\langle\mathbf 1\rvert\psi^R_j .
\label{eq:S-identity-gluing}
\end{equation}
The thermofield-double state is prepared by Euclidean evolution,
\begin{equation}
\lvert\mathrm{TFD}\rangle
=
\frac{1}{\sqrt{Z_\beta}}
e^{-\frac{\beta}{4}(H^R+H^L)}
\lvert\mathbf 1\rangle .
\label{eq:S-tfd-state}
\end{equation}
This representation fixes the endpoint conditions
\begin{equation}
\begin{aligned}
\psi^L_j(t_0+i\beta/4)
&=
-i\psi^R_j(t_0+i\beta/4),
\\
\psi^L_j(t_0-i\beta/4)
&=
i\psi^R_j(t_0-i\beta/4).
\end{aligned}
\label{eq:S-tfd-gluing}
\end{equation}

Define the contour-ordered two-point function by
\begin{equation}
G^\psi_{ab}(z_1,z_2)
=
-i\left\langle
\mathcal T_{\mathcal C}
\psi_a(z_1)\psi_b(z_2)
\right\rangle .
\label{eq:S-contour-two-point}
\end{equation}
The field-level conditions in Eq.~\eqref{eq:S-tfd-gluing} imply, for example,
\begin{equation}
\begin{aligned}
G^\psi_{aL}(z,t_0+i\beta/4)
&=
-iG^\psi_{aR}(z,t_0+i\beta/4),
\\
G^\psi_{La}(t_0-i\beta/4,z)
&=
iG^\psi_{Ra}(t_0-i\beta/4,z).
\end{aligned}
\label{eq:S-two-point-gluing}
\end{equation}
These relations provide the thermal boundary data for the real-time
evolution.

\paragraph*{Disorder-averaged effective action.}

At the path-integral level, introduce the microscopic collective fields
\begin{equation}
\widehat G^\psi_{ab}(z_1,z_2)
=
-\frac{i}{N_\psi}
\sum_{j=1}^{N_\psi}
\psi^a_j(z_1)\psi^b_j(z_2),
\qquad
\psi=\eta,\chi .
\label{eq:S-bilocal-def}
\end{equation}
Independent bilocal variables \(\mathcal G^\psi\) are constrained to equal
\(\widehat G^\psi\), with \(\Sigma^\psi\) acting as the corresponding
Lagrange multipliers. This is done by inserting the identity
\begin{equation}
\label{eq:bilocal-constraint}
\begin{aligned}
1
&=
\int
D\Sigma^\eta D\Sigma^\chi
D\mathcal{G}^\eta D\mathcal{G}^\chi
\exp\Bigg[
-\frac{(N_\eta+N_\chi)}{2}
\int_{\mathcal{C}}\mathrm{d}z_1\,\mathrm{d}z_2
\sum_{a,b}
\Big\{
s\,\Sigma^\eta_{ab}
\left(
\mathcal{G}^\eta_{ab}
-
\widehat{G}^\eta_{ab}
\right)
\\
&\hspace{5.1cm}
+
(1-s)\,\Sigma^\chi_{ab}
\left(
\mathcal{G}^\chi_{ab}
-
\widehat{G}^\chi_{ab}
\right)
\Big\}
\Bigg].
\end{aligned}
\end{equation}
The factors \(s\) and \(1-s\) appear because
\(N_\eta=s(N_\eta+N_\chi)\) and \(N_\chi=(1-s)(N_\eta+N_\chi)\). After imposing the constraint, we drop the calligraphic
notation; the saddle value \(G^\psi\) is the contour correlator defined in
Eq.~\eqref{eq:S-contour-two-point}.

After disorder averaging and imposing the bilocal constraint, the fermion path integral is Gaussian.  For a Majorana
quadratic form,
\begin{equation}
\label{eq:majorana-trace-log}
  \int D\psi\,
  \exp\left[
  \frac{i}{2}
  \int_{\mathcal{C}}\mathrm{d}z_1\,\mathrm{d}z_2\,
  \psi(z_1)
  \left(G_0^{-1}-\Sigma\right)(z_1,z_2)
  \psi(z_2)
  \right]
  =
  \exp\left[
  \frac{1}{2}
  \operatorname{Tr}\log
  \left[-i\left(G_0^{-1}-\Sigma\right)\right]
  \right].
\end{equation}
Applying this to the \(\eta\) and
\(\chi\) fermions gives the two trace-log terms in the effective action.

The interaction terms follow from the Gaussian disorder average
\begin{equation}
\label{eq:gaussian-disorder-average}
  \int
  \frac{\mathrm{d}J}{\sqrt{2\pi\left\langle J^2\right\rangle}}
  \exp\left[
  -\frac{J^2}{2\left\langle J^2\right\rangle}
  +
  JA
  \right]
  =
  \exp\left[
  \frac{\left\langle J^2\right\rangle}{2}A^2
  \right].
\end{equation}
At leading order in \(N_\psi\), the sum over increasing \(p\)-tuples gives
\begin{equation}
\label{eq:large-N-tuple-sum}
  \sum_{I_p}
  \psi^a_{I_p}(z_1)\psi^b_{I_p}(z_2)
  \simeq
  \frac{N_\psi^p}{p!}
  \left(
  G^\psi_{ab}(z_1,z_2)
  \right)^p .
\end{equation}

The disorder average of the mixed interactions produces the binomial
combination
\begin{equation}
\begin{aligned}
&l_c(z_1)l_c(z_2)
\sum_{m=1}^{p-1}
\binom{p}{m}s^m(1-s)^{p-m}
\left(2G^\eta_{ab}\right)^m
\left(2G^\chi_{ab}\right)^{p-m}
\\
&\quad =
l_c(z_1)l_c(z_2)
\Bigg[
\left(
2\left[sG^\eta_{ab}+(1-s)G^\chi_{ab}\right]
\right)^p
-s^p\left(2G^\eta_{ab}\right)^p
-(1-s)^p\left(2G^\chi_{ab}\right)^p
\Bigg].
\end{aligned}
\label{eq:S-mixed-binomial}
\end{equation}
The last two terms subtract the pure \(m=p\) and \(m=0\) interactions, which
are already included separately.

After integrating out the fermions, the disorder-averaged partition function
takes the saddle form
\begin{equation}
\overline Z
=
\int DG\,D\Sigma\,
\exp\left[
i(N_\eta+N_\chi)S[G,\Sigma]
\right].
\label{eq:S-bilocal-saddle}
\end{equation}
The effective action is
\begin{equation}
\begin{aligned}
S[G,\Sigma]
&=
-s\frac{i}{2}
\operatorname{Tr}\log
\left[
-i\left(G_0^{-1}-\Sigma^\eta\right)
\right]
-(1-s)\frac{i}{2}
\operatorname{Tr}\log
\left[
-i\left(G_0^{-1}-\Sigma^\chi\right)
\right]
\\
&\quad
+\frac{i}{2}
\int_{\mathcal C}dz_1\,dz_2
\sum_{a,b}
\left[
s\Sigma^\eta_{ab}G^\eta_{ab}
+(1-s)\Sigma^\chi_{ab}G^\chi_{ab}
\right]
\\
&\quad
-\frac{i\mathcal J^2}{4p^2}
\int_{\mathcal C}dz_1\,dz_2
\sum_{a,b}
(-1)^{1+p\delta_{ab}/2}
\Bigg[
l_\eta(z_1)l_\eta(z_2)
\left(2G^\eta_{ab}\right)^p
\\
&\hspace{3.0cm}
+l_\chi(z_1)l_\chi(z_2)
\left(2G^\chi_{ab}\right)^p
\\
&\hspace{3.0cm}
+l_c(z_1)l_c(z_2)
\bigg\{
\left(
2\left[sG^\eta_{ab}+(1-s)G^\chi_{ab}\right]
\right)^p
\\
&\hspace{5.2cm}
-s^p\left(2G^\eta_{ab}\right)^p
-(1-s)^p\left(2G^\chi_{ab}\right)^p
\bigg\}
\Bigg]
\\
&\quad
+\frac{1}{2p}
\int_{\mathcal C}dz_1\,dz_2\,
\delta_{\mathcal C}(z_1,z_2)
\Big[
s\mu_\eta(z_1)
\left(G^\eta_{LR}-G^\eta_{RL}\right)
\\
&\hspace{5.0cm}
+(1-s)\mu_\chi(z_1)
\left(G^\chi_{LR}-G^\chi_{RL}\right)
\Big].
\end{aligned}
\label{eq:S-effective-action}
\end{equation}
Here \(\operatorname{Tr}\) includes the trace over the \(R,L\) indices and
integration along the contour. All bilocals in
Eq.~\eqref{eq:S-effective-action} are evaluated at \((z_1,z_2)\) unless their
arguments are displayed explicitly. The factor
\((-1)^{1+p\delta_{ab}/2}\) incorporates the Majorana phases and the
common-disorder reflected-copy convention used in Methods. The contour
coupling \(\mu_\psi(z)\) restricts to the protocol in
Eq.~\eqref{eq:S-bilinear-protocol} on the real-time branches and vanishes on
the Euclidean preparation segments.

The bare inverse propagator is
\begin{equation}
(G_0^{-1})_{ab}(z_1,z_2)
=
i\delta_{ab}\partial_{z_1}
\delta_{\mathcal C}(z_1,z_2).
\label{eq:S-bare-inverse-propagator}
\end{equation}
Varying Eq.~\eqref{eq:S-effective-action} with respect to \(G^\psi\) gives
\begin{equation}
\begin{aligned}
\Sigma^\psi_{ab}(z_1,z_2)
&=
(-1)^{1+p\delta_{ab}/2}
\frac{\mathcal J^2}{p}
\Bigg[
\alpha_\psi(z_1,z_2)
\left(2G^\psi_{ab}(z_1,z_2)\right)^{p-1}
\\
&\hspace{2.7cm}
+l_c(z_1)l_c(z_2)
\left(
2\left[
sG^\eta_{ab}(z_1,z_2)
+(1-s)G^\chi_{ab}(z_1,z_2)
\right]
\right)^{p-1}
\Bigg]
\\
&\quad
-i\epsilon_{ab}
\frac{\mu_\psi(z_1)}{p}
\delta_{\mathcal C}(z_1,z_2),
\end{aligned}
\label{eq:S-self-energy}
\end{equation}
where
\begin{equation}
\epsilon :=
\begin{pmatrix}
\epsilon_{RR} & \epsilon_{RL} \\
\epsilon_{LR}  & \epsilon_{LL}
\end{pmatrix}
=
\begin{pmatrix}
0&1\\
-1&0
\end{pmatrix}
\label{eq:S-epsilon}
\end{equation}
and
\begin{equation}
\begin{aligned}
\alpha_\eta(z_1,z_2)
&=
s^{-1}l_\eta(z_1)l_\eta(z_2)
-s^{p-1}l_c(z_1)l_c(z_2),
\\
\alpha_\chi(z_1,z_2)
&=
(1-s)^{-1}l_\chi(z_1)l_\chi(z_2)
-(1-s)^{p-1}l_c(z_1)l_c(z_2).
\end{aligned}
\label{eq:S-alpha}
\end{equation}

For the sharp protocol in Eq.~\eqref{eq:S-l-protocol}, when both contour
arguments lie before evaporation one has
\(\alpha_\eta=a^2\), \(\alpha_\chi=b^2\), and \(l_c=0\). When both arguments
lie after evaporation, \(\alpha_\eta=\alpha_\chi=0\) and \(l_c=1\), leaving
only the common self-energy built from
\(sG^\eta+(1-s)G^\chi\). This reproduces the independent pre-evaporation
saddles and the combined post-evaporation saddle. Expanding the latter term
gives the weighted sum of products \(G_\eta^kG_\chi^{p-1-k}\) summarized
schematically in the main text and Methods.

\paragraph*{Kadanoff-Baym equations.}

In this paragraph \(G^\psi\) and \(\Sigma^\psi\) denote \(2\times2\) matrices
in the \(R,L\) copy indices. Matrix multiplication in this space and
convolution along \(\mathcal C\) are implicit.

Varying with respect to \(\Sigma^\psi\) gives the contour Dyson equation
\begin{equation}
\Sigma^\psi(z_1,z_2)
=
G_0^{-1}(z_1,z_2)
-
\left(G^\psi\right)^{-1}(z_1,z_2),
\label{eq:S-contour-dyson}
\end{equation}
where the inverse is defined by
\begin{equation}
\int_{\mathcal C}dz_3\,
\left(G^\psi\right)^{-1}(z_1,z_3)
G^\psi(z_3,z_2)
=
\delta_{\mathcal C}(z_1,z_2)\mathbb I .
\label{eq:S-contour-inverse}
\end{equation}
Multiplying Eq.~\eqref{eq:S-contour-dyson} by \(G^\psi\) from the right or
left gives the two contour Kadanoff-Baym equations,
\begin{equation}
\begin{aligned}
i\partial_{z_1}G^\psi(z_1,z_2)
&=
\delta_{\mathcal C}(z_1,z_2)\mathbb I
+
\int_{\mathcal C}dz_3\,
\Sigma^\psi(z_1,z_3)G^\psi(z_3,z_2),
\\
-i\partial_{z_2}G^\psi(z_1,z_2)
&=
\delta_{\mathcal C}(z_1,z_2)\mathbb I
+
\int_{\mathcal C}dz_3\,
G^\psi(z_1,z_3)\Sigma^\psi(z_3,z_2).
\end{aligned}
\label{eq:S-contour-KB}
\end{equation}
The two equations are related by complex conjugation.
With our conventions, complex conjugation reverses the orientation of the
contour.  In components,
\((i)^*=-i\),
\(\delta_{\mathcal{C}}(z_1,z_2)^*
=-\delta_{\mathcal{C}}(z_1^*,z_2^*)\), and
\(\left(\int_{\mathcal{C}}\mathrm{d}z\right)^*
=-\int_{\mathcal{C}}\mathrm{d}z^*\).  The bilocals obey
\[
  \left(G^\psi_{ab}(z_1,z_2)\right)^*
  =
  -G^\psi_{ba}(z_2^*,z_1^*),
  \qquad
  \left(\Sigma^\psi_{ab}(z_1,z_2)\right)^*
  =
  -\Sigma^\psi_{ba}(z_2^*,z_1^*) .
\]
Taking the complex conjugate of the first contour Eq.~\eqref{eq:S-contour-KB} and then relabeling
\(z_1\leftrightarrow z_2\) gives the second equation.

For real times, define
\begin{equation}
\begin{aligned}
G^{\psi,>}_{ab}(t_1,t_2)
&=
-i\left\langle
\psi_a(t_1)\psi_b(t_2)
\right\rangle,
\\
G^{\psi,<}_{ab}(t_1,t_2)
&=
i\left\langle
\psi_b(t_2)\psi_a(t_1)
\right\rangle
=
\left(
G^{\psi,>}_{ab}(t_1,t_2)
\right)^* .
\end{aligned}
\label{eq:S-greater-lesser}
\end{equation}
The smooth interaction self-energy obeys the analogous relation
\(\Sigma^{\psi,<}=(\Sigma^{\psi,>})^*\), component by component.
Superscripts \({\rm R},{\rm A}\) below denote retarded and advanced
components and should not be confused with the \(R,L\) copy labels.

Applying the Langreth rules to the first contour KB equation gives
\begin{equation}
\begin{aligned}
i\partial_{t_1}G^{\psi,>}(t_1,t_2)
&=
\int_{t_0}^{\infty}dt_3
\left[
\Sigma^{\psi,{\rm R}}(t_1,t_3)
G^{\psi,>}(t_3,t_2)
+
\Sigma^{\psi,>}(t_1,t_3)
G^{\psi,{\rm A}}(t_3,t_2)
\right]
\\
&\quad
+
\int_{\mathcal C^+_{\beta/4}\cup\mathcal C^-_{\beta/4}}
dz_3\,
\Sigma^\psi(t_1,z_3)G^\psi(z_3,t_2).
\end{aligned}
\label{eq:S-kb-greater-Langreth}
\end{equation}
The required retarded and advanced components are
\begin{equation}
\begin{aligned}
G^{\psi,{\rm R}}(t_1,t_2)
&=
\theta(t_1-t_2)
\left[
G^{\psi,>}(t_1,t_2)
-
G^{\psi,<}(t_1,t_2)
\right],
\\
G^{\psi,{\rm A}}(t_1,t_2)
&=
-\theta(t_2-t_1)
\left[
G^{\psi,>}(t_1,t_2)
-
G^{\psi,<}(t_1,t_2)
\right],
\\
\Sigma^{\psi,{\rm R}}(t_1,t_2)
&=
\theta(t_1-t_2)
\left[
\Sigma^{\psi,>}(t_1,t_2)
-
\Sigma^{\psi,<}(t_1,t_2)
\right]
-i\frac{\mu_\psi(t_1)}{p}
\delta(t_1-t_2)\epsilon,
\\
\Sigma^{\psi,{\rm A}}(t_1,t_2)
&=
-\theta(t_2-t_1)
\left[
\Sigma^{\psi,>}(t_1,t_2)
-
\Sigma^{\psi,<}(t_1,t_2)
\right]
-i\frac{\mu_\psi(t_1)}{p}
\delta(t_1-t_2)\epsilon .
\end{aligned}
\label{eq:S-retarded-advanced}
\end{equation}
For real \(t_1,t_2\), the Wightman components obey
\begin{equation}
\label{eq:lesser-greater-conjugation}
  G^{\psi,<}(t_1,t_2)
  =
  \left(G^{\psi,>}(t_1,t_2)\right)^*,
  \qquad
  \Sigma^{\psi,<}(t_1,t_2)
  =
  \left(\Sigma^{\psi,>}(t_1,t_2)\right)^*,
\end{equation}
where complex conjugation is taken componentwise.  Therefore
\begin{equation}
\label{eq:retarded-advanced-imaginary}
\begin{aligned}
  \Sigma^{\psi,R}(t_1,t_2)
  &=
  2i\,\theta(t_1-t_2)\,
  \operatorname{Im}\Sigma^{\psi,>}(t_1,t_2)
  -
  i\frac{\mu_\psi(t_1)}{p}\delta(t_1-t_2)\epsilon,
  \\
  G^{\psi,A}(t_1,t_2)
  &=
  -2i\,\theta(t_2-t_1)\,
  \operatorname{Im}G^{\psi,>}(t_1,t_2).
\end{aligned}
\end{equation}
Substituting these relations into
Eq.~\eqref{eq:S-kb-greater-Langreth} yields
\begin{equation}
\begin{aligned}
\partial_{t_1}G^{\psi,>}(t_1,t_2)
&=
2\int_{t_0}^{t_1}dt_3\,
\operatorname{Im}[\Sigma^{\psi,>}(t_1,t_3)]
G^{\psi,>}(t_3,t_2)
\\
&\quad
-2\int_{t_0}^{t_2}dt_3\,
\Sigma^{\psi,>}(t_1,t_3)
\operatorname{Im}[G^{\psi,>}(t_3,t_2)]
\\
&\quad
-\frac{\mu_\psi(t_1)}{p}
\epsilon\,G^{\psi,>}(t_1,t_2)
-i
\int_{\mathcal C^+_{\beta/4}\cup\mathcal C^-_{\beta/4}}
dz_3\,
\Sigma^\psi(t_1,z_3)G^\psi(z_3,t_2).
\end{aligned}
\label{eq:S-real-time-KB}
\end{equation}
Here \(\Sigma^{\psi,>}\) in the real-time memory integrals denotes the smooth
interaction part of the self-energy; the local bilinear contribution has
already been written separately. The last line,
\begin{equation}
\mathcal I^\psi_{\rm th}(t_1,t_2)
=
-i
\int_{\mathcal C^+_{\beta/4}\cup\mathcal C^-_{\beta/4}}
dz_3\,
\Sigma^\psi(t_1,z_3)G^\psi(z_3,t_2),
\label{eq:S-thermal-memory}
\end{equation}
contains the thermal-preparation data and generally depends on both
\(t_1\) and \(t_2\). The companion equation with a derivative acting on
\(t_2\) follows from the second contour KB equation.

\paragraph*{Large-\(p\) simplification.}

We now take the large-\(N\), large-\(p\) limit with
\begin{equation}
\frac{p^2}{N_\eta+N_\chi}
\longrightarrow0,
\label{eq:S-large-p-control}
\end{equation}
while keeping \(s\) finite at this stage. For the reflection-symmetric matrix
form used here, we further take \(p/2\) even. The reflection, conjugation, and
normalization conventions are collected in Sec.~S6. The probe limit and its analytic two-point functions are treated separately
in Sec.~S3.

Using the large-\(p\) parametrization of \(G^{\psi,>}\) introduced in the main
text, define
\begin{equation}
\mathcal K^\psi_a(t_1,t_2)
=
\alpha_\psi(t_1,t_2)e^{g^\psi_a(t_1,t_2)}
+
l_c(t_1)l_c(t_2)
e^{
s g^\eta_a(t_1,t_2)
+
(1-s)g^\chi_a(t_1,t_2)
},
\qquad
a=R,L .
\label{eq:S-large-p-kernel}
\end{equation}
The smooth interaction part of the greater self-energy becomes
\begin{equation}
\Sigma^{\psi,>}(t_1,t_2)
=
-\frac{\mathcal J^2}{p}
\begin{pmatrix}
i\mathcal K^\psi_R(t_1,t_2)
&
-\mathcal K^\psi_L(t_1,t_2)
\\
\mathcal K^\psi_L(t_1,t_2)
&
i\mathcal K^\psi_R(t_1,t_2)
\end{pmatrix}.
\label{eq:S-large-p-self-energy}
\end{equation}
The local bilinear term is excluded from
Eq.~\eqref{eq:S-large-p-self-energy} and appears explicitly in
Eq.~\eqref{eq:S-real-time-KB}.

At leading order in \(1/p\), the first real-time KB equation gives
\begin{equation}
\begin{aligned}
\partial_{t_1}g^\psi_R(t_1,t_2)
&=
-2\mathcal J^2
\int_{t_0}^{t_1}dt_3
\left[
\operatorname{Re}\mathcal K^\psi_R(t_1,t_3)
-i\operatorname{Im}\mathcal K^\psi_L(t_1,t_3)
\right]
\\
&\quad
+2\mathcal J^2
\int_{t_0}^{t_2}dt_3\,
\mathcal K^\psi_R(t_1,t_3)
-i\mu_\psi(t_1)
+\mathcal V^\psi_R(t_1),
\\
\partial_{t_1}g^\psi_L(t_1,t_2)
&=
-2\mathcal J^2
\int_{t_0}^{t_1}dt_3
\left[
\operatorname{Re}\mathcal K^\psi_R(t_1,t_3)
-i\operatorname{Im}\mathcal K^\psi_L(t_1,t_3)
\right]
\\
&\quad
-2\mathcal J^2
\int_{t_0}^{t_2}dt_3\,
\mathcal K^\psi_L(t_1,t_3)
-i\mu_\psi(t_1)
+\mathcal V^\psi_L(t_1).
\end{aligned}
\label{eq:S-first-order-large-p-KB}
\end{equation}
The one-time functions generated by the Euclidean preparation segments are
\begin{equation}
\begin{aligned}
\mathcal V^\psi_R(t_1)
&=
\mathcal J^2
\int_{\mathcal C^+_{\beta/4}}dz_3
\left[
\mathcal K^\psi_R(t_1,z_3)
+
\mathcal K^\psi_L(t_1,z_3)
\right]
\\
&\quad
+
\mathcal J^2
\int_{\mathcal C^-_{\beta/4}}dz_3
\left[
\mathcal K^\psi_R(z_3,t_1)
+
\mathcal K^\psi_L(z_3,t_1)
\right],
\\
\mathcal V^\psi_L(t_1)
&=
-\mathcal J^2
\int_{\mathcal C^+_{\beta/4}}dz_3
\left[
\mathcal K^\psi_R(t_1,z_3)
+
\mathcal K^\psi_L(t_1,z_3)
\right]
\\
&\quad
+
\mathcal J^2
\int_{\mathcal C^-_{\beta/4}}dz_3
\left[
\mathcal K^\psi_R(z_3,t_1)
+
\mathcal K^\psi_L(z_3,t_1)
\right].
\end{aligned}
\label{eq:S-large-p-vertical}
\end{equation}
Here \(\mathcal K^\psi_{R,L}\) is analytically continued to the Euclidean
segments where necessary.

Unlike the finite-\(p\) thermal term
\(\mathcal I^\psi_{\rm th}(t_1,t_2)\), its leading large-\(p\) projections
\(\mathcal V^\psi_{R,L}(t_1)\) depend only on \(t_1\). Taking a
\(t_2\) derivative of Eq.~\eqref{eq:S-first-order-large-p-KB} therefore gives
the local Liouville-type equations
\begin{equation}
\begin{aligned}
\partial_{t_1}\partial_{t_2}
g^\psi_R(t_1,t_2)
&=
2\mathcal J^2
\mathcal K^\psi_R(t_1,t_2),
\\
\partial_{t_1}\partial_{t_2}
g^\psi_L(t_1,t_2)
&=
-2\mathcal J^2
\mathcal K^\psi_L(t_1,t_2).
\end{aligned}
\label{eq:S-large-p-Liouville-general}
\end{equation}
The companion first-order equations with a derivative acting on \(t_2\)
follow by complex conjugation and exchange of the two time arguments.

The thermal gluing conditions become
\begin{equation}
g^\psi_L(t,t_0+i\beta/4)
=
g^\psi_R(t,t_0+i\beta/4),
\qquad
g^\psi_L(t_0-i\beta/4,t)
=
g^\psi_R(t_0-i\beta/4,t).
\label{eq:S-large-p-gluing}
\end{equation}
The first-order equations also supply boundary data that are lost upon taking
the second derivative. The ultraviolet normalization
\(g^\psi_R(t,t)=0\) implies
\begin{equation}
\left.
\left(
\partial_{t_1}
+
\partial_{t_2}
\right)
g^\psi_R(t_1,t_2)
\right|_{t_1=t_2=t}
=
0,
\label{eq:S-equal-time-R}
\end{equation}
while the off-diagonal component obeys
\begin{equation}
\left.
\left(
\partial_{t_1}
-
\partial_{t_2}
\right)
g^\psi_L(t_1,t_2)
\right|_{t_1=t_2=t}
=
-2i\mu_\psi(t).
\label{eq:S-equal-time-L}
\end{equation}
At a discontinuous quench, these equal-time relations are understood as
one-sided limits. Thus the local Liouville equations must be supplemented by
the thermal gluing conditions and the first-order KB boundary data.

\section*{S3. Probe limit and analytic two-point functions}

\paragraph*{Probe limit and reduced kernels.}

We now specialize the large-\(p\) equations of Sec.~S2 to the probe limit
used in the main text and Methods,
\begin{equation}
p\longrightarrow\infty,
\qquad
s\longrightarrow0,
\qquad
sp\longrightarrow0,
\qquad
\frac{p^2}{N_\eta}\longrightarrow0,
\qquad
b=d=1 .
\label{eq:S3-probe-limit}
\end{equation}
The small-sector condition \(p^2/N_\eta\to0\) controls the
pre-evaporation \(\eta\) SYK expansion even as \(s\to0\), and also
implies \(p^2/(N_\eta+N_\chi)\to0\). The \(\chi\) sector is
therefore unaffected by the \(\eta\) sector, whereas
the \(\eta\) self-energy after evaporation is determined by the fixed
\(\chi\) background. We retain the parameters \(a\) and \(\nu\) defined in
Sec.~S1; the main calculation uses
\(a=b=d=1\) and \(c=\nu=0\). The formulas below assume thermal \(\eta\) data at \(t=0\). A
formation stage with \(c\ne0\) can supply these data at leading large
\(p\) when its initial wormhole is tuned to the thermal angle specified
below; a general formation history requires its own matching data.

For \(t\geq0\), the protocol in Sec.~S1 reduces to
\begin{equation}
\begin{aligned}
l_\eta(t)
&\longrightarrow
a\,s^{1/2}\theta(t_{\rm ev}-t),
&
l_\chi(t)
&\longrightarrow1,
\\
l_c(t)
&=
\theta(t-t_{\rm ev}),
&
\mu_\eta(t)
&=
\nu\mu_0\,\theta(t-t_{\rm ev}),
\\
\mu_\chi(t)
&=
\mu_0 .
\end{aligned}
\label{eq:S3-probe-protocol}
\end{equation}
It is convenient to denote the post-evaporation difference of the bilinear
couplings by
\begin{equation}
\Delta\mu
\equiv
\mu_\chi-\mu_\eta
=
\mu_0(1-\nu).
\label{eq:S3-delta-mu}
\end{equation}

Equation~\eqref{eq:S-alpha} gives
\begin{equation}
\begin{aligned}
\alpha_\eta(t_1,t_2)
&=
a^2
\theta(t_{\rm ev}-t_1)
\theta(t_{\rm ev}-t_2),
\\
\alpha_\chi(t_1,t_2)
&=
1-
\theta(t_1-t_{\rm ev})
\theta(t_2-t_{\rm ev}).
\end{aligned}
\label{eq:S3-probe-alpha}
\end{equation}
Consequently, the kernels in Eq.~\eqref{eq:S-large-p-kernel} become
\begin{equation}
\begin{aligned}
\mathcal K^\chi_X(t_1,t_2)
&=
e^{g^\chi_X(t_1,t_2)},
\\
\mathcal K^\eta_X(t_1,t_2)
&=
a^2e^{g^\eta_X(t_1,t_2)}
\theta(t_{\rm ev}-t_1)
\theta(t_{\rm ev}-t_2)
\\
&\quad+
e^{g^\chi_X(t_1,t_2)}
\theta(t_1-t_{\rm ev})
\theta(t_2-t_{\rm ev}),
\qquad
X=R,L .
\end{aligned}
\label{eq:S3-probe-kernels}
\end{equation}
In particular,
\(\mathcal K^\eta_X(t_1,t_2)=0\) when the two arguments lie on opposite
sides of \(t_{\rm ev}\). For the main-text choice \(a=1\), substitution into
Eq.~\eqref{eq:S-large-p-Liouville-general} gives
Eqs.~\eqref{eq:large_p_chi} and \eqref{eq:large_p_eta}.

\paragraph*{Bath and pre-evaporation solutions.}

The fixed \(\chi\) background is the time-translation-invariant
traversable-wormhole solution
\cite{maldacena2018eternal,lensky2021rescuing}. Define
\(0<r_G<1\), \(\phi_G\), and \(V_G\) by
\begin{equation}
\begin{aligned}
r_G^2
=
e^{-2\phi_G}
&=
\frac{\mu_0}{2\mathcal J}
\left[
\sqrt{1+\left(\frac{\mu_0}{4\mathcal J}\right)^2}
-
\frac{\mu_0}{4\mathcal J}
\right],
\\
V_G
&=
\frac{r_G}{\sqrt{1-r_G^2}} .
\end{aligned}
\label{eq:S3-bath-parameters}
\end{equation}
Equivalently,
\[
\mu_0
=
2\mathcal J
\frac{e^{-2\phi_G}}{\sqrt{1-e^{-2\phi_G}}}.
\]
The bath correlators are
\begin{equation}
\begin{aligned}
e^{g^\chi_R(t_1,t_2)}
&=
e^{-i\pi}
\left[
\frac{V_G}{
\sin\left(
\mathcal J V_G(t_1-t_2)
-i\tanh^{-1}r_G
\right)}
\right]^2,
\\
e^{g^\chi_L(t_1,t_2)}
&=
\left[
\frac{V_G}{
\cos\left(
\mathcal J V_G(t_1-t_2)
-i\tanh^{-1}r_G
\right)}
\right]^2 .
\end{aligned}
\label{eq:S3-bath-solution}
\end{equation}
These expressions obey
\(g^\chi_R(t,t)=0\) and
\(g^\chi_L(t,t)=-2\phi_G\). Whenever \(g^\chi_{R,L}\) rather than its
exponential appears below, the continuous logarithm is chosen consistently
with these equal-time values and the conjugation relations of Sec.~S6. The
Lensky-Qi parametrization underlying this solution is also reviewed there.
This stationary bath is chosen independently of the \(\eta\) temperature.
At leading large \(p\), its initial correlators agree with TFD data tuned to
\(\vartheta_\chi=\arcsin r_G\) and
\(\beta_\chi\mathcal J r_G=\pi-2\arcsin r_G\); in general
\(\beta_\chi\ne\beta_\eta\). The \(\beta\) in the pre-evaporation
\(\eta\) solution below denotes \(\beta_\eta\). The common-temperature
finite-\(p\) preparation of Sec.~S5 is evolved numerically and is not
assumed to be an exact stationary bath saddle.

Before evaporation, with the \(\eta\) bilinear turned off, the solution for
\(0<t_1,t_2<t_{\rm ev}\) is
\begin{equation}
\begin{aligned}
e^{g^\eta_R(t_1,t_2)}
&=
e^{-i\pi}
\left[
\frac{\sin\vartheta}{
\sinh\left(
a\mathcal J\sin\vartheta\,(t_1-t_2)
-i\vartheta
\right)}
\right]^2,
\\
e^{g^\eta_L(t_1,t_2)}
&=
\left[
\frac{\sin\vartheta}{
\cosh\left(
a\mathcal J\sin\vartheta\,(t_1+t_2)
\right)}
\right]^2 .
\end{aligned}
\label{eq:S3-eta-pre}
\end{equation}
The thermal angle \(0<\vartheta<\pi/2\) is fixed by
\begin{equation}
\beta a\mathcal J\sin\vartheta
=
\pi-2\vartheta .
\label{eq:S3-thermal-angle}
\end{equation}
We use \(\vartheta\), rather than \(\epsilon\), to distinguish this thermal
parameter from the antisymmetric \(R,L\) matrix in
Eq.~\eqref{eq:S-epsilon}.

\paragraph*{Matching across the evaporation surface.}

Consider first \(t_2<t_{\rm ev}<t_1\). Because the mixed-region
\(\eta\) kernel vanishes, the local Liouville equation is homogeneous there.
The first-order KB equation \eqref{eq:S-first-order-large-p-KB} fixes the two
single-time functions that are left undetermined by the second-order
equation. The result is
\begin{equation}
\begin{aligned}
g^\eta_X(t_1,t_2)
&=
g^\eta_X(t_{\rm ev},t_2)
+
\operatorname{Re}g^\chi_R(t_1,t_{\rm ev})
+i\operatorname{Im}g^\chi_L(t_1,t_{\rm ev})
+i\Delta\mu\,(t_1-t_{\rm ev}),
\qquad
X=R,L .
\end{aligned}
\label{eq:S3-cross-solution}
\end{equation}
The single-time functions \(f_{1X}\) and \(f_{2X}\) appearing in
Eq.~\eqref{eq:regionII} can therefore be read off directly from
Eq.~\eqref{eq:S3-cross-solution}.

The region \(t_1<t_{\rm ev}<t_2\) follows from
\begin{equation}
g^\eta_X(t_1,t_2)
=
\left[g^\eta_X(t_2,t_1)\right]^* .
\label{eq:S3-conjugation}
\end{equation}

Equation~\eqref{eq:S3-cross-solution} makes the factorization of the
mixed-region correlator explicit:
\begin{equation}
\left|e^{g^\eta_X(t_1,t_2)}\right|
=
\left|e^{g^\eta_X(t_{\rm ev},t_2)}\right|
\left|e^{g^\chi_R(t_1,t_{\rm ev})}\right|,
\qquad
X=R,L .
\label{eq:S3-cross-factorization}
\end{equation}
The same factorization holds with every exponent replaced by \(g/p\), as in
the physical Wightman functions. Since the bath correlator is bounded in the traversable-wormhole background, i.e. $ \left|e^{g^\chi_R(t_1,t_{\mathrm{ev}})}\right|
  \leq
  1$,
we obtain

\begin{equation}
  \left|e^{g^\eta_X(t_1,t_2)}\right|
  \leq
  \left|e^{g^\eta_X(t_{\rm ev},t_2)}\right|,
  \qquad
  X=R,L .
\label{eq:S3-cross-region-bound}
\end{equation}
Thus the simple \(\eta\) correlator cannot grow in magnitude across the
evaporation surface.

\paragraph*{Post-evaporation solution.}

For \(t_1,t_2>t_{\rm ev}\), the probe kernel is sourced entirely by the bath,
\begin{equation}
\mathcal K^\eta_X(t_1,t_2)
=
e^{g^\chi_X(t_1,t_2)},
\qquad
X=R,L .
\end{equation}
The second-order Liouville equation determines
\(\partial_{t_1}\partial_{t_2}g^\eta_X\), but by itself leaves undetermined
functions of the individual time arguments. These functions are fixed by the
first-order KB equation. In the leading large-\(p\) probe equations, the KB
evolution is causal in its first time argument and can be integrated from
\(t_1=t_{\rm ev}\) in the form
\begin{equation}
g^\eta_X(t_1,t_2)
=
g^\eta_X(t_{\rm ev},t_2)
+
\mathcal F^\chi_X(t_1,t_2),
\qquad
\mathcal F^\chi_X(t_{\rm ev},t_2)=0 ,
\qquad
X=R,L ,
\label{eq:S3-post-causal-form}
\end{equation}
where \(\mathcal F^\chi_X\) is completely determined by the known bath
correlator and the prescribed post-evaporation bilinear couplings.

Setting \(t_2=t_{\rm ev}\) in
Eq.~\eqref{eq:S3-post-causal-form}, and then using the real-time conjugation
relation in Eq.~\eqref{eq:S3-conjugation}, gives
\begin{equation}
\begin{aligned}
g^\eta_X(t_1,t_{\rm ev})
&=
g^\eta_X(t_{\rm ev},t_{\rm ev})
+
\mathcal F^\chi_X(t_1,t_{\rm ev}),
\\
g^\eta_X(t_{\rm ev},t_2)
&=
g^\eta_X(t_{\rm ev},t_{\rm ev})
+
\left[
\mathcal F^\chi_X(t_2,t_{\rm ev})
\right]^* .
\end{aligned}
\label{eq:S3-post-boundary-data}
\end{equation}
Here we used the reality of the equal-time values. Substitution back into
Eq.~\eqref{eq:S3-post-causal-form} yields
\begin{equation}
g^\eta_X(t_1,t_2)
=
g^\eta_X(t_{\rm ev},t_{\rm ev})
+
\mathcal F^\chi_X(t_1,t_2)
+
\left[
\mathcal F^\chi_X(t_2,t_{\rm ev})
\right]^* .
\label{eq:S3-post-causal-closure}
\end{equation}
Thus no independent function on the matching surface is required: the
first-order evolution and conjugation reduce the \(\eta\) matching data to
their coincident values at \((t_{\rm ev},t_{\rm ev})\). This closure is a
feature of the leading large-\(p\), probe-limit equations used here and would
not follow from the second-order Liouville equation alone.

Evaluating the bath-dependent functional explicitly, define
\begin{equation}
\begin{aligned}
f_{3R}(t)
&=
i\operatorname{Im}
\left[
-g^\chi_R(t,t_{\rm ev})
+g^\chi_L(t,t_{\rm ev})
\right]
+i\Delta\mu\,(t-t_{\rm ev}),
\\
f_{3L}(t)
&=\frac{1}{2}
\left[
g^\eta_L(t_{\rm ev},t_{\rm ev})
+
g^\chi_L(t_{\rm ev},t_{\rm ev})
\right]+
\operatorname{Re}
\left[
g^\chi_R(t,t_{\rm ev})
-g^\chi_L(t,t_{\rm ev})
\right]
+i\Delta\mu\,(t-t_{\rm ev}).
\end{aligned}
\label{eq:S3-post-functions}
\end{equation}
The complete post-evaporation solution is therefore
\begin{equation}
g^\eta_X(t_1,t_2)
=
g^\chi_X(t_1,t_2)
+
f_{3X}(t_1)
+
\left[f_{3X}(t_2)\right]^*,
\qquad
X=R,L ,
\label{eq:S3-post-solution}
\end{equation}
which is the explicit form of Eq.~\eqref{eq:regionIII}. The required
equal-time matching data follow from the pre-evaporation solution:
\begin{equation}
g^\eta_R(t_{\rm ev},t_{\rm ev})=0,
\qquad
\left|e^{g^\eta_L(t_{\rm ev},t_{\rm ev})}\right|
=
\left[
\frac{\sin\vartheta}{
\cosh\left(
2a\mathcal Jt_{\rm ev}\sin\vartheta
\right)}
\right]^2 .
\label{eq:S3-matching-data}
\end{equation}

Since \(f_{3R}\) is purely imaginary,
\begin{equation}
\left|e^{g^\eta_R(t_1,t_2)}\right|
=
\left|e^{g^\chi_R(t_1,t_2)}\right|.
\label{eq:S3-post-R-magnitude}
\end{equation}

\paragraph*{Post-evaporation left-right bound.}

For the left-right component, Eq.~\eqref{eq:S3-post-solution} gives
\begin{equation}
\left|e^{g^\eta_L(t_1,t_2)}\right|
=
h_\chi(t_1,t_2;t_{\rm ev})
\left|e^{g^\eta_L(t_{\rm ev},t_{\rm ev})}\right|,
\label{eq:S3-post-L-magnitude}
\end{equation}
where the nonnegative bath-dependent transfer factor is
\begin{equation}
\begin{aligned}
h_\chi(t_1,t_2;t_{\rm ev})
&\equiv
\left|
e^{
g^\chi_L(t_1,t_2)
+
g^\chi_L(t_{\rm ev},t_{\rm ev})
}
\times
\frac{e^{
g^\chi_R(t_1,t_{\rm ev})
+
g^\chi_R(t_{\rm ev},t_2)
}
}{
e^{
g^\chi_L(t_1,t_{\rm ev})
+
g^\chi_L(t_{\rm ev},t_2)
}
}
\right| .
\end{aligned}
\label{eq:S3-post-L-ratio}
\end{equation}

\textit{Post-evaporation bound:}
Suppose that the bath correlators admit the normalized physical Lensky-Qi
parametrization with a common differentiable function \(y(t)\) satisfying
\begin{equation}
\operatorname{Im}y(t)<0,
\qquad
|y'(t)|
=
\sinh\left(
2\mathcal J|\operatorname{Im}y(t)|
\right),
\label{eq:S3-LQ-bound-assumptions}
\end{equation}
and that the post-evaporation matching form
\eqref{eq:S3-post-solution} applies. Then
\begin{equation}
0
\leq
h_\chi(t_1,t_2;t_{\rm ev})
\leq
1,
\qquad
t_1,t_2\geq t_{\rm ev}.
\label{eq:S3-post-L-ratio-bound}
\end{equation}
The proof is given in Sec.~S6 using the disk representation
\(w(t)=e^{-2i\mathcal J y(t)}\). The disk inequality does not require the
bath trajectory to be a fixed point or to be time-translation invariant;
its scope is the normalized single-function Lensky-Qi class, conditional on
the post-evaporation matching form derived above. The phase proportional to
the real parameter \(\Delta\mu\) drops out of the modulus.

For the stationary bath used in the main text, define
\(x_i=\mathcal J V_G(t_i-t_{\rm ev})\). Direct substitution of
Eq.~\eqref{eq:S3-bath-solution} gives
\begin{equation}
\begin{aligned}
h_\chi(t_1,t_2;t_{\rm ev})
&=
\frac{
\left(V_G^2+\cos^2x_1\right)
\left(V_G^2+\cos^2x_2\right)
}{
\left(V_G^2+\sin^2x_1\right)
\left(V_G^2+\sin^2x_2\right)
}
\times
\frac{V_G^4}{
\left[V_G^2+\cos^2(x_1-x_2)\right]
\left(V_G^2+1\right)
}
\leq1 .
\end{aligned}
\label{eq:S3-post-L-bound-factor}
\end{equation}
The non-strict inequality is essential: it is saturated at \(t_1=t_2=t_{\rm ev}\), and whenever both disk points
\(w(t_1)\) and \(w(t_2)\) return to \(w(t_{\rm ev})\). A single
argument on the matching surface does not in general saturate the bound.

Combining Eqs.~\eqref{eq:S3-post-L-magnitude} and
\eqref{eq:S3-post-L-ratio-bound}, we obtain
\begin{equation}
\left|e^{g^\eta_L(t_1,t_2)}\right|
\leq
\left|e^{g^\eta_L(t_{\rm ev},t_{\rm ev})}\right|,
\qquad
t_1,t_2\geq t_{\rm ev}.
\label{eq:S3-post-L-bound}
\end{equation}
Since \(p>0\), the corresponding physical Wightman factors obey
\begin{equation}
\left|e^{g^\eta_L(t_1,t_2)/p}\right|
\leq
\left|e^{g^\eta_L(t_{\rm ev},t_{\rm ev})/p}\right|.
\end{equation}
Thus the post-evaporation left-right correlator cannot exceed its value on
the evaporation surface. Its dependence on the pre-evaporation history is
carried entirely by this matching value, which becomes parametrically small
at late \(t_{\rm ev}\) according to Eq.~\eqref{eq:S3-matching-data}.

\paragraph*{Anticommutator matrix.}
For bulk reconstruction it is useful to package the simple-channel
anticommutators into a \(2\times2\) matrix.  Flavor symmetry gives
\begin{equation}
\label{eq:S3-anticommutator-flavor}
  A_{aj,bk}(t_1,t_2)
  =
  \left\langle
  \left\{
  \eta^a_j(t_1),\eta^b_k(t_2)
  \right\}
  \right\rangle
  =
  \delta_{jk}\,A_{ab}(t_1,t_2),
  \qquad
  a,b\in\{R,L\}.
\end{equation}
In terms of Wightman functions,
\begin{equation}
\label{eq:S3-anticommutator-wightman}
\begin{aligned}
  A_{ab}(t_1,t_2)
  &=
  i\left[
  G^{\eta,>}_{ab}(t_1,t_2)
  +
  G^{\eta,>}_{ba}(t_2,t_1)
  \right]
  \\
  &=
  i\left[
  G^{\eta,>}_{ab}(t_1,t_2)
  -
  \left(G^{\eta,>}_{ab}(t_1,t_2)\right)^*
  \right]
  =
  -2\,\operatorname{Im}G^{\eta,>}_{ab}(t_1,t_2).
\end{aligned}
\end{equation}
Using the large-\(p\) ansatz,
\begin{equation}
\label{eq:S3-anticommutator-components}
  A_{RR}(t_1,t_2)
  =
  \operatorname{Re}\left[
  e^{g^\eta_R(t_1,t_2)/p}
  \right],
  \qquad
  A_{RL}(t_1,t_2)
  =
  \operatorname{Im}\left[
  e^{g^\eta_L(t_1,t_2)/p}
  \right].
\end{equation}
Together with \(A_{LL}=A_{RR}\) and \(A_{LR}=-A_{RL}\), this gives
\begin{equation}
\label{eq:S3-anticommutator-matrix}
  A(t_1,t_2)
  =
  \begin{pmatrix}
    \operatorname{Re}\left[e^{g^\eta_R(t_1,t_2)/p}\right]
    &
    \operatorname{Im}\left[e^{g^\eta_L(t_1,t_2)/p}\right]
    \\
    -\operatorname{Im}\left[e^{g^\eta_L(t_1,t_2)/p}\right]
    &
    \operatorname{Re}\left[e^{g^\eta_R(t_1,t_2)/p}\right]
  \end{pmatrix}.
\end{equation}
The conjugation properties of \(g^\eta_R\) and \(g^\eta_L\) imply
\begin{equation}
\label{eq:S3-anticommutator-symmetry}
  A_{RR}(t_2,t_1)=A_{RR}(t_1,t_2),
  \qquad
  A_{RL}(t_2,t_1)=-A_{RL}(t_1,t_2).
\end{equation}
This anticommutator matrix supplies the boundary inner product used in the
bulk reconstruction summarized in Eq.~\eqref{eq:HKLL_kernel}.

\section*{S4. Operator size calculation}

\noindent
Methods states the folded-contour response used to compute \(\Delta L_\eta\)
and \(\Delta L_\chi\). Here we derive the response functions and list their
region-by-region forms.

We study the fate of a simple \(\eta\) excitation created before evaporation
and ask how it is detected by the size operators in the \(\eta\) and \(\chi\)
sectors after evaporation. We use \(\psi\in\{\eta,\chi\}\) to denote the
sector whose size operator is measured.

Using the creation and annihilation operators defined in
Eq.~\eqref{eq:S-c-def}, the single-mode number operator is
\begin{equation}
  \hat n^\psi_j
  =
  \left(c^\psi_j\right)^\dagger c^\psi_j
  =
  \frac{1}{2}
  +
  i\psi^L_j\psi^R_j .
\label{eq:S4-number-operator}
\end{equation}
Thus
\begin{equation}
  \widehat{\mathcal N}_\psi
  =
  \sum_{j=1}^{N_\psi}\hat n^\psi_j
  =
  \frac{N_\psi}{2}
  +
  \hat L_\psi,
  \qquad
  \hat L_\psi
  =
  i\sum_{j=1}^{N_\psi}\psi^L_j\psi^R_j .
\label{eq:S4-total-number}
\end{equation}
The constant \(N_\psi/2\) cancels in differences of expectation values, so
\(\Delta\mathcal N_\psi=\Delta L_\psi\).

Let \(\lvert\Phi\rangle\) denote the unperturbed state. A normalized simple
\(\eta\) excitation inserted from the right boundary at
\(0<t_b<t_{\rm ev}\) is
\begin{equation}
  \lvert\Phi_{b,j}\rangle
  =
  \sqrt{2}\,\eta^R_j(t_b)\lvert\Phi\rangle .
\label{eq:S4-boundary-excitation}
\end{equation}
Its background-subtracted size at a measurement time \(t_0>t_{\rm ev}\) is
\begin{equation}
\begin{aligned}
  \Delta L^{\rm bdry}_\psi(t_b;t_0)
  &=
  \frac{1}{N_\eta}
  \sum_{j=1}^{N_\eta}
  \left[
  \langle\Phi_{b,j}\rvert
  \hat L_\psi(t_0)
  \lvert\Phi_{b,j}\rangle
  -
  \langle\Phi\rvert
  \hat L_\psi(t_0)
  \lvert\Phi\rangle
  \right].
\end{aligned}
\label{eq:S4-physical-boundary-size}
\end{equation}
The normalization is by \(N_\eta\), because the inserted fermion belongs to
the \(\eta\) sector.

To compute this response, introduce the \(RR\) generating correlator
\begin{equation}
\begin{aligned}
  f_\psi(t_1,t_2;t_0;\lambda)
  &=
  -\frac{2i}{N_\eta}
  \sum_{j=1}^{N_\eta}
  \langle\Phi\rvert
  e^{i\lambda\hat L_\psi(t_0)/p}
  \eta^R_j(t_1)
  e^{-i\lambda\hat L_\psi(t_0)/p}
  \eta^R_j(t_2)
  \lvert\Phi\rangle .
\end{aligned}
\label{eq:S4-generating-correlator}
\end{equation}
For equal insertion times,
\begin{equation}
  \Delta L^{\rm bdry}_\psi(t_b;t_0)
  =
  -p\,
  \left.
  \partial_\lambda
  f_\psi(t_b,t_b;t_0;\lambda)
  \right|_{\lambda=0}.
\label{eq:S4-response-from-generating}
\end{equation}

The conjugations in Eq.~\eqref{eq:S4-generating-correlator} are represented
by folding the real-time contour at \(t_0\). On the unfolded contour the
Hamiltonian is
\begin{equation}
  H^{(\psi)}_\lambda(t)
  =
  \theta(t_0-t)H(t)
  -
  \theta(t-t_0)H(2t_0-t)
  +
  \frac{\lambda}{p}
  \delta(t-t_0)\hat L_\psi .
\label{eq:S4-folded-Hamiltonian}
\end{equation}
For a physical time \(t_1<t_0\) on the return branch, define
\begin{equation}
  \widetilde t_1
  =
  2t_0-t_1,
  \qquad
  \widetilde t_{\rm ev}
  =
  2t_0-t_{\rm ev}.
\label{eq:S4-reflected-coordinates}
\end{equation}
When the Lensky--Qi formula is expressed in unfolded contour times, its
right-hand side includes the fixed orientation factor
\(\varsigma(u)\varsigma(v)\), where \(\varsigma=+1\) on the outward
branch and \(\varsigma=-1\) on the return branch. This compensates the
reversed derivative of the folded trajectory and preserves the equal-physical-time
normalization. It is independent of \(\lambda\) and therefore cancels from
all logarithmic responses below.
Thus \(t_1,t_2\) below denote physical boundary times, whereas
\(\widetilde t_1>t_0>t_2\) is the corresponding independent contour
ordering. In this ordering,
\begin{equation}
  f_\psi(t_1,t_2;t_0;\lambda)
  =
  2G^{\eta,>}_{RR,\psi}
  (\widetilde t_1,t_2;t_0,\lambda),
\label{eq:S4-generating-G}
\end{equation}
where the subscript \(\psi\) records the sector in which the source is
inserted; the external fermions remain \(\eta\) fermions. Using
\begin{equation}
  G^{\eta,>}_{RR,\psi}
  (\widetilde t_1,t_2;t_0,\lambda)
  =
  -\frac{i}{2}
  \exp\left[
  \frac{
  g^\eta_{R,\psi}(\widetilde t_1,t_2;t_0,\lambda)
  }{p}
  \right],
\label{eq:S4-source-deformed-large-p}
\end{equation}
define
\begin{equation}
\begin{aligned}
  g^\eta_{R,0}(t_1,t_2)
  &\equiv
  g^\eta_{R,\psi}(\widetilde t_1,t_2;t_0,0),
  \\
  \mathfrak D_\psi(t_1,t_2;t_0)
  &\equiv
  i\left.
  \partial_\lambda
  g^\eta_{R,\psi}(\widetilde t_1,t_2;t_0,\lambda)
  \right|_{\lambda=0}.
\end{aligned}
\label{eq:S4-D-definition}
\end{equation}
It follows that the \(RR\) bilocal response is
\begin{equation}
  \Delta L_\psi(t_1,t_2;t_0)
  =
  \exp\left[
  \frac{g^\eta_{R,0}(t_1,t_2)}{p}
  \right]
  \mathfrak D_\psi(t_1,t_2;t_0).
\label{eq:S4-response-D-relation}
\end{equation}
For unequal \(t_1,t_2\), this is a folded-contour bilocal response rather
than an individual real size expectation value. For \(t_1=t_2=t_b\), the
unperturbed folded evolution cancels and
\(g^\eta_{R,0}(t_b,t_b)=0\). Therefore
\begin{equation}
  \Delta L^{\rm bdry}_\psi(t_b;t_0)
  =
  \mathfrak D_\psi(t_b,t_b;t_0).
\label{eq:S4-physical-size-D}
\end{equation}

\paragraph*{Boundary data at the reflected evaporation surface.}

We first determine the data that the kick at \(t_0\) induces at
\(\widetilde t_{\rm ev}\). The natural variables are the Lensky--Qi variables
reviewed in Sec.~S6. In particular, Eqs.~\eqref{eq:S6-y-phase-parametrization}
and \eqref{eq:S6-LQ-Hamiltonian} give
\begin{equation}
  y'(t)
  =
  \frac{e^{ip(t)}}{\sqrt{e^{2\phi(t)}-1}},
  \qquad
  \mathcal H_Q
  =
  -2\mathcal J_{\rm eff}
  \sqrt{1-e^{-2\phi}}\cos p
  +
  \mu_{\rm eff}(t)\phi .
\label{eq:S4-LQ-reminder}
\end{equation}
Here, as in Sec.~S6, \(p(t)\) is the Lensky--Qi canonical phase and should not
be confused with the interaction order \(p\).

The delta-function source in Eq.~\eqref{eq:S4-folded-Hamiltonian} shifts the
bilinear coupling and hence produces the canonical jump
\begin{equation}
  p(t_0^+)-p(t_0^-)
  =
  -\lambda .
\label{eq:S4-canonical-kick}
\end{equation}
Writing \(\delta\equiv\partial_\lambda|_{\lambda=0}\), the directly kicked
sector therefore obeys
\begin{equation}
  \delta\phi(t_0)=0,
  \qquad
  \delta p(t_0)=-1.
\label{eq:S4-local-kick-data}
\end{equation}

The post-\(t_0\) segment transports this kick to
\(\widetilde t_{\rm ev}\). We denote the matching data induced in the
pre-evaporation \(\eta\) trajectory by
\(\delta\phi_\psi^{\rm eff}(\widetilde t_{\rm ev})\) and
\(\delta p_\psi^{\rm eff}(\widetilde t_{\rm ev})\). The superscript
``eff'' distinguishes these data from the actual bath fluctuation produced
by a \(\chi\)-sector kick. Both sources give
\begin{equation}
  \delta\phi_\psi^{\rm eff}(\widetilde t_{\rm ev})=0,
  \qquad
  \delta p_\psi^{\rm eff}(\widetilde t_{\rm ev})\in\mathbb R,
\label{eq:S4-effective-data-form}
\end{equation}
with
\begin{equation}
\begin{aligned}
  \delta p_\eta^{\rm eff}(\widetilde t_{\rm ev})
  &=
  -1,
  \\
  \delta p_\chi^{\rm eff}(\widetilde t_{\rm ev})
  &=
  \frac{1-r_G^2}{2-r_G^2}
  \left[
  1-
  \cos\left(
  \omega_G(t_0-t_{\rm ev})
  \right)
  \right].
\end{aligned}
\label{eq:S4-effective-matching-summary}
\end{equation}
Here
\begin{equation}
  \omega_G^2
  =
  4\mathcal J^2
  \frac{r_G^2(2-r_G^2)}{1-r_G^2},
\label{eq:S4-bath-frequency}
\end{equation}
and \(r_G,\phi_G\), and \(V_G\) are defined in
Eq.~\eqref{eq:S3-bath-parameters}. We derive
Eqs.~\eqref{eq:S4-effective-data-form} and
\eqref{eq:S4-effective-matching-summary} below.

\paragraph*{Pre-evaporation propagation.}

The effect of the post-evaporation source is summarized by the effective data
at \(\widetilde t_{\rm ev}\), while propagation from \(t_{\rm ev}\) back to
the insertion time is a purely pre-evaporation \(\eta\) problem. This step is
the same for the \(\eta\)- and \(\chi\)-size sources; the only difference is
the value of \(\delta p_\psi^{\rm eff}(\widetilde t_{\rm ev})\).

Set
\begin{equation}
  \mathcal J_\eta
  \equiv
  a\mathcal J .
\label{eq:S4-Jeta}
\end{equation}
In terms of the thermal angle \(\vartheta\) defined in
Eq.~\eqref{eq:S3-thermal-angle}, the unperturbed pre-evaporation trajectory is
\begin{equation}
  y_{0,\eta}(t)
  =
  \frac{1}{\mathcal J_\eta}
  \tan^{-1}
  \tanh\left(
  \mathcal J_\eta t\sin\vartheta
  -
  \frac{i\vartheta}{2}
  \right),
\label{eq:S4-unperturbed-eta-trajectory}
\end{equation}
in agreement with Eq.~\eqref{eq:S6-eta-trajectory}. The linearized solution
generated by a \(\psi\)-sector source can be written as
\begin{equation}
\begin{aligned}
  \delta_\psi y_\eta(t)
  &=
  \delta\gamma_\psi
  +
  y'_{0,\eta}(t)
  \left[
  -\delta t_{\star,\psi}
  +
  \delta\vartheta_\psi
  \frac{
  2\mathcal J_\eta t\cos\vartheta-i
  }{
  2\mathcal J_\eta\sin\vartheta
  }
  \right].
\end{aligned}
\label{eq:S4-linearized-eta-trajectory}
\end{equation}
These three terms are obtained by varying the additive \(y\)-shift, time
origin, and thermal angle of the exact trajectory, and hence solve the
linearized Liouville equation. Imposing the data in
Eq.~\eqref{eq:S4-effective-data-form}, together with coordinate continuity
\(\delta_\psi y_\eta(t_{\rm ev})=0\), fixes all three constants and gives
\begin{equation}
\begin{aligned}
  \delta\vartheta_\psi
  &=
  \delta p_\psi^{\rm eff}(\widetilde t_{\rm ev})
  \tanh\left(
  2\mathcal J_\eta t_{\rm ev}\sin\vartheta
  \right),
  \\
  \delta t_{\star,\psi}
  &=
  \delta p_\psi^{\rm eff}(\widetilde t_{\rm ev})
  \cot\vartheta
  \left[
  t_{\rm ev}
  \tanh\left(
  2\mathcal J_\eta t_{\rm ev}\sin\vartheta
  \right)
  -
  \frac{\csc\vartheta}{2\mathcal J_\eta}
  \right],
  \\
  \delta\gamma_\psi
  &=
  -\frac{
  \delta p_\psi^{\rm eff}(\widetilde t_{\rm ev})
  }{
  2\mathcal J_\eta
  }
  \csc\vartheta\,
  \operatorname{sech}\left(
  2\mathcal J_\eta t_{\rm ev}\sin\vartheta
  \right).
\end{aligned}
\label{eq:S4-linearized-eta-parameters}
\end{equation}

For two physical times \(t_1,t_2<t_{\rm ev}\), variation of the Lensky--Qi
form in Eq.~\eqref{eq:S6-LQ-parametrization} gives
\begin{equation}
\begin{aligned}
  \mathfrak D^{\rm pre}_\psi(t_1,t_2;t_0)
  &=
  i
  \Bigg[
  \frac{
  \delta_\psi y'_\eta(t_1)
  }{
  y'_{0,\eta}(t_1)
  }
  -
  2\mathcal J_\eta
  \delta_\psi y_\eta(t_1)
  \cot\left(
  \mathcal J_\eta
  \left[
  y_{0,\eta}(t_1)
  -
  y_{0,\eta}(t_2)^*
  \right]
  \right)
  \Bigg].
\end{aligned}
\label{eq:S4-pre-bilocal-response}
\end{equation}
All times on the right-hand side are reflected
physical times; this convention accounts for the orientation of the return
branch.

For the equal-time boundary insertion, substituting
Eq.~\eqref{eq:S4-linearized-eta-parameters} into
Eq.~\eqref{eq:S4-pre-bilocal-response} gives
\begin{equation}
  \mathfrak D_\psi(t_b,t_b;t_0)
  =
  -\delta p_\psi^{\rm eff}(\widetilde t_{\rm ev})\,
  \mathcal E(t_b,t_{\rm ev}),
\label{eq:S4-universal-physical-response}
\end{equation}
where
\begin{equation}
\begin{aligned}
  \mathcal E(t_b,t_{\rm ev})
  &=
  1
  -
  \csc^2\vartheta
  \left[
  1-
  \frac{
  \cosh\left(
  2\mathcal J_\eta t_b\sin\vartheta
  \right)
  }{
  \cosh\left(
  2\mathcal J_\eta t_{\rm ev}\sin\vartheta
  \right)
  }
  \right]
  \\
  &\quad
  +
  2\mathcal J_\eta(t_{\rm ev}-t_b)
  \frac{\cos^2\vartheta}{\sin\vartheta}
  \tanh\left(
  2\mathcal J_\eta t_{\rm ev}\sin\vartheta
  \right).
\end{aligned}
\label{eq:S4-enhancement-factor}
\end{equation}

\paragraph*{Propagating the kick to the reflected evaporation surface.}

We now derive the effective data quoted in
Eq.~\eqref{eq:S4-effective-matching-summary}. For a source in the \(\eta\)
sector, the bath is unperturbed in the probe limit. On the return branch,
for \(t_0<u<\widetilde t_{\rm ev}\) and \(t_2<t_{\rm ev}\), the mixed-region
solution following from Eq.~\eqref{eq:S3-cross-solution} is
\begin{equation}
\begin{aligned}
  g_R^\eta(u,t_2;t_0,\lambda)
  &=
  g_R^\eta(t_{\rm ev},t_2)
  +
  \operatorname{Re}g_R^\chi(u,t_{\rm ev})
  +
  i\,\operatorname{Im}g_L^\chi(u,t_{\rm ev})
  \\
  &\quad
  +
  i\Delta\mu\,
  (2t_0-u-t_{\rm ev})
  -
  i\lambda ,
\end{aligned}
\label{eq:S4-eta-kick-mixed-region}
\end{equation}
where \(\Delta\mu\) is defined in Eq.~\eqref{eq:S3-delta-mu}. Setting
\(u=\widetilde t_{\rm ev}\) and \(t_2=t_{\rm ev}\) gives
\begin{equation}
  \left.
  \partial_\lambda
  g_R^\eta(
  \widetilde t_{\rm ev},t_{\rm ev};t_0,\lambda)
  \right|_{\lambda=0}
  =
  -i.
\label{eq:S4-eta-boundary-variation}
\end{equation}

From Eq.~\eqref{eq:S4-LQ-reminder},
\begin{equation}
  \frac{\delta y'}{y'}
  =
  i\,\delta p
  -
  \frac{\delta\phi}{1-e^{-2\phi}}.
\label{eq:S4-yprime-variation}
\end{equation}
At the matching surface
\(\delta_\eta y_\eta(t_{\rm ev})=0\) and
\(\delta\phi_\eta^{\rm eff}(\widetilde t_{\rm ev})=0\). Hence
Eq.~\eqref{eq:S4-eta-boundary-variation} implies
\begin{equation}
  \delta p_\eta^{\rm eff}(\widetilde t_{\rm ev})
  =
  -1.
\label{eq:S4-eta-effective-momentum}
\end{equation}

For a source in the \(\chi\) sector, the kick first excites the bath. The
unperturbed bath is the fixed point
\begin{equation}
  \phi_\chi=\phi_G,
  \qquad
  p_\chi=0,
\label{eq:S4-chi-fixed-point}
\end{equation}
as in Eq.~\eqref{eq:S6-chi-fixed-point}. To distinguish the actual bath
fluctuation from the effective pre-evaporation matching datum, write it as
\((\delta\phi_\chi^{\rm bath},\delta p_\chi^{\rm bath})\). Its initial data
are
\begin{equation}
  \delta\phi_\chi^{\rm bath}(t_0)=0,
  \qquad
  \delta p_\chi^{\rm bath}(t_0)=-1.
\label{eq:S4-chi-local-kick}
\end{equation}
Linearizing the Hamilton equations in
Eq.~\eqref{eq:S6-LQ-Hamilton-equations} on the return branch gives
\begin{equation}
\begin{aligned}
  -\delta\dot\phi_\chi^{\rm bath}
  &=
  2\mathcal J\sqrt{1-r_G^2}\,
  \delta p_\chi^{\rm bath},
  \\
  -\delta\dot p_\chi^{\rm bath}
  &=
  -2\mathcal J
  \frac{
  r_G^2(2-r_G^2)
  }{
  (1-r_G^2)^{3/2}
  }
  \delta\phi_\chi^{\rm bath}.
\end{aligned}
\label{eq:S4-bath-linearized-equations}
\end{equation}
Equivalently,
\(\delta\ddot\phi_\chi^{\rm bath}
+\omega_G^2\delta\phi_\chi^{\rm bath}=0\), with \(\omega_G\) as in
Eq.~\eqref{eq:S4-bath-frequency}. The solution is
\begin{equation}
\begin{aligned}
  \delta\phi_\chi^{\rm bath}(t)
  &=
  \frac{
  2\mathcal J\sqrt{1-r_G^2}
  }{
  \omega_G
  }
  \sin\left[
  \omega_G(t-t_0)
  \right],
  \\
  \delta p_\chi^{\rm bath}(t)
  &=
  -\cos\left[
  \omega_G(t-t_0)
  \right],
  \qquad
  t>t_0.
\end{aligned}
\label{eq:S4-bath-kick-solution}
\end{equation}
Substituting this bath fluctuation into the linearized mixed-region matching
at \(t=\widetilde t_{\rm ev}\) gives
\begin{equation}
\begin{aligned}
  \delta\phi_\chi^{\rm eff}(\widetilde t_{\rm ev})
  &=
  0,
  \\
  \delta p_\chi^{\rm eff}(\widetilde t_{\rm ev})
  &=
  \frac{1-r_G^2}{2-r_G^2}
  \left[
  1-
  \cos\left(
  \omega_G(t_0-t_{\rm ev})
  \right)
  \right],
\end{aligned}
\label{eq:S4-chi-effective-momentum}
\end{equation}
which completes the derivation of
Eq.~\eqref{eq:S4-effective-matching-summary}.

Combining Eqs.~\eqref{eq:S4-physical-size-D},
\eqref{eq:S4-universal-physical-response},
\eqref{eq:S4-eta-effective-momentum}, and
\eqref{eq:S4-chi-effective-momentum}, one obtains
\begin{equation}
\begin{aligned}
  \Delta L^{\rm bdry}_\eta(t_b;t_0)
  &=
  \mathcal E(t_b,t_{\rm ev}),
  \\
  \Delta L^{\rm bdry}_\chi(t_b;t_0)
  &=
  -\frac{1-r_G^2}{2-r_G^2}
  \left[
  1-
  \cos\left(
  \omega_G(t_0-t_{\rm ev})
  \right)
  \right]
  \mathcal E(t_b,t_{\rm ev}).
\end{aligned}
\label{eq:S4-boundary-size-results}
\end{equation}
The negative \(\chi\) response is consistent with this definition: it is a change relative to the background size, rather than an absolute occupation number.

\paragraph*{Bilocal size response in all regions.}

We now keep the two physical insertion times \(t_1,t_2<t_0\) independent.
The generating correlator remains Eq.~\eqref{eq:S4-generating-correlator},
and on the unfolded contour we work in the independent ordering
\begin{equation}
  \widetilde t_1=2t_0-t_1>t_0>t_2.
\label{eq:S4-bilocal-domain}
\end{equation}
The corresponding response is given by
Eq.~\eqref{eq:S4-response-D-relation}. The exponential prefactor equals one
for \(t_1=t_2=t_b\), but is nontrivial for a general bilocal.

\paragraph*{Linearized bath solution.}

For a \(\chi\)-sector source we require the variation of the bath functions
that enter the \(RR\) probe solution. The unperturbed folded bath trajectory
is
\begin{equation}
  y^{\rm fold}_{0,\chi}(t)
  =
  \begin{cases}
  \displaystyle
  V_Gt
  -
  \frac{i}{2\mathcal J}\tanh^{-1}(r_G),
  &
  t<t_0,
  \\[6pt]
  \displaystyle
  V_G(2t_0-t)
  -
  \frac{i}{2\mathcal J}\tanh^{-1}(r_G),
  &
  t>t_0.
  \end{cases}
\label{eq:S4-folded-bath-trajectory}
\end{equation}
For \(t>t_0\), the kick-induced variation of \(y\) is
\begin{equation}
\begin{aligned}
  \delta_\chi y_{\rm bath}(t)
  &=
  \frac{iV_G}{\omega_G}
  \sin\left[
  \omega_G(t-t_0)
  \right]+
  \frac{
  2\mathcal J V_G
  }{
  \omega_G^2\sqrt{1-r_G^2}
  }
  \left[
  1-
  \cos\left(
  \omega_G(t-t_0)
  \right)
  \right].
\end{aligned}
\label{eq:S4-folded-bath-variation}
\end{equation}

For \(u>t_0>v\), define
\begin{equation}
  \mathcal B_X(u,v;t_0)
  \equiv
  \left.
  \partial_\lambda
  g^\chi_X(u,v;t_0,\lambda)
  \right|_{\lambda=0},
  \qquad
  X=R,L .
\label{eq:S4-bath-response-definition}
\end{equation}
The Lensky--Qi parametrization gives
\begin{equation}
\begin{aligned}
  \mathcal B_R(u,v;t_0)
  &=
  \frac{
  \delta_\chi y'_{\rm bath}(u)
  }{
  \left(y^{\rm fold}_{0,\chi}\right)'(u)
  }
  -
  2\mathcal J\,
  \delta_\chi y_{\rm bath}(u)
  \cot\left(
  \mathcal J
  \left[
  y^{\rm fold}_{0,\chi}(u)
  -
  y^{\rm fold}_{0,\chi}(v)^*
  \right]
  \right),
  \\
  \mathcal B_L(u,v;t_0)
  &=
  \frac{
  \delta_\chi y'_{\rm bath}(u)
  }{
  \left(y^{\rm fold}_{0,\chi}\right)'(u)
  }
  +
  2\mathcal J\,
  \delta_\chi y_{\rm bath}(u)
  \tan\left(
  \mathcal J
  \left[
  y^{\rm fold}_{0,\chi}(u)
  -
  y^{\rm fold}_{0,\chi}(v)^*
  \right]
  \right).
\end{aligned}
\label{eq:S4-bath-response-functions}
\end{equation}
Here
\(\left(y^{\rm fold}_{0,\chi}\right)'(u)=-V_G\) for \(u>t_0\).
The quantity \(\mathcal B_L\) is not an \(RL\) size response: it is an
internal bath function required because the \(RR\) probe solution in Sec.~S3
contains \(\operatorname{Im}g_L^\chi\). For an \(\eta\)-sector source the
bath is unperturbed in the probe limit, and both bath variations vanish.

\paragraph*{\(\boldsymbol{t_1,t_2<t_{\rm ev}}\).}

Both physical insertion times lie before evaporation. The response is
\begin{equation}
  \mathfrak D_\psi(t_1,t_2;t_0)
  =
  \mathfrak D^{\rm pre}_\psi(t_1,t_2;t_0),
  \qquad
  t_1,t_2<t_{\rm ev},
\label{eq:S4-bilocal-pre-region}
\end{equation}
where the right-hand side is given in
Eq.~\eqref{eq:S4-pre-bilocal-response}, with the parameters in
Eq.~\eqref{eq:S4-linearized-eta-parameters}.

\paragraph*{\(\boldsymbol{t_2<t_{\rm ev}<t_1<t_0}\).}

The mixed-region solution is
\begin{equation}
\begin{aligned}
  g_R^\eta(\widetilde t_1,t_2;t_0,\lambda)
  &=
  g_R^\eta(t_{\rm ev},t_2)
  +
  \operatorname{Re}
  g_R^\chi(\widetilde t_1,t_{\rm ev};t_0,\lambda)
  +
i\,\operatorname{Im}
  g_L^\chi(\widetilde t_1,t_{\rm ev};t_0,\lambda)
  \\
  &\quad
  +
  i\Delta\mu(t_1-t_{\rm ev})
  +
  \begin{cases}
  -i\lambda,
  &
  \psi=\eta,
  \\
  +i\lambda,
  &
  \psi=\chi .
  \end{cases}
\end{aligned}
\label{eq:S4-bilocal-region-II-integrated}
\end{equation}
Therefore
\begin{equation}
\begin{aligned}
  \mathfrak D_\eta(t_1,t_2;t_0)
  &=
  1,
  \\
  \mathfrak D_\chi(t_1,t_2;t_0)
  &=
  -1
  +
  i\,\operatorname{Re}
  \mathcal B_R(\widetilde t_1,t_{\rm ev};t_0)
  -
  \operatorname{Im}
  \mathcal B_L(\widetilde t_1,t_{\rm ev};t_0),
\end{aligned}
\qquad
t_2<t_{\rm ev}<t_1<t_0 .
\label{eq:S4-bilocal-region-II-result}
\end{equation}

\paragraph*{\(\boldsymbol{t_{\rm ev}<t_1,t_2<t_0}\).}

When both physical insertion times lie after evaporation,
\begin{equation}
\begin{aligned}
  g_R^\eta(\widetilde t_1,t_2;t_0,\lambda)
  &=
  g_R^\chi(\widetilde t_1,t_2;t_0,\lambda)
  \\
  &\quad
  +
  i\,\operatorname{Im}
  \left[
  -g_R^\chi(\widetilde t_1,t_{\rm ev};t_0,\lambda)
  -g_R^\chi(t_{\rm ev},t_2)
+g_L^\chi(\widetilde t_1,t_{\rm ev};t_0,\lambda)
  +g_L^\chi(t_{\rm ev},t_2)
  \right]
  \\
  &\quad
  +
  i\Delta\mu(t_1-t_2)
  +
  \begin{cases}
  -i\lambda,
  &
  \psi=\eta,
  \\
  +i\lambda,
  &
  \psi=\chi .
  \end{cases}
\end{aligned}
\label{eq:S4-bilocal-region-III-integrated}
\end{equation}
The terms whose two arguments lie before \(t_0\) are independent of
\(\lambda\). Hence
\begin{equation}
\begin{aligned}
  \mathfrak D_\eta(t_1,t_2;t_0)
  &=
  1,
  \\
  \mathfrak D_\chi(t_1,t_2;t_0)
  &=
  -1
  +
  i\,\mathcal B_R(\widetilde t_1,t_2;t_0)
  \\
  &\quad
  +
  \operatorname{Im}
  \left[
  \mathcal B_R(\widetilde t_1,t_{\rm ev};t_0)
  -
  \mathcal B_L(\widetilde t_1,t_{\rm ev};t_0)
  \right],
\end{aligned}
\qquad
t_{\rm ev}<t_1,t_2<t_0 .
\label{eq:S4-bilocal-region-III-result}
\end{equation}

\paragraph*{\(\boldsymbol{t_1<t_{\rm ev}<t_2<t_0}\).}

In this ordering, differentiation of the \(t_2\)-evolution equation gives
\begin{equation}
  \partial_{t_2}
  \left[
  \left.
  \partial_\lambda
  g_R^\eta(\widetilde t_1,t_2;t_0,\lambda)
  \right|_{\lambda=0}
  \right]
  =
  \partial_{t_2}
  \left[
  \left.
  \partial_\lambda
  g_R^\chi(\widetilde t_{\rm ev},t_2;t_0,\lambda)
  \right|_{\lambda=0}
  \right].
\label{eq:S4-bilocal-region-IV-equation}
\end{equation}
Integrating from \(t_{\rm ev}\) to \(t_2\) yields
\begin{equation}
\begin{aligned}
  \mathfrak D_\eta(t_1,t_2;t_0)
  &=
  \mathfrak D^{\rm pre}_\eta(t_1,t_{\rm ev};t_0),
  \\
  \mathfrak D_\chi(t_1,t_2;t_0)
  &=
  \mathfrak D^{\rm pre}_\chi(t_1,t_{\rm ev};t_0)
  \\
  &\quad
  +
  i
  \left[
  \mathcal B_R(\widetilde t_{\rm ev},t_2;t_0)
  -
  \mathcal B_R(\widetilde t_{\rm ev},t_{\rm ev};t_0)
  \right],
\end{aligned}
\qquad
t_1<t_{\rm ev}<t_2<t_0 .
\label{eq:S4-bilocal-region-IV-result}
\end{equation}

\paragraph*{Summary.}

The direct \(\eta\)-sector response is
\begin{equation}
\boxed{
  \mathfrak D_\eta(t_1,t_2;t_0)
  =
  \begin{cases}
  \mathfrak D^{\rm pre}_\eta(t_1,t_2;t_0),
  &
  t_1,t_2<t_{\rm ev},
  \\[4pt]
  1,
  &
  t_2<t_{\rm ev}<t_1<t_0,
  \\[4pt]
  1,
  &
  t_{\rm ev}<t_1,t_2<t_0,
  \\[4pt]
  \mathfrak D^{\rm pre}_\eta(t_1,t_{\rm ev};t_0),
  &
  t_1<t_{\rm ev}<t_2<t_0 .
  \end{cases}
}
\label{eq:S4-Deta-regions}
\end{equation}
The \(\chi\)-sector response is
\begin{equation}
\boxed{
  \mathfrak D_\chi(t_1,t_2;t_0)
  =
  \begin{cases}
  \mathfrak D^{\rm pre}_\chi(t_1,t_2;t_0),
  &
  t_1,t_2<t_{\rm ev},
  \\[4pt]
  \text{the second line of
  Eq.~\eqref{eq:S4-bilocal-region-II-result},}
  &
  t_2<t_{\rm ev}<t_1<t_0,
  \\[4pt]
  \text{the second line of
  Eq.~\eqref{eq:S4-bilocal-region-III-result},}
  &
  t_{\rm ev}<t_1,t_2<t_0,
  \\[4pt]
  \text{the second line of
  Eq.~\eqref{eq:S4-bilocal-region-IV-result},}
  &
  t_1<t_{\rm ev}<t_2<t_0 .
  \end{cases}
}
\label{eq:S4-Dchi-regions}
\end{equation}
Together with Eq.~\eqref{eq:S4-response-D-relation}, these formulas determine
the independent folded-contour \(RR\) response in every physical time region.
The physical boundary size is recovered by setting \(t_1=t_2=t_b\). The
response relevant to a reconstructed right-moving bulk mode is obtained by
taking the corresponding linear superposition with the right-boundary
reconstruction kernel in Eq.~\eqref{eq:HKLL_kernel}. This gives the
operator-size response shown in Fig.~\ref{fig:operator_size}.

\section*{S5. Two-replica calculation and R\'enyi mutual information}

\noindent
Methods summarizes the two-replica KB equation and the readout of
\(e^{-I^{(2)}}\) and \(S^{(2)}\). Here we give the reduced-state identities,
fermionic signs, numerical discretization, and on-shell-action formula.

\paragraph*{Reduced-state overlaps and transition matrices.}

We now give the two-replica construction underlying the R\'enyi-2 mutual
information in Eq.~\eqref{eq:renyi2_MI}. The ordinary real-time evolution is
performed on the folded contour \(\mathcal C\) defined in
Eq.~\eqref{eq:S-kb-contour}; the replica construction uses two copies of this
contour. In this section \(\alpha,\beta\in\{1,2\}\) denote replica indices,
while \(a,b\in\{R,L\}\) denote the physical left-right copies, as in Sec.~S2.

Let \(\lvert\Phi_{\rm in}\rangle\) be the initial doubled state prepared by
the Euclidean parts of the contour. The unperturbed state at the final
replica-gluing time \(t_f\) is
\begin{equation}
  \lvert\Phi(t_f)\rangle
  =
  U(t_f,0)\lvert\Phi_{\rm in}\rangle .
\label{eq:S5-unperturbed-state}
\end{equation}
The boundary-fermion building block used below is the time-evolved version of
the excitation in Eq.~\eqref{eq:S4-boundary-excitation},
\begin{equation}
  \lvert\Phi_{b,j}(t_f)\rangle
  =
  \sqrt{2}\,
  U(t_f,t_b)\eta^R_j
  U(t_b,0)\lvert\Phi_{\rm in}\rangle,
  \qquad
  0<t_b<t_{\rm ev}<t_f .
\label{eq:S5-boundary-excitation}
\end{equation}
The factor \(\sqrt{2}\) gives unit norm because
\((\eta^R_j)^2=1/2\). Define the corresponding reduced density matrices by
\begin{equation}
\begin{aligned}
  \rho_\eta(t_f)
  &=
  \operatorname{Tr}_{\chi}
  \left[
  \lvert\Phi(t_f)\rangle\langle\Phi(t_f)\rvert
  \right],
  &
  \rho_\chi(t_f)
  &=
  \operatorname{Tr}_{\eta}
  \left[
  \lvert\Phi(t_f)\rangle\langle\Phi(t_f)\rvert
  \right],
  \\
  \sigma_{\eta,j}(t_b;t_f)
  &=
  \operatorname{Tr}_{\chi}
  \left[
  \lvert\Phi_{b,j}(t_f)\rangle
  \langle\Phi_{b,j}(t_f)\rvert
  \right],
  &
  \sigma_{\chi,j}(t_b;t_f)
  &=
  \operatorname{Tr}_{\eta}
  \left[
  \lvert\Phi_{b,j}(t_f)\rangle
  \langle\Phi_{b,j}(t_f)\rvert
  \right].
\end{aligned}
\label{eq:S5-reduced-density-matrices}
\end{equation}
Because the unperturbed full state is pure, the two unperturbed reduced
density matrices have the same nonzero Schmidt eigenvalues. Their purities
are therefore equal:
\begin{equation}
  \mathcal P_\eta(t_f)
  \equiv
  \operatorname{Tr}_{\eta}\rho_\eta(t_f)^2
  =
  \operatorname{Tr}_{\chi}\rho_\chi(t_f)^2
  \equiv
  \mathcal P_\chi(t_f).
\label{eq:S5-unperturbed-purity}
\end{equation}

\begin{figure}
    \centering

\tikzset{every picture/.style={line width=0.75pt}}

\begin{tikzpicture}[x=0.75pt,y=0.75pt,yscale=-1,xscale=1]

\draw    (90,30) -- (340,30) ;

\draw    (340,30) -- (340,60) ;

\draw    (90,60) -- (340,60) ;

\draw [color={rgb, 255:red, 208; green, 2; blue, 27 }  ,draw opacity=1 ]   (90,80) -- (340,80) ;

\draw [color={rgb, 255:red, 208; green, 2; blue, 27 }  ,draw opacity=1 ]   (340,80) -- (340,110) ;

\draw [color={rgb, 255:red, 208; green, 2; blue, 27 }  ,draw opacity=1 ]   (90,110) -- (340,110) ;

\draw    (90,130) -- (340,130) ;

\draw    (340,130) -- (340,160) ;

\draw    (90,160) -- (340,160) ;

\draw [color={rgb, 255:red, 208; green, 2; blue, 27 }  ,draw opacity=1 ]   (90,180) -- (340,180) ;

\draw [color={rgb, 255:red, 208; green, 2; blue, 27 }  ,draw opacity=1 ]   (340,180) -- (340,210) ;

\draw [color={rgb, 255:red, 208; green, 2; blue, 27 }  ,draw opacity=1 ]   (90,210) -- (340,210) ;

\draw [color={rgb, 255:red, 74; green, 144; blue, 226 }  ,draw opacity=1 ]   (230,50) -- (230,90) ;

\draw [color={rgb, 255:red, 74; green, 144; blue, 226 }  ,draw opacity=1 ]   (250,50) -- (250,90) ;

\draw [color={rgb, 255:red, 74; green, 144; blue, 226 }  ,draw opacity=1 ]   (270,50) -- (270,90) ;

\draw [color={rgb, 255:red, 74; green, 144; blue, 226 }  ,draw opacity=1 ]   (290,50) -- (290,90) ;

\draw [color={rgb, 255:red, 74; green, 144; blue, 226 }  ,draw opacity=1 ]   (310,50) -- (310,90) ;

\draw [color={rgb, 255:red, 74; green, 144; blue, 226 }  ,draw opacity=1 ]   (330,50) -- (330,90) ;

\draw [color={rgb, 255:red, 74; green, 144; blue, 226 }  ,draw opacity=1 ]   (230,100) -- (230,140) ;

\draw [color={rgb, 255:red, 74; green, 144; blue, 226 }  ,draw opacity=1 ]   (250,100) -- (250,140) ;

\draw [color={rgb, 255:red, 74; green, 144; blue, 226 }  ,draw opacity=1 ]   (270,100) -- (270,140) ;

\draw [color={rgb, 255:red, 74; green, 144; blue, 226 }  ,draw opacity=1 ]   (290,100) -- (290,140) ;

\draw [color={rgb, 255:red, 74; green, 144; blue, 226 }  ,draw opacity=1 ]   (310,100) -- (310,140) ;

\draw [color={rgb, 255:red, 74; green, 144; blue, 226 }  ,draw opacity=1 ]   (330,100) -- (330,140) ;

\draw [color={rgb, 255:red, 74; green, 144; blue, 226 }  ,draw opacity=1 ]   (230,150) -- (230,190) ;

\draw [color={rgb, 255:red, 74; green, 144; blue, 226 }  ,draw opacity=1 ]   (250,150) -- (250,190) ;

\draw [color={rgb, 255:red, 74; green, 144; blue, 226 }  ,draw opacity=1 ]   (270,150) -- (270,190) ;

\draw [color={rgb, 255:red, 74; green, 144; blue, 226 }  ,draw opacity=1 ]   (290,150) -- (290,190) ;

\draw [color={rgb, 255:red, 74; green, 144; blue, 226 }  ,draw opacity=1 ]   (310,150) -- (310,190) ;

\draw [color={rgb, 255:red, 74; green, 144; blue, 226 }  ,draw opacity=1 ]   (330,150) -- (330,190) ;

\draw [color={rgb, 255:red, 74; green, 144; blue, 226 }  ,draw opacity=1 ]   (230,200) -- (230,220) ;

\draw [color={rgb, 255:red, 74; green, 144; blue, 226 }  ,draw opacity=1 ]   (250,200) -- (250,220) ;

\draw [color={rgb, 255:red, 74; green, 144; blue, 226 }  ,draw opacity=1 ]   (270,200) -- (270,220) ;

\draw [color={rgb, 255:red, 74; green, 144; blue, 226 }  ,draw opacity=1 ]   (290,200) -- (290,220) ;

\draw [color={rgb, 255:red, 74; green, 144; blue, 226 }  ,draw opacity=1 ]   (310,200) -- (310,220) ;

\draw [color={rgb, 255:red, 74; green, 144; blue, 226 }  ,draw opacity=1 ]   (330,200) -- (330,220) ;

\draw [color={rgb, 255:red, 74; green, 144; blue, 226 }  ,draw opacity=1 ]   (230,20) -- (230,40) ;

\draw [color={rgb, 255:red, 74; green, 144; blue, 226 }  ,draw opacity=1 ]   (250,20) -- (250,40) ;

\draw [color={rgb, 255:red, 74; green, 144; blue, 226 }  ,draw opacity=1 ]   (270,20) -- (270,40) ;

\draw [color={rgb, 255:red, 74; green, 144; blue, 226 }  ,draw opacity=1 ]   (290,20) -- (290,40) ;

\draw [color={rgb, 255:red, 74; green, 144; blue, 226 }  ,draw opacity=1 ]   (310,20) -- (310,40) ;

\draw [color={rgb, 255:red, 74; green, 144; blue, 226 }  ,draw opacity=1 ]   (330,20) -- (330,40) ;

\draw (61,32) node [anchor=north west][inner sep=0.75pt]   [align=left] {$\displaystyle \chi^{1}$};

\draw (61,82) node [anchor=north west][inner sep=0.75pt]   [align=left] {$\displaystyle \textcolor[rgb]{0.82,0.01,0.11}{\eta^{2}}$};

\draw (61,132) node [anchor=north west][inner sep=0.75pt]   [align=left] {$\displaystyle \chi^{2}$};

\draw (61,182) node [anchor=north west][inner sep=0.75pt]   [align=left] {$\displaystyle \textcolor[rgb]{0.82,0.01,0.11}{\eta^{1}}$};

\draw (132,17) node [anchor=north west][inner sep=0.75pt]  [font=\scriptsize] [align=left] {$\displaystyle \mathcal{C}^{+}$};

\draw (132,47) node [anchor=north west][inner sep=0.75pt]  [font=\scriptsize] [align=left] {$\displaystyle \mathcal{C}^{-}$};

\draw (132,67) node [anchor=north west][inner sep=0.75pt]  [font=\scriptsize] [align=left] {$\displaystyle \mathcal{C}^{-}$};

\draw (132,97) node [anchor=north west][inner sep=0.75pt]  [font=\scriptsize] [align=left] {$\displaystyle \mathcal{C}^{+}$};

\draw (132,117) node [anchor=north west][inner sep=0.75pt]  [font=\scriptsize] [align=left] {$\displaystyle \mathcal{C}^{+}$};

\draw (132,147) node [anchor=north west][inner sep=0.75pt]  [font=\scriptsize] [align=left] {$\displaystyle \mathcal{C}^{-}$};

\draw (132,167) node [anchor=north west][inner sep=0.75pt]  [font=\scriptsize] [align=left] {$\displaystyle \mathcal{C}^{-}$};

\draw (132,197) node [anchor=north west][inner sep=0.75pt]  [font=\scriptsize] [align=left] {$\displaystyle \mathcal{C}^{+}$};

\end{tikzpicture}
    \caption{\textbf{From the \(\eta\)-twisted to the \(\chi\)-twisted
representation.}
The reduced-state contraction
\(\operatorname{Tr}_{\eta}(\sigma_{\eta,j}\rho_\eta)\) cyclically glues the
\(\eta\) indices between replicas, while the \(\chi\) indices close within
each replica. Regrouping the same wave-function coefficients as in
Eq.~\eqref{eq:S5-transition-identity} gives the identical contraction
\(\operatorname{Tr}_{\chi}
(\tau_{\chi,j}^{\dagger}\tau_{\chi,j})\), in which the \(\eta\) indices close
within each transition strip and the signed replica twist is carried by
\(\chi\). After the branch-dependent rotation in
Eq.~\eqref{eq:S5-branch-rotation}, this equivalence is represented by the
replica-diagonal pairings
\((\chi^{1},\eta^{1})\) and \((\chi^{2},\eta^{2})\) on
\(\mathcal C^{+}\), and the crossed pairings
\((\chi^{1},\eta^{2})\) and \((\chi^{2},\eta^{1})\) on
\(\mathcal C^{-}\). Red and black curves denote the \(\eta\) and \(\chi\)
contours, respectively, while the blue links indicate their mixed-interaction
pairing. Here \(R,L\) copy indices are suppressed.}
\label{fig:S5-eta-chi-twist}
\end{figure}
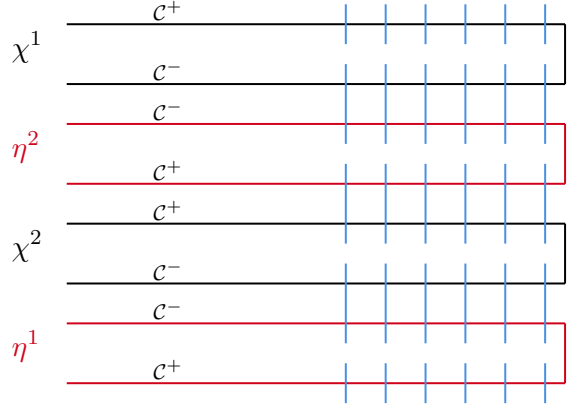

The first normalized contraction required for the mutual-information
calculation is
\begin{equation}
  F_2(t_b;t_f)
  =
  \frac{1}{N_\eta}
  \sum_{j=1}^{N_\eta}
  \frac{
  \operatorname{Tr}_{\eta}
  \left[
  \sigma_{\eta,j}(t_b;t_f)\rho_\eta(t_f)
  \right]
  }{
  \mathcal P_\eta(t_f)
  } .
\label{eq:S5-F2-density}
\end{equation}
It is convenient to express the numerator as a transition-matrix
contraction in the bath Hilbert space. Define
\begin{equation}
  \tau_{\chi,j}(t_b;t_f)
  =
  \operatorname{Tr}_{\eta}
  \left[
  \lvert\Phi_{b,j}(t_f)\rangle
  \langle\Phi(t_f)\rvert
  \right].
\label{eq:S5-transition-matrix}
\end{equation}
For an ordered fermionic occupation basis
\(\{\lvert i_\eta,x_\chi\rangle\}\), write
\begin{equation}
  (\Phi_{b,j})_{ix}
  =
  \langle i_\eta,x_\chi\vert\Phi_{b,j}(t_f)\rangle,
  \qquad
  \Phi_{ix}
  =
  \langle i_\eta,x_\chi\vert\Phi(t_f)\rangle .
\label{eq:S5-wavefunction-components}
\end{equation}
A direct contraction of these coefficients gives
\begin{equation}
\begin{aligned}
  \operatorname{Tr}_{\eta}
  \left[
  \sigma_{\eta,j}\rho_\eta
  \right]
  &=
  \sum_{i,i',x,y}
  (\Phi_{b,j})_{ix}
  (\Phi_{b,j})^*_{i'x}
  \Phi_{i'y}
  \Phi^*_{iy}
  \\
  &=
  \operatorname{Tr}_{\chi}
  \left[
  \tau_{\chi,j}^{\dagger}
  \tau_{\chi,j}
  \right].
\end{aligned}
\label{eq:S5-transition-identity}
\end{equation}
No additional fermion sign is generated in this regrouping: after an ordered
occupation basis has been fixed, the wavefunction components are ordinary
complex coefficients. Although \(\tau_{\chi,j}\) is fermion odd,
\(\tau_{\chi,j}^{\dagger}\tau_{\chi,j}\) is fermion even. Consequently,
\begin{equation}
  F_2(t_b;t_f)
  =
  \frac{
  \displaystyle
  \frac{1}{N_\eta}
  \sum_{j=1}^{N_\eta}
  \operatorname{Tr}_{\chi}
  \left[
  \tau_{\chi,j}^{\dagger}(t_b;t_f)
  \tau_{\chi,j}(t_b;t_f)
  \right]
  }{
  \mathcal P_\chi(t_f)
  } .
\label{eq:S5-F2-transition}
\end{equation}
In this representation the \(\eta\) indices close within each transition
strip, while the \(\chi\) indices are cyclically glued between the two
strips. Thus the replica twist may equivalently be placed on the \(\chi\)
fields; its contour realization in the branch-dependent replica frame is
shown in Fig.~\ref{fig:S5-eta-chi-twist}.

We choose replica \(1\) to represent
\(\tau_{\chi,j}^{\dagger}\) and replica \(2\) to represent
\(\tau_{\chi,j}\). The insertion from
\(\langle\Phi_{b,j}(t_f)\rvert\) therefore lies on the lower branch of
replica \(1\), whereas the insertion from
\(\lvert\Phi_{b,j}(t_f)\rangle\) lies on the upper branch of replica \(2\).
Define the normalized two-replica Wightman function in this
\(\chi\)-twisted representation by
\begin{equation}
  G^{(2),\eta;\chi\text{-tw},>}_{RR;\alpha\beta}
  (t_1,t_2;t_f)
  =
  -\frac{i}{N_\eta}
  \sum_{j=1}^{N_\eta}
  \left\langle
  \eta^\alpha_{R,j}(t_{1,-})
  \eta^\beta_{R,j}(t_{2,+})
  \right\rangle_{\chi\text{-twisted}} .
\label{eq:S5-two-replica-greater}
\end{equation}
Here the twisted expectation value is normalized by the unperturbed
two-replica partition function, equivalently by
\(\mathcal P_\chi(t_f)\). With the branch and replica assignments above,
Eq.~\eqref{eq:S5-F2-transition} becomes
\begin{equation}
  F_2(t_b;t_f)
  =
  2i\,
  G^{(2),\eta;\chi\text{-tw},>}_{RR;12}
  (t_b,t_b;t_f).
\label{eq:S5-F2-green}
\end{equation}

\paragraph*{Signed fermionic swap and the untwisted bath frame.}

The replica gluing must retain the sign associated with a fermionic trace.
For a fermionic operator \(O\), the coherent-state representation of the
trace is
\begin{equation}
  \operatorname{Tr}_{\chi}O
  =
  \int
  \mathrm d\bar\xi\,\mathrm d\xi\,
  e^{-\bar\xi\xi}
  \langle-\xi\vert O\vert\xi\rangle .
\label{eq:S5-fermion-trace}
\end{equation}
Applying this identity to
\(\operatorname{Tr}_{\chi}
[\tau_{\chi,j}^{\dagger}\tau_{\chi,j}]\)
and inserting a coherent-state resolution of the identity between the two
transition matrices gives
\begin{equation}
\begin{aligned}
  \operatorname{Tr}_{\chi}
  \left[
  \tau_{\chi,j}^{\dagger}\tau_{\chi,j}
  \right]
  &=
  \int
  \mathrm d\bar\xi_0\,\mathrm d\xi_0\,
  \mathrm d\bar\xi_1\,\mathrm d\xi_1\,
  e^{-\bar\xi_0\xi_0-\bar\xi_1\xi_1}
  \\
  &\quad\times
  \langle-\xi_0\vert
  \tau_{\chi,j}^{\dagger}
  \vert\xi_1\rangle
  \langle\xi_1\vert
  \tau_{\chi,j}
  \vert\xi_0\rangle .
\end{aligned}
\label{eq:S5-transition-coherent-trace}
\end{equation}
The endpoints of the two transition strips are consequently identified as
\begin{equation}
  \chi^1(t_{f,+})
  =
  -\chi^2(t_{f,-}),
  \qquad
  \chi^2(t_{f,+})
  =
  \chi^1(t_{f,-}).
\label{eq:S5-chi-endpoint-gluing}
\end{equation}
Writing the two replica fields as a column vector, this becomes
\begin{equation}
  \begin{pmatrix}
    \chi^1(t_{f,+})\\
    \chi^2(t_{f,+})
  \end{pmatrix}
  =
  \mathsf S
  \begin{pmatrix}
    \chi^1(t_{f,-})\\
    \chi^2(t_{f,-})
  \end{pmatrix},
  \qquad
  \mathsf S
  =
  \begin{pmatrix}
    0&-1\\
    1&0
  \end{pmatrix}.
\label{eq:S5-signed-swap}
\end{equation}
The minus sign in the first row originates from the anti-periodic trace in
Eq.~\eqref{eq:S5-fermion-trace}. The signed swap obeys
\begin{equation}
  \mathsf S^\dagger
  =
  \mathsf S^T
  =
  -\mathsf S,
  \qquad
  \mathsf S^\dagger\mathsf S
  =
  \mathbf 1,
  \qquad
  \mathsf S^2
  =
  -\mathbf 1 .
\label{eq:S5-signed-swap-properties}
\end{equation}

For later use, divide the folded contour into its upper and lower folds,
\begin{equation}
  \mathcal C_{\rm u}
  =
  \mathcal C^+_{\beta/4}\cup\mathcal C^+,
  \qquad
  \mathcal C_{\rm l}
  =
  \mathcal C^-\cup\mathcal C^-_{\beta/4}.
\label{eq:S5-upper-lower-folds}
\end{equation}
Introduce the branch-frame rotation
\begin{equation}
  \mathsf U(z)
  =
  \begin{cases}
    \mathbf 1,
    &z\in\mathcal C_{\rm u},
    \\[2pt]
    \mathsf S,
    &z\in\mathcal C_{\rm l}.
  \end{cases}
\label{eq:S5-branch-rotation}
\end{equation}
The symbol \(P\) is therefore reserved exclusively for the auxiliary
fermionic reference mode introduced below.

Define the untwisted bath field by
\begin{equation}
  \widetilde\chi(z)
  =
  \mathsf U(z)\chi(z),
  \qquad
  \chi(z)
  =
  \mathsf U^\dagger(z)\widetilde\chi(z).
\label{eq:S5-untwisted-bath-field}
\end{equation}
At the final-time endpoint,
\[
  \widetilde\chi(t_{f,+})
  =
  \chi(t_{f,+}),
  \qquad
  \widetilde\chi(t_{f,-})
  =
  \mathsf S\chi(t_{f,-}),
\]
so Eq.~\eqref{eq:S5-signed-swap} implies the ordinary gluing condition
\begin{equation}
  \widetilde\chi(t_{f,+})
  =
  \widetilde\chi(t_{f,-}).
\label{eq:S5-untwisted-bath-gluing}
\end{equation}
Thus the twist has been moved from the endpoint condition into the
branch-dependent replica frame.

The two-replica bilocals in this frame are
\begin{equation}
\begin{aligned}
  (G^\eta_{ab})_{\alpha\beta}(z_1,z_2)
  &=
  -\frac{i}{N_\eta}
  \sum_{j=1}^{N_\eta}
  \eta^\alpha_{a,j}(z_1)
  \eta^\beta_{b,j}(z_2),
  \\
  (\widetilde G^\chi_{ab})_{\alpha\beta}(z_1,z_2)
  &=
  -\frac{i}{N_\chi}
  \sum_{k=1}^{N_\chi}
  \widetilde\chi^\alpha_{a,k}(z_1)
  \widetilde\chi^\beta_{b,k}(z_2).
\end{aligned}
\label{eq:S5-two-replica-bilocals}
\end{equation}
The physical \(\chi\) bilocal in the twisted frame is therefore
\begin{equation}
  G^{\chi,\mathrm{tw}}_{ab}(z_1,z_2)
  =
  \mathsf U^\dagger(z_1)
  \widetilde G^\chi_{ab}(z_1,z_2)
  \mathsf U(z_2).
\label{eq:S5-twisted-bath-bilocal}
\end{equation}
We rotate the bath self-energy in the same way,
\begin{equation}
  \Sigma^{\chi,\mathrm{tw}}_{ab}(z_1,z_2)
  =
  \mathsf U^\dagger(z_1)
  \widetilde\Sigma^\chi_{ab}(z_1,z_2)
  \mathsf U(z_2).
\label{eq:S5-twisted-bath-self-energy}
\end{equation}
Since \(\mathsf U\) is a real orthogonal signed-permutation matrix, the
componentwise bilocal pairing is invariant:
\begin{equation}
\begin{aligned}
  \sum_{\alpha,\beta}
  (\Sigma^{\chi,\mathrm{tw}}_{ab})_{\alpha\beta}
  (G^{\chi,\mathrm{tw}}_{ab})_{\alpha\beta}
  =
  \sum_{\alpha,\beta}
  (\widetilde\Sigma^\chi_{ab})_{\alpha\beta}
  (\widetilde G^\chi_{ab})_{\alpha\beta}.
\end{aligned}
\label{eq:S5-SigmaG-invariance}
\end{equation}

The kinetic and local left-right bilinear terms are also invariant under
Eq.~\eqref{eq:S5-untwisted-bath-field}. The pure \(\chi\) interaction is
invariant because left and right multiplication by \(\mathsf U\) only
permutes matrix entries and changes some signs, while \(p\) is even:
\begin{equation}
\begin{aligned}
  \sum_{\alpha,\beta}
  \left[
  2\left(
  \mathsf U^\dagger(z_1)
  \widetilde G^\chi_{ab}(z_1,z_2)
  \mathsf U(z_2)
  \right)_{\alpha\beta}
  \right]^p
  =
  \sum_{\alpha,\beta}
  \left[
  2(\widetilde G^\chi_{ab})_{\alpha\beta}(z_1,z_2)
  \right]^p .
\end{aligned}
\label{eq:S5-pure-bath-invariance}
\end{equation}
The rotation cannot be removed from the mixed interaction because the
\(\eta\) bilocal remains in the original replica frame. The
post-evaporation mixed combination is therefore
\begin{equation}
  s(G^\eta_{ab})_{\alpha\beta}
  +
  (1-s)
  (G^{\chi,\mathrm{tw}}_{ab})_{\alpha\beta}.
\label{eq:S5-mixed-bilocal}
\end{equation}

\paragraph*{Two-replica effective action and probe limit.}

The two-replica bare inverse propagator is the replica-diagonal lift of
Eq.~\eqref{eq:S-bare-inverse-propagator},
\begin{equation}
  (G_{0,2}^{-1})_{\alpha\beta;ab}(z_1,z_2)
  =
  \delta_{\alpha\beta}\,
  i\delta_{ab}\partial_{z_1}
  \delta_{\mathcal C}(z_1,z_2).
\label{eq:S5-bare-two-replica-propagator}
\end{equation}
Using the normalization of the one-replica action
\eqref{eq:S-effective-action}, the finite-\(s\), finite-\(p\) action in the
\(\chi\)-twisted representation is
\begin{equation}
\begin{aligned}
  \mathcal S_{\mathrm{2rep}}
  &=
  -s\frac{i}{2}
  \operatorname{Tr}\log
  \left[
  -i\left(G_{0,2}^{-1}-\Sigma^\eta\right)
  \right]
  -(1-s)\frac{i}{2}
  \operatorname{Tr}\log
  \left[
  -i\left(G_{0,2}^{-1}-\widetilde\Sigma^\chi\right)
  \right]
  \\
  &\quad
  +
  \frac{i}{2}
  \int_{\mathcal C}
  \mathrm dz_1\,\mathrm dz_2
  \sum_{a,b}
  \sum_{\alpha,\beta=1}^{2}
  \Big[
  s(\Sigma^\eta_{ab})_{\alpha\beta}
  (G^\eta_{ab})_{\alpha\beta}
  +
  (1-s)(\widetilde\Sigma^\chi_{ab})_{\alpha\beta}
  (\widetilde G^\chi_{ab})_{\alpha\beta}
  \Big]
  \\
  &\quad
  -
  \frac{i\mathcal J^2}{4p^2}
  \int_{\mathcal C}
  \mathrm dz_1\,\mathrm dz_2
  \sum_{a,b}
  \sum_{\alpha,\beta=1}^{2}
  (-1)^{1+p\delta_{ab}/2}
  \Bigg[
  l_\eta(z_1)l_\eta(z_2)
  \left[
  2(G^\eta_{ab})_{\alpha\beta}
  \right]^p
  \\
  &\hspace{3.2cm}
  +
  l_\chi(z_1)l_\chi(z_2)
  \left[
  2(\widetilde G^\chi_{ab})_{\alpha\beta}
  \right]^p
  \\
  &\hspace{3.2cm}
  +
  l_c(z_1)l_c(z_2)
  \Bigg\{
  \left[
  2\left(
  s(G^\eta_{ab})_{\alpha\beta}
  +
  (1-s)
  (G^{\chi,\mathrm{tw}}_{ab})_{\alpha\beta}
  \right)
  \right]^p
  \\
  &\hspace{5.5cm}
  -
  s^p
  \left[
  2(G^\eta_{ab})_{\alpha\beta}
  \right]^p
  -
  (1-s)^p
  \left[
  2(\widetilde G^\chi_{ab})_{\alpha\beta}
  \right]^p
  \Bigg\}
  \Bigg]
  \\
  &\quad
  +
  \frac{1}{2p}
  \int_{\mathcal C}
  \mathrm dz_1\,\mathrm dz_2\,
  \delta_{\mathcal C}(z_1,z_2)
  \sum_{\alpha=1}^{2}
  \Bigg[
  s\mu_\eta(z_1)
  \left(
  (G^\eta_{LR})_{\alpha\alpha}
  -
  (G^\eta_{RL})_{\alpha\alpha}
  \right)
  \\
  &\hspace{5.5cm}
  +
  (1-s)\mu_\chi(z_1)
  \left(
  (\widetilde G^\chi_{LR})_{\alpha\alpha}
  -
  (\widetilde G^\chi_{RL})_{\alpha\alpha}
  \right)
  \Bigg].
\end{aligned}
\label{eq:S5-two-replica-action}
\end{equation}
All bilocals in Eq.~\eqref{eq:S5-two-replica-action} are evaluated at
\((z_1,z_2)\) unless their arguments are shown explicitly. The trace in the
first line includes the contour, the \(R,L\) indices, and the replica
indices. The profiles \(l_\eta,l_\chi,l_c\) and
\(\mu_\eta,\mu_\chi\) are precisely those defined in
Eqs.~\eqref{eq:S-l-protocol} and \eqref{eq:S-bilinear-protocol}; no additional
bilinear-profile functions are introduced here.

For the two-replica calculation below, we work in the large-\(N\)
probe, or large-bath, limit
\begin{equation}
  N_\eta,N_\chi\longrightarrow\infty,
  \qquad
  s\longrightarrow0,
  \qquad
  b=d=1,
  \qquad
  p\ \text{fixed},\qquad p/2\ \text{even}.
\label{eq:S5-finite-p-probe-limit}
\end{equation}
At fixed \(p\), \(sp\to0\) follows from \(s\to0\), and the \(\chi\) saddle
has no backreaction from the \(\eta\) sector. Throughout the remainder of
this section, the two-replica equations are studied at finite \(p\).
An analytic treatment of the two-replica problem in the additional
large-\(p\) limit is left for future work.

In the untwisted frame the bath saddle is replica diagonal:
\begin{equation}
  (\widetilde G^\chi_{ab})_{\alpha\beta}(z_1,z_2)
  =
  \delta_{\alpha\beta}\,
  G^\chi_{ab}(z_1,z_2),
\label{eq:S5-replica-diagonal-bath}
\end{equation}
where \(G^\chi_{ab}\) is the ordinary one-replica bath solution of
Eqs.~\eqref{eq:S-contour-dyson} and \eqref{eq:S-self-energy}. Transforming
back to the physical twisted frame gives
\begin{equation}
  (G^{\chi,\mathrm{tw}}_{ab})_{\alpha\beta}(z_1,z_2)
  =
  G^\chi_{ab}(z_1,z_2)\,
  \Pi_{\alpha\beta}(z_1,z_2),
  \qquad
  \Pi(z_1,z_2)
  \equiv
  \mathsf U^\dagger(z_1)\mathsf U(z_2).
\label{eq:S5-twist-kernel}
\end{equation}
Explicitly,
\begin{equation}
  \Pi(z_1,z_2)
  =
  \begin{cases}
    \mathbf 1,
    &z_1,z_2\in\mathcal C_{\rm u}
    \ \text{or}\
    z_1,z_2\in\mathcal C_{\rm l},
    \\[4pt]
    \mathsf S,
    &z_1\in\mathcal C_{\rm u},
    \quad
    z_2\in\mathcal C_{\rm l},
    \\[4pt]
    \mathsf S^\dagger=-\mathsf S,
    &z_1\in\mathcal C_{\rm l},
    \quad
    z_2\in\mathcal C_{\rm u}.
  \end{cases}
\label{eq:S5-twist-kernel-cases}
\end{equation}
Every entry of \(\Pi\) is \(0\) or \(\pm1\). Since \(p-1\) is odd,
the elementwise power appearing in the self-energy satisfies
\begin{equation}
  \left[
  2G^\chi_{ab}(z_1,z_2)
  \Pi(z_1,z_2)
  \right]^{\circ(p-1)}
  =
  \left[
  2G^\chi_{ab}(z_1,z_2)
  \right]^{p-1}
  \Pi(z_1,z_2),
\label{eq:S5-twist-power}
\end{equation}
where the superscript \(\circ(p-1)\) denotes an elementwise power in replica
space.

For the sharp protocol, let \(\mathcal C_<\) denote the Euclidean preparation
segments together with the real-time portions before \(t_{\rm ev}\), and let
\(\mathcal C_>\) denote the real-time portions after \(t_{\rm ev}\). Varying
Eq.~\eqref{eq:S5-two-replica-action} and taking the limit
\eqref{eq:S5-finite-p-probe-limit} gives
\begin{equation}
\begin{aligned}
  (\Sigma^\eta_{ab})_{\alpha\beta}(z_1,z_2)
  &=
  (-1)^{1+p\delta_{ab}/2}
  \frac{\mathcal J^2}{p}
  \begin{cases}
    a^2
    \left[
    2(G^\eta_{ab})_{\alpha\beta}(z_1,z_2)
    \right]^{p-1},
    &z_1,z_2\in\mathcal C_<,
    \\[6pt]
    \left[
    2G^\chi_{ab}(z_1,z_2)
    \right]^{p-1}
    \Pi_{\alpha\beta}(z_1,z_2),
    &z_1,z_2\in\mathcal C_>,
    \\[6pt]
    0,
    &z_1\in\mathcal C_<,\ z_2\in\mathcal C_>
    \ \text{or vice versa}
  \end{cases}
  \\
  &\quad
  -
  i\epsilon_{ab}\,
  \delta_{\alpha\beta}\,
  \frac{\mu_\eta(z_1)}{p}
  \delta_{\mathcal C}(z_1,z_2),
\end{aligned}
\label{eq:S5-probe-self-energy}
\end{equation}
where \(\epsilon_{ab}\) is defined in Eq.~\eqref{eq:S-epsilon}. The
pre-evaporation replica problem retains the nonlinear \(\eta\) self-energy,
whereas after evaporation the \(\eta\) probe is driven by the known scalar
bath solution multiplied by the signed replica kernel \(\Pi\). There is no
interaction self-energy when the two arguments lie on opposite sides of the
evaporation surface. The probe calculation therefore requires no new
nonlinear bath saddle.

\paragraph*{Numerical two-replica equations.}

The scalar bath correlator \(G^\chi\) is first obtained from the ordinary
one-replica equations \eqref{eq:S-contour-dyson} and
\eqref{eq:S-self-energy}. The resulting bath solution is then held fixed
while solving the two-replica \(\eta\) equation.

Discretize the contour by points \(z_i\), ordered so that
\(i>j\) means \(z_i>_{\mathcal C}z_j\). Each fold contains a Euclidean
preparation segment of target length \(\beta/4\) and a real-time segment
ending at \(t_f\). The implementation uses real-time spacing \(\Delta t\)
and chooses the integer node counts as follows:

\begin{equation}
  N_\beta
  =
  \operatorname{round}\!\left(\frac{\beta}{4\Delta t}\right),
  \qquad
  N_t
  =
  \operatorname{round}\!\left(\frac{t_f}{\Delta t}\right),
  \qquad
  N_{\rm fold}
  =
  N_\beta+N_t,
  \qquad
  N_{\mathcal C}
  =
  2N_{\rm fold}.
\label{eq:S5-grid-sizes}
\end{equation}
The plotted final times are grid aligned, so \(N_t\Delta t=t_f\).
The Euclidean spacing is chosen separately as
\(\Delta\tau=\beta/(4N_\beta)\), so that each preparation segment
has length \(N_\beta\Delta\tau=\beta/4\) and
\(\beta_{\rm grid}=4N_\beta\Delta\tau=\beta\), up to floating-point precision.
For Fig.~\ref{fig:finite_p}, \(\beta=5.7355465869\) and
\(\Delta t=0.5\) give \(N_\beta=3\) and
\(\Delta\tau\simeq0.4779622156\). The contour measure is represented
by the diagonal weight matrix
\begin{equation}
  W^{ab}_{ij}
  =
  \delta_{ab}\delta_{ij}w_i,
  \qquad
  w_i
  =
  \begin{cases}
    -i\Delta\tau,
    &0\leq i<N_\beta,
    \\[2pt]
    +\Delta t,
    &N_\beta\leq i<N_{\rm fold},
    \\[2pt]
    -\Delta t,
    &N_{\rm fold}\leq i<N_{\rm fold}+N_t,
    \\[2pt]
    -i\Delta\tau,
    &N_{\rm fold}+N_t\leq i<2N_{\rm fold}.
  \end{cases}
\label{eq:S5-contour-weights}
\end{equation}

Using the composite index
\[
  \mathcal A=(\alpha,a,i),
  \qquad
  \alpha=1,2,
  \quad
  a\in\{R,L\},
  \quad
  i=0,\ldots,N_{\mathcal C}-1,
\]
the replica-diagonal free propagator and weight matrix are
\begin{equation}
\begin{aligned}
  (\mathbf G_{0,2})_{\alpha a i,\beta b j}
  &=
  \delta_{\alpha\beta}
  (\mathbf G_0)_{a i,b j},
  \\
  (\mathbf W_2)_{\alpha a i,\beta b j}
  &=
  \delta_{\alpha\beta}\delta_{ab}\delta_{ij}w_i .
\end{aligned}
\label{eq:S5-discrete-replica-lift}
\end{equation}
At the discrete level the signed twist is
\begin{equation}
  \Pi_{ij}
  =
  \mathsf U^\dagger(z_i)\mathsf U(z_j).
\label{eq:S5-discrete-twist}
\end{equation}

Let \(T_i\) be the real time associated with \(z_i\), with Euclidean
preparation points assigned to the pre-evaporation region. Define the masks
\begin{equation}
\begin{aligned}
  M_{\rm I}(i,j)
  &=
  \mathbf 1_{T_i\leq t_{\rm ev}}
  \mathbf 1_{T_j\leq t_{\rm ev}},
  \\
  M_{\rm III}(i,j)
  &=
  \mathbf 1_{T_i>t_{\rm ev}}
  \mathbf 1_{T_j>t_{\rm ev}},
  \\
  M_{\rm II}(i,j)
  &=
  1-M_{\rm I}(i,j)-M_{\rm III}(i,j).
\end{aligned}
\label{eq:S5-region-masks}
\end{equation}
Here \(\mathbf 1\) is an indicator function. A grid point exactly at
\(t_{\rm ev}\) belongs to the pre-evaporation region, so the masks form a
partition also on the quench slice.
The pre-evaporation self-energy is nonlinear in the unknown two-replica
correlator:
\begin{equation}
  \left(
  \Sigma^{\eta,\rm I}_{ab}
  \right)_{\alpha\beta;ij}
  =
  M_{\rm I}(i,j)
  (-1)^{1+p\delta_{ab}/2}
  \frac{a^2\mathcal J^2}{p}
  \left[
  2(G^\eta_{ab})_{\alpha\beta;ij}
  \right]^{p-1}.
\label{eq:S5-discrete-sigma-I}
\end{equation}
The post-evaporation self-energy is fixed by the bath:
\begin{equation}
  \left(
  \Sigma^{\eta,\rm III}_{ab}
  \right)_{\alpha\beta;ij}
  =
  M_{\rm III}(i,j)
  (-1)^{1+p\delta_{ab}/2}
  \frac{\mathcal J^2}{p}
  \left[
  2G^\chi_{ab;ij}
  \right]^{p-1}
  (\Pi_{ij})_{\alpha\beta}.
\label{eq:S5-discrete-sigma-III}
\end{equation}
There is no interaction contribution in region II. A local \(\eta\)
bilinear, if retained, is represented by
\begin{equation}
  (\mathbf V_\eta)_{\alpha a i,\beta b j}
  =
  -i\epsilon_{ab}\,
  \delta_{\alpha\beta}\delta_{ij}
  \frac{\mu_\eta(z_i)}{p}.
\label{eq:S5-discrete-bilinear}
\end{equation}
For the protocol used in the numerical results,
\(\mu_\eta=0\), so \(\mathbf V_\eta=0\).

The discrete two-replica Dyson equation is
\begin{equation}
\begin{aligned}
  \mathbf G_\eta
  &=
  \mathbf G_{0,2}
  +
  \mathbf G_{0,2}
  \Big[
  \mathbf W_2\mathbf V_\eta
  +
  \mathbf W_2
  \left(
  \boldsymbol\Sigma^{\eta,\rm I}[\mathbf G_\eta]
  +
  \boldsymbol\Sigma^{\eta,\rm III}_{\chi}
  \right)
  \mathbf W_2
  \Big]
  \mathbf G_\eta .
\end{aligned}
\label{eq:S5-discrete-Dyson}
\end{equation}
The local vertex carries one contour weight, whereas a nonlocal bilocal
self-energy carries one weight for each contour argument.

Because the post-evaporation bath source is fixed, define
\begin{equation}
  \mathbf A_\chi
  =
  \mathbf 1
  -
  \mathbf G_{0,2}
  \left[
  \mathbf W_2\mathbf V_\eta
  +
  \mathbf W_2
  \boldsymbol\Sigma^{\eta,\rm III}_{\chi}
  \mathbf W_2
  \right].
\label{eq:S5-fixed-bath-operator}
\end{equation}
The remaining nonlinear residual is
\begin{equation}
\begin{aligned}
  R_\eta(\mathbf G_\eta)
  &=
  \mathbf A_\chi\mathbf G_\eta
  -
  \mathbf G_{0,2}
  -
  \mathbf G_{0,2}
  \mathbf W_2
  \boldsymbol\Sigma^{\eta,\rm I}[\mathbf G_\eta]
  \mathbf W_2
  \mathbf G_\eta .
\end{aligned}
\label{eq:S5-replica-residual}
\end{equation}
We solve \(R_\eta=0\) using a damped Newton--Krylov method. The
Jacobian-vector product is
\begin{equation}
\begin{aligned}
  J_\eta[\mathbf G_\eta]\mathbf X
  &=
  \mathbf A_\chi\mathbf X
  -
  \mathbf G_{0,2}\mathbf W_2
  \delta\boldsymbol\Sigma^{\eta,\rm I}[\mathbf X]
  \mathbf W_2\mathbf G_\eta
  -
  \mathbf G_{0,2}\mathbf W_2
  \boldsymbol\Sigma^{\eta,\rm I}[\mathbf G_\eta]
  \mathbf W_2\mathbf X ,
\end{aligned}
\label{eq:S5-replica-JVP}
\end{equation}
where
\begin{equation}
\begin{aligned}
  \left(
  \delta\Sigma^{\eta,\rm I}_{ab}[\mathbf X]
  \right)_{\alpha\beta;ij}
  &=
  M_{\rm I}(i,j)
  (-1)^{1+p\delta_{ab}/2}
  \frac{a^2\mathcal J^2}{p}
  (p-1)\,2
  \left[
  2(G^\eta_{ab})_{\alpha\beta;ij}
  \right]^{p-2}
  (X_{ab})_{\alpha\beta;ij}.
\end{aligned}
\label{eq:S5-replica-delta-sigma}
\end{equation}
At each Newton step, we use the generalized minimal residual method (GMRES) to solve the linearized equation
\(J_\eta[\mathbf G_\eta]\delta\mathbf G_\eta
=-R_\eta(\mathbf G_\eta)\). The solver requires only the Jacobian-vector
product in Eq.~\eqref{eq:S5-replica-JVP}, so the full Jacobian is never
constructed. Starting from the linearized residual \(\mathbf r\), GMRES
builds the Krylov subspace
\(\operatorname{span}\{\mathbf r,J_\eta\mathbf r,\ldots,
J_\eta^{m-1}\mathbf r\}\) and chooses the correction that minimizes the
residual norm within this subspace. In practice, \(\mathbf A_\chi^{-1}\) is used as a preconditioner: it
approximately removes the stiff linear contribution from the fixed bath,
improving the conditioning of the linear system and accelerating GMRES
without changing the desired solution.

If \(i_-(t_b)\) and \(i_+(t_b)\) are the lower- and upper-branch indices
corresponding to \(t_b\), the normalized overlap is read off as
\begin{equation}
  F_2(t_b;t_f)
  =
  2i\,
  (\mathbf G_\eta)_{
  (1,R,i_-(t_b)),
  (2,R,i_+(t_b))
  }.
\label{eq:S5-discrete-F2-readout}
\end{equation}

\paragraph*{R\'enyi-2 mutual information.}

For compactness, define the branch-resolved contractions
\begin{equation}
  C_{\alpha\beta}^{\lambda_1\lambda_2}(t_b;t_f)
  \equiv
  2i\,
  G^{(2),\eta;\chi\text{-tw}}_{RR;\alpha\beta}
  (t_{b,\lambda_1},t_{b,\lambda_2};t_f),
  \qquad
  \lambda_1,\lambda_2\in\{+,-\}.
\label{eq:S5-branch-contractions}
\end{equation}
Thus
\begin{equation}
  F_2(t_b;t_f)
  =
  C_{12}^{-+}(t_b;t_f).
\label{eq:S5-F2-C}
\end{equation}

The second contraction required below is the fixed-flavor perturbed purity,
averaged over the flavor index:
\begin{equation}
  F_4(t_b;t_f)
  =
  \frac{1}{N_\eta}
  \sum_{j=1}^{N_\eta}
  \frac{
  \operatorname{Tr}_{\eta}
  \left[
  \sigma_{\eta,j}(t_b;t_f)^2
  \right]
  }{
  \mathcal P_\eta(t_f)
  } .
\label{eq:S5-F4-density}
\end{equation}
At leading order in large \(N_\eta\), the corresponding four-point
function factorizes into two-replica two-point functions. Keeping the
fermionic Wick signs gives
\begin{equation}
\begin{aligned}
  F_4(t_b;t_f)
  &=
  C_{11}^{-+}(t_b;t_f)
  C_{22}^{-+}(t_b;t_f)
  -
  C_{12}^{-+}(t_b;t_f)
  C_{21}^{-+}(t_b;t_f)
  -
  C_{12}^{--}(t_b;t_f)
  C_{12}^{++}(t_b;t_f).
\end{aligned}
\label{eq:S5-F4-green}
\end{equation}
The last term pairs the two lower-branch insertions with each other and the
two upper-branch insertions with each other. Thus no independent four-point
saddle is required.

The corresponding \(\chi\)-side cross contraction is
\begin{equation}
\begin{aligned}
  K_2(t_b;t_f)
  &=
  \frac{1}{N_\eta}
  \sum_{j=1}^{N_\eta}
  \frac{
  \operatorname{Tr}_{\chi}
  \left[
  \sigma_{\chi,j}(t_b;t_f)
   \rho_\chi(t_f)
  \right]
  }{
  \mathcal P_\chi(t_f)
  }
  \\
  &=
  C_{11}^{-+}(t_b;t_f)
  =
  2i\,
  G^{(2),\eta;\chi\text{-tw},>}_{RR;11}
  (t_b,t_b;t_f).
\end{aligned}
\label{eq:S5-K2}
\end{equation}

To retain the coherence of the infalling excitation, introduce an auxiliary
two-dimensional reference system \(P\) and purify the choice of whether or
not the simple fermion was inserted:
\begin{equation}
  \lvert\Omega_j(t_f)\rangle_{P\eta\chi}
  =
  \frac{1}{\sqrt{2}}
  \left[
  \lvert0\rangle_P
  \lvert\Phi(t_f)\rangle_{\eta\chi}
  +
  \lvert1\rangle_P
  \lvert\Phi_{b,j}(t_f)\rangle_{\eta\chi}
  \right].
\label{eq:S5-reference-state}
\end{equation}
Since \(\eta^R_j(t_b)\) is fermion odd, the states
\(\lvert\Phi(t_f)\rangle\) and
\(\lvert\Phi_{b,j}(t_f)\rangle\) have opposite fermion parity and are
therefore orthogonal. It follows that the reference system is maximally
mixed:
\begin{equation}
  \rho_P
  =
  \frac{1}{2}
  \left(
  \lvert0\rangle\langle0\rvert
  +
  \lvert1\rangle\langle1\rvert
  \right),
  \qquad
  S_P^{(2)}
  =
  -\log\operatorname{Tr}\rho_P^2
  =
  \log2 .
\label{eq:S5-reference-entropy}
\end{equation}

The R\'enyi-2 mutual-information diagnostic between \(P\) and \(\eta\) is
\begin{equation}
  I^{(2)}(P:\eta)
  =
  S_P^{(2)}
  +
  S_\eta^{(2)}
  -
  S_{P\eta}^{(2)}.
\label{eq:S5-MI-definition}
\end{equation}
The full \(P\eta\chi\) state is pure, so
\(S_{P\eta}^{(2)}=S_\chi^{(2)}\). Therefore
\begin{equation}
\begin{aligned}
  e^{-I^{(2)}(P:\eta)}
  &=
  \frac{
  \operatorname{Tr}_{\eta}
  \left[
  \left(
  \rho_\eta+\sigma_{\eta,j}
  \right)^2
  \right]
  }{
  2\operatorname{Tr}_{\chi}
  \left[
  \left(
  \rho_\chi+\sigma_{\chi,j}
  \right)^2
  \right]
  }
  \\
  &=
  \frac{1}{2}
  \frac{
  1+2F_2+F_4
  }{
  1+2K_2+F_4
  }.
\end{aligned}
\label{eq:S5-MI-ratio}
\end{equation}
The first line is precisely the density-matrix expression quoted in
Eq.~\eqref{eq:renyi2_MI}. The second line uses flavor symmetry at the
leading disorder-averaged large-\(N\) saddle, where the fixed-flavor
contractions equal \(F_2,F_4,K_2\); it is not an identity between averages
of ratios in an arbitrary finite-\(N\) realization. In the ideal limit in which \(P\) and the retained
\(\eta\) information are maximally entangled, this diagnostic equals
\(2\log2\).

\paragraph*{Projection onto the reconstructed bulk mode.}

Equations~\eqref{eq:S5-F2-green}, \eqref{eq:S5-F4-green}, and
\eqref{eq:S5-K2} were written first for a simple boundary insertion. The
main-text result instead uses the reconstructed infalling mode
\(\psi_{R,f}(u,t_{\rm max})\). According to
Eq.~\eqref{eq:HKLL_kernel}, this operator is a linear combination of
boundary operators over the reconstruction interval.

Let \(\mathbf K_f(u,t_{\rm max})\) denote the corresponding discretized
reconstruction row, including the boundary-copy components used in the
reconstruction. For an insertion on the ket or upper branch and on the bra
or lower branch, respectively, use
\begin{equation}
  \mathbf K_f^{(+)}
  =
  \mathbf K_f,
  \qquad
  \mathbf K_f^{(-)}
  =
  \mathbf K_f^*.
\label{eq:S5-kernel-branch-rule}
\end{equation}
If
\(\mathbf C_{\alpha\beta}^{\lambda_1\lambda_2}\)
denotes the branch-resolved two-replica correlator matrix assembled from the
\(RR,RL,LR,LL\) boundary blocks, define its bulk projection by
\begin{equation}
  C_{\alpha\beta}^{\lambda_1\lambda_2}
  [\mathbf K_f]
  =
  \mathbf K_f^{(\lambda_1)}
  \mathbf C_{\alpha\beta}^{\lambda_1\lambda_2}
  \left(
  \mathbf K_f^{(\lambda_2)}
  \right)^T .
\label{eq:S5-kernel-projection}
\end{equation}
A boundary insertion at \(t_b\) is recovered by taking
\(\mathbf K_f\) to be the corresponding delta-function row. For the
reconstructed bulk mode, the quantities entering the mutual information are
\begin{equation}
\begin{aligned}
  F_2[\mathbf K_f]
  &=
  C_{12}^{-+}[\mathbf K_f],
  \\
  K_2[\mathbf K_f]
  &=
  C_{11}^{-+}[\mathbf K_f],
  \\
  F_4[\mathbf K_f]
  &=
  C_{11}^{-+}[\mathbf K_f]
  C_{22}^{-+}[\mathbf K_f]
  -
  C_{12}^{-+}[\mathbf K_f]
  C_{21}^{-+}[\mathbf K_f]
  -
  C_{12}^{--}[\mathbf K_f]
  C_{12}^{++}[\mathbf K_f].
\end{aligned}
\label{eq:S5-bulk-contractions}
\end{equation}
Substitution of these kernel-projected contractions into
Eq.~\eqref{eq:S5-MI-ratio} gives the bulk-mode R\'enyi-2 mutual information
shown in Fig.~\ref{fig:finite_p}.

\paragraph*{On-shell action and second R\'enyi entropy.}

The unperturbed two-replica saddles also determine the second R\'enyi entropy
of the full \(\eta\) sector. Let the statistical on-shell action be
\begin{equation}
  \mathcal I
  =
  -i\,
  \mathcal S_{\mathrm{2rep,on\mbox{-}shell}},
\label{eq:S5-statistical-action}
\end{equation}
where the contour action includes both Euclidean preparation segments and
both real-time branches. The contour weights in
Eq.~\eqref{eq:S5-contour-weights} already include all branch orientations.

The purity is the ratio of the \(\chi\)-twisted partition function to the
disconnected two-replica partition function:
\begin{equation}
  \operatorname{Tr}\rho_\eta(t_f)^2
  =
  \frac{
  Z_{\chi\text{-tw}}(t_f)
  }{
  Z_{\rm disc}(t_f)
  }
  \sim
  \exp
  \left[
  -(N_\eta+N_\chi)
  \left(
  \mathcal I_{\chi\text{-tw}}
  -
  \mathcal I_{\rm disc}
  \right)
  \right].
\label{eq:S5-purity-action-ratio}
\end{equation}
In the probe limit the statistical action has the expansion
\begin{equation}
  \mathcal I
  =
  \mathcal I_\chi
  +
  s\,i_\eta
  +
  O(s^2),
  \qquad
  s
  =
  \frac{N_\eta}{N_\eta+N_\chi}.
\label{eq:S5-probe-action-expansion}
\end{equation}
The bath saddle is common to the twisted and disconnected solutions at this
order, so its contribution cancels. Define
\begin{equation}
  \Delta i_\eta(t_f)
  =
  i_{\eta,\chi\text{-tw}}(t_f)
  -
  i_{\eta,\rm disc}(t_f).
\label{eq:S5-delta-action}
\end{equation}
The entropy normalization is then
\begin{equation}
  \frac{
  S_\eta^{(2)}(t_f)
  }{
  N_\eta
  }
  =
  \operatorname{Re}\Delta i_\eta(t_f),
  \qquad
  \frac{
  S_\eta^{(2)}(t_f)
  }{
  N_\eta+N_\chi
  }
  =
  s\,
  \operatorname{Re}\Delta i_\eta(t_f).
\label{eq:S5-entropy-normalization}
\end{equation}
Here
\(S_\eta^{(2)}(t_f)
=-\log\operatorname{Tr}\rho_\eta(t_f)^2\).
The imaginary part of \(\Delta i_\eta\) is a numerical convergence
diagnostic and should vanish for a converged physical saddle.

The doubled \(\eta\) system contains \(2N_\eta\) Majorana fermions and has
Hilbert-space dimension \(2^{N_\eta}\). Consequently,
\begin{equation}
  0
  \leq
  \frac{S_\eta^{(2)}(t_f)}{N_\eta}
  =
  \operatorname{Re}\Delta i_\eta(t_f)
  \leq
  \log2.
\label{eq:S5-entropy-bound}
\end{equation}

For the \(\chi\)-twisted saddle, the post-evaporation source is the one in
Eq.~\eqref{eq:S5-discrete-sigma-III}. The disconnected saddle uses the same
scalar bath correlator but no replica mixing:
\begin{equation}
  \left(
  \Sigma^{\eta,\rm III}_{\rm disc,ab}
  \right)_{\alpha\beta;ij}
  =
  M_{\rm III}(i,j)
  (-1)^{1+p\delta_{ab}/2}
  \frac{\mathcal J^2}{p}
  \left[
  2G^\chi_{ab;ij}
  \right]^{p-1}
  \delta_{\alpha\beta}.
\label{eq:S5-disconnected-source}
\end{equation}
For the convention used in the plots, \(p/2\) is even, so
\((-1)^{1+p\delta_{ab}/2}=-1\) in every \(R,L\) block. We also use
\(\mathbf V_\eta=0\).

The free determinant is identical for the twisted and disconnected saddles.
Writing
\[
  G_{0,2}^{-1}-\Sigma^\eta
  =
  G_{0,2}^{-1}
  \left(
  \mathbf 1-G_{0,2}\Sigma^\eta
  \right)
\]
makes this cancellation explicit. The UV-subtracted discrete
log-determinant contribution is
\begin{equation}
\begin{aligned}
  i_{\eta,\log}
  &=
  -\frac{1}{2}
  \log\det
  \left[
  \mathbf 1
  -
  \mathbf G_{0,2}
  \mathbf W_2
  \left(
  \boldsymbol\Sigma^{\eta,\rm I}
  +
  \boldsymbol\Sigma^{\eta,\rm III}
  \right)
  \mathbf W_2
  \right].
\end{aligned}
\label{eq:S5-relative-logdet}
\end{equation}
Both contour arguments of a nonlocal self-energy carry a quadrature weight.

At leading order in the probe expansion, the explicit region-III
\(\Sigma G\) contribution cancels the region-III mixed potential. The
post-evaporation self-energy nevertheless remains in the determinant and in
the saddle equation. The surviving pre-evaporation interaction contribution
is
\begin{equation}
\begin{aligned}
  i_{\eta,\rm I}
  &=
  \frac{a^2\mathcal J^2}{4p^2}
  \sum_{\alpha,\beta=1}^{2}
  \sum_{a,b=R,L}
  \sum_{i,j}
  w_iw_j\,
  M_{\rm I}(i,j)
  (-1)^{1+p\delta_{ab}/2}
  \left[
  2(G^\eta_{ab})_{\alpha\beta;ij}
  \right]^p .
\end{aligned}
\label{eq:S5-early-interaction-action}
\end{equation}
The corresponding constraint contribution is
\begin{equation}
\begin{aligned}
  i_{\Sigma_{\rm I}G}
  &=
  \frac{1}{2}
  \sum_{\alpha,\beta=1}^{2}
  \sum_{a,b=R,L}
  \sum_{i,j}
  w_iw_j\,
  \left(
  \Sigma^{\eta,\rm I}_{ab}
  \right)_{\alpha\beta;ij}
  \left(
  G^\eta_{ab}
  \right)_{\alpha\beta;ij}.
\end{aligned}
\label{eq:S5-SigmaG-action}
\end{equation}
This is an elementwise bilocal contraction; it should not be replaced by an
ordinary matrix trace, which reverses one of the matrix indices.

The \(\eta\)-normalized probe action evaluated on either saddle is therefore
\begin{equation}
  i_\eta
  =
  i_{\eta,\log}
  +
  i_{\Sigma_{\rm I}G}
  -
  i_{\eta,\rm I}.
\label{eq:S5-probe-action}
\end{equation}
For each \(t_f\), the required action difference is
\begin{equation}
\begin{aligned}
  \Delta i_\eta(t_f)
  &=
  i_\eta
  \left[
  G^\eta_{\chi\text{-tw}};
  \Sigma^{\eta,\rm III}_{\chi\text{-tw}}
  \right]
  -
  i_\eta
  \left[
  G^\eta_{\rm disc};
  \Sigma^{\eta,\rm III}_{\rm disc}
  \right].
\end{aligned}
\label{eq:S5-final-action-difference}
\end{equation}

Finally, at fixed contour extent and fixed discretization,
\(\Sigma^\eta=O(p^{-1})\) gives the formal large-\(p\) expansion of the
relative determinant (the series converges when the spectral radius of
\(\mathbf A\) is less than one):
\begin{equation}
  -\frac{1}{2}\log\det(\mathbf 1-\mathbf A)
  =
  \frac{1}{2}\operatorname{Tr}\mathbf A
  +
  \frac{1}{4}\operatorname{Tr}\mathbf A^2
  +
  \cdots,
  \qquad
  \mathbf A
  =
  \mathbf G_{0,2}
  \mathbf W_2
  \boldsymbol\Sigma^\eta
  \mathbf W_2 .
\label{eq:S5-logdet-expansion}
\end{equation}
Accordingly,
\begin{equation}
  \Delta i_\eta(t_f)
  =
  \frac{c_1(t_f)}{p}
  +
  \frac{c_2(t_f)}{p^2}
  +
  O(p^{-3}),
\label{eq:S5-large-p-action-scaling}
\end{equation}
unless an additional symmetry removes the leading coefficient. At fixed final time, this gives
the large-\(p\) scaling
\(S_\eta^{(2)}=O(N_\eta/p)\) quoted in the main text. The numerical
consistency checks are
\begin{equation}
  \operatorname{Re}\Delta i_\eta\geq0,
  \qquad
  \operatorname{Im}\Delta i_\eta\simeq0,
  \qquad
  \Delta i_\eta(t_f\leq t_{\rm ev})=0,
  \qquad
  \operatorname{Re}\Delta i_\eta\leq\log2.
\label{eq:S5-action-checks}
\end{equation}

\section*{S6. Large-\(p\) conventions, Lensky-Qi parametrization, and the left-right bound}
\paragraph*{Reflection symmetry and the large-\(p\) ansatz.}

We first collect the symmetry, normalization, and branch conventions used in
Secs.~S2-S5. For either species
\(\psi\in\{\eta,\chi\}\), and for \(p/2\) even, the doubled Hamiltonian is invariant under
\begin{equation}
\mathbf R[\psi_R]=-\psi_L,
\qquad
\mathbf R[\psi_L]=\psi_R .
\label{eq:S6-reflection}
\end{equation}
The bilinear \(\psi_L\psi_R\) is invariant under the same transformation.

For real times, define
\begin{equation}
G^{\psi,>}_{ab}(t_1,t_2)
=
-i\left\langle
\psi_a(t_1)\psi_b(t_2)
\right\rangle,
\qquad
a,b\in\{R,L\}.
\label{eq:S6-Wightman-definition}
\end{equation}
Hermiticity and reflection symmetry imply
\begin{equation}
\begin{aligned}
\left[
G^{\psi,>}_{ab}(t_1,t_2)
\right]^*
&=
-G^{\psi,>}_{ba}(t_2,t_1),
\\
G^{\psi,>}_{LL}(t_1,t_2)
&=
G^{\psi,>}_{RR}(t_1,t_2),
\qquad
G^{\psi,>}_{LR}(t_1,t_2)
=
-G^{\psi,>}_{RL}(t_1,t_2).
\end{aligned}
\label{eq:S6-Wightman-symmetries}
\end{equation}
Consequently,
\begin{equation}
\begin{aligned}
G^{\psi,>}_{RR}(t_1,t_2)
&=
-\left[
G^{\psi,>}_{RR}(t_2,t_1)
\right]^*,
\\
G^{\psi,>}_{RL}(t_1,t_2)
&=
\left[
G^{\psi,>}_{RL}(t_2,t_1)
\right]^* .
\end{aligned}
\label{eq:S6-independent-components}
\end{equation}
For complex contour times, Hermiticity becomes
\begin{equation}
\left[
G^{\psi,>}_{ab}(z_1,z_2)
\right]^*
=
-G^{\psi,>}_{ba}(z_2^*,z_1^*) .
\label{eq:S6-complex-Wightman-conjugation}
\end{equation}

The normalization and form of the large-\(p\) ansatz can be motivated from the free
bilinear problem
\begin{equation}
\label{eq:app-free-bilinear-hamiltonian}
  H_{\mathrm{bil}}=i\mu \psi_L\psi_R .
\end{equation}
Here \(\mu\) is the physical single-pair gap; in the large-\(p\) scaling of the model it is replaced by \(\mu_\psi/p\).
With
\begin{equation}
\label{eq:app-free-c}
  c=\frac{\psi_L+i\psi_R}{\sqrt{2}},
  \qquad
  c^\dagger=\frac{\psi_L-i\psi_R}{\sqrt{2}},
\end{equation}
one has
\begin{equation}
\label{eq:app-free-spectrum}
  H_{\mathrm{bil}}
  =
  \mu\left(c^\dagger c-\frac{1}{2}\right).
\end{equation}
In a thermal state \(\rho=e^{-\beta H_{\mathrm{bil}}}/Z_\beta\),
\begin{equation}
\label{eq:app-free-GRR}
  G^{>}_{RR}(t,0)
  =
  -\frac{i}{2}
  \frac{
  \cosh\left[\frac{\mu}{2}(\beta-2it)\right]
  }{
  \cosh\left(\frac{\beta\mu}{2}\right)
  }
  \xrightarrow{\beta\mu\to\infty}
  -\frac{i}{2}e^{-i\mu t},
\end{equation}
and
\begin{equation}
\label{eq:app-free-GRL}
  G^{>}_{RL}(t,0)
  =
  -\frac{1}{2}
  \frac{
  \sinh\left[\frac{\mu}{2}(\beta-2it)\right]
  }{
  \cosh\left(\frac{\beta\mu}{2}\right)
  }
  \xrightarrow{\beta\mu\to\infty}
  -\frac{1}{2}e^{-i\mu t}.
\end{equation}
This motivates the large-\(p\) parametrization
\begin{equation}
\begin{aligned}
G^{\psi,>}_{RR}(t_1,t_2)
&=
-\frac{i}{2}
e^{g^\psi_R(t_1,t_2)/p}
=
G^{\psi,>}_{LL}(t_1,t_2),
\\
G^{\psi,>}_{RL}(t_1,t_2)
&=
-\frac{1}{2}
e^{g^\psi_L(t_1,t_2)/p}
=
-G^{\psi,>}_{LR}(t_1,t_2).
\end{aligned}
\label{eq:S6-large-p-components}
\end{equation}
Thus the subscript \(R\) on \(g^\psi_R\) denotes the diagonal
\(RR/LL\) channel, whereas \(L\) on \(g^\psi_L\) denotes the off-diagonal
\(RL/LR\) channel.

We choose the continuous logarithmic branch for which
\(g^\psi_{R,L}=O(1)\) as \(p\to\infty\). Equations
\eqref{eq:S6-independent-components} and
\eqref{eq:S6-large-p-components} then give
\begin{equation}
g^\psi_X(t_1,t_2)
=
\left[
g^\psi_X(t_2,t_1)
\right]^*,
\qquad
X=R,L .
\label{eq:S6-real-time-g-conjugation}
\end{equation}
On the complex contour,
\begin{equation}
\left[
g^\psi_X(z_1,z_2)
\right]^*
=
g^\psi_X(z_2^*,z_1^*),
\qquad
X=R,L .
\label{eq:S6-complex-time-g-conjugation}
\end{equation}
In particular,
\begin{equation}
g^\psi_R(t,t)=0,
\qquad
g^\psi_L(t,t)\in\mathbb R .
\label{eq:S6-equal-time-g}
\end{equation}
These relations fix the conjugation and branch conventions used in the
matching formulas of Sec.~S3.

\paragraph*{Lensky-Qi parametrization.}

Consider a real-time region in which the correlators satisfy the homogeneous
Liouville equations with effective interaction scale
\(\mathcal J_{\mathrm{eff}}\),
\begin{equation}
\partial_{t_1}\partial_{t_2}g_X(t_1,t_2)
=
2\sigma_X\mathcal J_{\mathrm{eff}}^{\,2}
e^{g_X(t_1,t_2)},
\qquad
\sigma_R=1,
\qquad
\sigma_L=-1 .
\label{eq:S6-LQ-Liouville}
\end{equation}
For the bath,
\(\mathcal J_{\mathrm{eff}}=\mathcal J\), whereas for the
pre-evaporation \(\eta\) solution,
\(\mathcal J_{\mathrm{eff}}=a\mathcal J\).

Following Lensky and Qi \cite{lensky2021rescuing}, the physical solution
selected by the thermal gluing conditions
\eqref{eq:S-large-p-gluing} and the equal-time data
\eqref{eq:S-equal-time-R} and \eqref{eq:S-equal-time-L} may be described by
a single complex function \(y(t)\):
\begin{equation}
\begin{aligned}
e^{g_R(t_1,t_2)}
&=
-
\frac{
y'(t_1)y'(t_2)^*
}{
\sin^2\left(
\mathcal J_{\mathrm{eff}}
\left[y(t_1)-y(t_2)^*\right]
\right)
},
\\
e^{g_L(t_1,t_2)}
&=
\frac{
y'(t_1)y'(t_2)^*
}{
\cos^2\left(
\mathcal J_{\mathrm{eff}}
\left[y(t_1)-y(t_2)^*\right]
\right)
}.
\end{aligned}
\label{eq:S6-LQ-parametrization}
\end{equation}
The variable denoted by \(y\) here is denoted by \(\psi\) in
Ref.~\cite{lensky2021rescuing}; we use \(y\) to avoid confusion with the
species label \(\psi\in\{\eta,\chi\}\). Its dimensions are
\begin{equation}
[y]=[t]=[\mathcal J_{\mathrm{eff}}]^{-1}.
\label{eq:S6-y-dimension}
\end{equation}

We choose the physical lower-half-plane branch
\begin{equation}
\operatorname{Im}y(t)<0 .
\label{eq:S6-physical-branch}
\end{equation}
The equal-time normalization \(g_R(t,t)=0\) then requires
\begin{equation}
|y'(t)|
=
\sinh\left(
2\mathcal J_{\mathrm{eff}}
|\operatorname{Im}y(t)|
\right).
\label{eq:S6-y-normalization}
\end{equation}

It is useful to introduce real variables \(\phi(t)\) and \(p(t)\) by
\begin{equation}
y'(t)
=
|y'(t)|e^{ip(t)}
=
\frac{
e^{ip(t)}
}{
\sqrt{e^{2\phi(t)}-1}
}.
\label{eq:S6-y-phase-parametrization}
\end{equation}
Here \(p(t)\) is the Lensky--Qi canonical phase conjugate to \(\phi(t)\); it
should not be confused with the interaction order \(p\). The equal-time
left-right correlator is
\begin{equation}
g_L(t,t)=-2\phi(t).
\label{eq:S6-gL-phi}
\end{equation}

For a real bilinear coupling \(\mu_{\mathrm{eff}}(t)\), the variables
\((\phi,p)\) evolve according to the classical Hamiltonian
\begin{equation}
\mathcal H_Q(\phi,p;t)
=
-2\mathcal J_{\mathrm{eff}}
\sqrt{1-e^{-2\phi}}\cos p
+
\mu_{\mathrm{eff}}(t)\phi,
\qquad
\{\phi,p\}_{\mathrm{P.B.}}=1 .
\label{eq:S6-LQ-Hamiltonian}
\end{equation}
The corresponding Hamilton equations are
\begin{equation}
\begin{aligned}
\dot\phi
&=
2\mathcal J_{\mathrm{eff}}
\sqrt{1-e^{-2\phi}}\sin p,
\\
\dot p
&=
-\mu_{\mathrm{eff}}(t)
+
2\mathcal J_{\mathrm{eff}}
\frac{
e^{-2\phi}
}{
\sqrt{1-e^{-2\phi}}
}
\cos p .
\end{aligned}
\label{eq:S6-LQ-Hamilton-equations}
\end{equation}

\paragraph*{Lensky-Qi trajectories used in Secs.~S3 and S4.}

For the pre-evaporation \(\eta\) segment,
\(\mathcal J_{\mathrm{eff}}=a\mathcal J\) and
\(\mu_{\mathrm{eff}}=0\). In terms of the thermal angle \(\vartheta\) defined
in Eq.~\eqref{eq:S3-thermal-angle}, the corresponding trajectory is
\begin{equation}
\begin{aligned}
\phi_\eta(t)
&=
\log\left[
\frac{
\cosh\left(
2a\mathcal Jt\sin\vartheta
\right)
}{
\sin\vartheta
}
\right],
\\
p_\eta(t)
&=
\tan^{-1}\left[
\tan\vartheta\,
\tanh\left(
2a\mathcal Jt\sin\vartheta
\right)
\right],
\\
y_\eta(t)
&=
\frac{1}{a\mathcal J}
\tan^{-1}
\tanh\left(
a\mathcal Jt\sin\vartheta
-\frac{i\vartheta}{2}
\right).
\end{aligned}
\label{eq:S6-eta-trajectory}
\end{equation}
The inverse functions and logarithms in this expression are taken on their
continuous branches. Substitution into
Eq.~\eqref{eq:S6-LQ-parametrization} reproduces the pre-evaporation
correlators in Eq.~\eqref{eq:S3-eta-pre}.

For the stationary traversable-wormhole bath,
\(\mathcal J_{\mathrm{eff}}=\mathcal J\) and
\(\mu_{\mathrm{eff}}=\mu_0\). It is the fixed point
\begin{equation}
\phi_\chi(t)=\phi_G,
\qquad
p_\chi(t)=0,
\label{eq:S6-chi-fixed-point}
\end{equation}
where \(\phi_G\) and \(V_G\) are defined in
Eq.~\eqref{eq:S3-bath-parameters}. A convenient representative of the
corresponding \(y\)-trajectory is
\begin{equation}
y_\chi(t)
=
V_G(t-t_\star)
-
\frac{i}{2\mathcal J}
\tanh^{-1}(r_G),
\qquad
V_G
=
\frac{r_G}{\sqrt{1-r_G^2}},
\label{eq:S6-chi-trajectory}
\end{equation}
where \(t_\star\) is an arbitrary time origin. Substitution into
Eq.~\eqref{eq:S6-LQ-parametrization} reproduces the stationary bath
correlators in Eq.~\eqref{eq:S3-bath-solution}.

\paragraph*{Disk-coordinate proof of the post-evaporation bound.}

We now prove Eq.~\eqref{eq:S3-post-L-ratio-bound} under the assumptions of
Eq.~\eqref{eq:S3-LQ-bound-assumptions}, with \(h_\chi\) defined in
Eq.~\eqref{eq:S3-post-L-ratio}. In this paragraph
\(g_{R,L}=g^\chi_{R,L}\), and therefore
\(\mathcal J_{\mathrm{eff}}=\mathcal J\).

Introduce the disk coordinate
\begin{equation}
w(t)
=
e^{-2i\mathcal J y(t)}.
\label{eq:S6-disk-coordinate}
\end{equation}
Because \(\operatorname{Im}y(t)<0\),
\begin{equation}
|w(t)|
=
e^{2\mathcal J\operatorname{Im}y(t)}
<1 .
\label{eq:S6-disk-interior}
\end{equation}
Equation~\eqref{eq:S6-y-normalization} becomes
\begin{equation}
|y'(t)|
=
\frac{
1-|w(t)|^2
}{
2|w(t)|
}.
\label{eq:S6-yprime-disk}
\end{equation}

For \(w_a=w(t_a)\) and \(w_b=w(t_b)\),
\begin{equation}
\begin{aligned}
\left|
\sin\left(
\mathcal J[y(t_a)-y(t_b)^*]
\right)
\right|^2
&=
\frac{
|1-w_aw_b^*|^2
}{
4|w_a||w_b|
},
\\
\left|
\cos\left(
\mathcal J[y(t_a)-y(t_b)^*]
\right)
\right|^2
&=
\frac{
|1+w_aw_b^*|^2
}{
4|w_a||w_b|
}.
\end{aligned}
\label{eq:S6-disk-trigonometric-identities}
\end{equation}
Define the disk invariant
\begin{equation}
\mathcal P(u,v)
\equiv
\frac{
(1-|u|^2)(1-|v|^2)
}{
|1-uv^*|^2
}.
\label{eq:S6-disk-kernel}
\end{equation}
It is invariant under simultaneous automorphisms of the unit disk.
Equations~\eqref{eq:S6-LQ-parametrization},
\eqref{eq:S6-yprime-disk}, and
\eqref{eq:S6-disk-trigonometric-identities} give
\begin{equation}
\left|e^{g^\chi_R(t_a,t_b)}\right|
=
\mathcal P(w_a,w_b),
\qquad
\left|e^{g^\chi_L(t_a,t_b)}\right|
=
\mathcal P(w_a,-w_b).
\label{eq:S6-correlators-disk}
\end{equation}

Let
\begin{equation}
w_i=w(t_i),
\qquad
w_{\rm ev}=w(t_{\rm ev}).
\end{equation}
Substitution into Eq.~\eqref{eq:S3-post-L-ratio} gives
\begin{equation}
h_\chi
=\mathcal P(w_1,-w_2)
\mathcal P(w_{\rm ev},-w_{\rm ev})
\frac{\mathcal P(w_1,w_{\rm ev})
\mathcal P(w_2,w_{\rm ev})
}{
\mathcal P(w_1,-w_{\rm ev})
\mathcal P(w_2,-w_{\rm ev})
}.
\label{eq:S6-h-disk}
\end{equation}
Here and below the arguments
\((t_1,t_2;t_{\rm ev})\) of \(h_\chi\) are suppressed.

Apply the disk automorphism
\begin{equation}
z(w)
=
\frac{
w-w_{\rm ev}
}{
1-w_{\rm ev}^*w
}.
\label{eq:S6-disk-automorphism}
\end{equation}
It obeys
\begin{equation}
z(w_{\rm ev})=0,
\qquad
z(-w_{\rm ev})=-v,
\qquad
z(-w)
=
-\frac{
z(w)+v
}{
1+v^*z(w)
}, \qquad {\text{where~}}
v
=
\frac{
2w_{\rm ev}
}{
1+|w_{\rm ev}|^2
}.
\label{eq:S6-disk-images}
\end{equation}
Because \(|w_{\rm ev}|<1\), one has \(|v|<1\).

Writing
\begin{equation}
z_j=z(w_j),
\qquad
z'_j=z(-w_j),
\end{equation}
the invariance of \(\mathcal P\) reduces Eq.~\eqref{eq:S6-h-disk} to
\begin{equation}
h_\chi
=
\frac{
\mathcal P(z_1,0)\mathcal P(z_2,0)
}{
\mathcal P(z'_1,0)\mathcal P(z'_2,0)
}
\mathcal P(z_1,z'_2)\mathcal P(v,0).
\label{eq:S6-h-transformed}
\end{equation}
Using Eq.~\eqref{eq:S6-disk-images} explicitly gives
\begin{equation}
h_\chi
=
\frac{
(1-|z_1|^2)(1-|z_2|^2)
|1+z_1v^*|^2
|1+z_2v^*|^2
}{
|1+z_1v^*+z_2^*v+z_1z_2^*|^2
}.
\label{eq:S6-h-z}
\end{equation}

Remove the phase of \(v\) by writing
\begin{equation}
v=\rho e^{i\alpha_0},
\qquad
\rho=|v|<1,
\qquad
\tau_j=e^{-i\alpha_0}z_j ,
\label{eq:S6-phase-rotation}
\end{equation}
with an arbitrary choice of \(\alpha_0\) if \(v=0\). Then
\begin{equation}
h_\chi
=
\frac{
(1-|\tau_1|^2)(1-|\tau_2|^2)
|1+\rho\tau_1|^2
|1+\rho\tau_2|^2
}{
\left|
1+\rho\tau_1+\rho\tau_2^*
+\tau_1\tau_2^*
\right|^2
},
\qquad
|\tau_1|,|\tau_2|<1 .
\label{eq:S6-h-canonical-disk-form}
\end{equation}
Notice that \(h_\chi\) is already a nonnegative modulus; the ratio in
Eq.~\eqref{eq:S6-h-canonical-disk-form} is \(h_\chi\), not \(h_\chi^2\).

To bound this expression, define
\begin{equation}
\kappa
=
\frac{
\rho+\tau_1
}{
1+\rho\tau_1
}.
\label{eq:S6-CS-kappa}
\end{equation}
Since \(\rho\) is real,
\begin{equation}
1-|\kappa|^2
=
\frac{
(1-\rho^2)(1-|\tau_1|^2)
}{
|1+\rho\tau_1|^2
}
>0,
\label{eq:S6-CS-kappa-disk}
\end{equation}
so \(\kappa\) lies in the open unit disk. Moreover,
\begin{equation}
1+\rho\tau_1+\rho\tau_2^*
+\tau_1\tau_2^*
=
(1+\rho\tau_1)
(1+\kappa\tau_2^*).
\label{eq:S6-CS-factorization}
\end{equation}
Consequently,
\begin{equation}
h_\chi
=
(1-|\tau_2|^2)
\left|F(\tau_2)\right|^2,
\label{eq:S6-CS-h-F}
\end{equation}
where
\begin{equation}
F(\lambda)
=
\sqrt{1-|\tau_1|^2}\,
\frac{
1+\rho\lambda
}{
1+\kappa^*\lambda
}.
\label{eq:S6-CS-function}
\end{equation}
Because \(|\kappa|<1\), \(F\) is holomorphic on the unit disk. Its Taylor
series is
\begin{equation}
F(\lambda)
=
\sqrt{1-|\tau_1|^2}
\left[
1+
(\rho-\kappa^*)
\sum_{n=1}^{\infty}
(-\kappa^*)^{n-1}\lambda^n
\right].
\label{eq:S6-CS-series}
\end{equation}
Writing
\begin{equation}
F(\lambda)=\sum_{n=0}^{\infty}d_n\lambda^n,
\end{equation}
the coefficients are
\begin{equation}
d_0
=
\sqrt{1-|\tau_1|^2},
\qquad
d_n
=
\sqrt{1-|\tau_1|^2}
(\rho-\kappa^*)
(-\kappa^*)^{n-1},
\quad n\geq1 .
\label{eq:S6-CS-coefficients}
\end{equation}

The associated Hardy norm is therefore finite and equals
\begin{equation}
\begin{aligned}
\|F\|_{H^2}^2
&\equiv
\sum_{n=0}^{\infty}|d_n|^2
\\
&=
(1-|\tau_1|^2)
\left[
1+
\frac{
|\rho-\kappa^*|^2
}{
1-|\kappa|^2
}
\right].
\end{aligned}
\label{eq:S6-CS-Hardy-norm}
\end{equation}
Since \(\rho\) is real,
\begin{equation}
\rho-\kappa
=
-\frac{
(1-\rho^2)\tau_1
}{
1+\rho\tau_1
},
\qquad
|\rho-\kappa^*|
=
|\rho-\kappa|.
\label{eq:S6-CS-rho-minus-kappa}
\end{equation}
Combining Eqs.~\eqref{eq:S6-CS-kappa-disk} and
\eqref{eq:S6-CS-rho-minus-kappa} gives
\begin{equation}
\|F\|_{H^2}^2
=
1-\rho^2|\tau_1|^2
\leq1 .
\label{eq:S6-CS-Hardy-norm-final}
\end{equation}

Because \(F\in H^2\), Cauchy--Schwarz applied to its Taylor series gives
\begin{equation}
\begin{aligned}
|F(\tau_2)|^2
&=
\left|
\sum_{n=0}^{\infty}
d_n\tau_2^n
\right|^2
\\
&\leq
\left(
\sum_{n=0}^{\infty}|d_n|^2
\right)
\left(
\sum_{n=0}^{\infty}|\tau_2|^{2n}
\right)
\\
&=
\frac{
\|F\|_{H^2}^2
}{
1-|\tau_2|^2
}.
\end{aligned}
\label{eq:S6-CS-evaluation}
\end{equation}
Using Eq.~\eqref{eq:S6-CS-h-F}, we obtain
\begin{equation}
h_\chi
\leq
\|F\|_{H^2}^2
=
1-\rho^2|\tau_1|^2
\leq1 .
\label{eq:S6-CS-first-bound}
\end{equation}
The canonical expression
\eqref{eq:S6-h-canonical-disk-form} is invariant under complex conjugation
combined with exchange of the two insertion points. Applying the same
argument after this exchange gives
\begin{equation}
h_\chi
\leq
1-\rho^2|\tau_2|^2
\leq1 .
\label{eq:S6-CS-second-bound}
\end{equation}
Since \(h_\chi\) is nonnegative by definition,
\begin{equation}
0
\leq
h_\chi(t_1,t_2;t_{\rm ev})
\leq
1 .
\label{eq:S6-h-bound}
\end{equation}
This proves Eq.~\eqref{eq:S3-post-L-ratio-bound}. Together with
Eq.~\eqref{eq:S3-post-L-magnitude}, it establishes
Eq.~\eqref{eq:S3-post-L-bound}.

\end{document}